\documentclass[trackchanges]{aastex7}

\submitjournal{AJ}
\accepted{December 22, 2025}

\shorttitle{Dark matter in LTGs}
\shortauthors{Nigoche-Netro et al.}
\graphicspath{{./}{img/}}

\begin{document}\label{firstpage}


\title{Tracing Galaxy Evolution in the Nearby Universe: The Role of Dark Matter}
\correspondingauthor{A. Nigoche-Netro}
\email{alberto.nigoche@academicos.udg.mx}

\author[orcid=0000-0002-8683-0982]{A. Nigoche-Netro}
\affiliation{Instituto de Astronom\'ia y Meteorolog\'ia,
CUCEI, Universidad de Guadalajara, 
Guadalajara, Jal. 44130, M\'exico.}
\email{alberto.nigoche@academicos.udg.mx}

\author[orcid=0000-0002-2321-8657]{P. Lagos}
\affiliation{Instituto de Astrof\'isica e Ci\^encias do
Espa\c{c}o, Universidade do Porto, CAUP, 
Rua das Estrelas, 4150-762 Porto, Portugal.}
\affiliation{Institute of Astrophysics, Facultad de Ciencias Exactas, 
Universidad Andr\'es Bello, 
Sede Concepci\'on, Talcahuano, Chile.}
\email{Patricio.Lagos@astro.up.pt}

\author[orcid=0000-0001-9716-5335]{R. J. Diaz}
\affiliation{Gemini Observatory, NSF NOIRLab,
950 N Cherry Ave, Tucson, AZ 85719, USA.}
\affiliation{Universidad Nacional de C\'ordoba
Laprida 854, C\'ordoba, X5000BGR, Argentina.}
\email{rdiaz@gemini.edu}

\author[orcid=0000-0001-9643-4134]{E. de la Fuente}
\affiliation{Departamento de F\'isica, CUCEI,
Universidad de Guadalajara, Guadalajara, Jal. 44130, M\'exico.}
\email{eduardo.delafuentea@academicos.udg.mx}

\author{M. P. Ag\"uero}
\affiliation{Universidad Nacional de C\'ordoba
Laprida 854, C\'ordoba, X5000BGR, Argentina.}
\affiliation{Consejo Nacional de Investigaciones Cient\'ificas
y T\'ecnicas (CONICET), Argentina.}
\email{mpaguero@unc.edu.ar}

\author[orcid=0000-0002-7800-1308]{A. Ruelas-Mayorga}
\affiliation{Instituto de Astronom\'ia, Universidad Nacional Aut\'onoma de M\'exico, Cd. Universitaria, M\'exico D.F. 04510, M\'exico.}
\email{rarm@astro.unam.mx}

\author{S. N. Kemp}
\affiliation{Instituto de Astronom\'ia y Meteorolog\'ia,
CUCEI, Universidad de Guadalajara, 
Guadalajara, Jal. 44130, M\'exico.}
\email{simon.kemp@academicos.udg.mx}

\author{R. A. Marquez-Lugo}
\affiliation{Instituto de Astronom\'ia y Meteorolog\'ia,
CUCEI, Universidad de Guadalajara, 
Guadalajara, Jal. 44130, M\'exico.}
\email{alejandro.marquez@academicos.udg.mx}

\author{R. Ibarra-Nuño}
\affiliation{Instituto de Astronom\'ia y Meteorolog\'ia,
CUCEI, Universidad de Guadalajara, 
Guadalajara, Jal. 44130, M\'exico.}
\email{rodolfoibarra776@gmail.com}


\begin{abstract}

Using a sample of $\sim$126{,}000 late-type galaxies (LTGs) from the Sloan Digital Sky Survey, we analyzed stellar mass as a function of the dynamical mass. Stellar masses are estimated using eight stellar population synthesis (SPS) models with constant initial mass functions (IMFs), while dynamical masses are derived from seven formulations based on Newtonian dynamics and virial equilibrium, incorporating both stellar and gas velocity dispersions. We account for key factors affecting dynamical mass estimation, including the inclination, color, concentration, and Sérsic index.

We find that the difference between dynamical and stellar mass ($\Delta \log \mathbf{M}$) ranges from nearly zero to $\sim$95\% of the dynamical mass, depending on mass and redshift. $\Delta \log \mathbf{M}$ appears to decrease with increasing redshift, but exhibits a saddlelike shape at low mass and low redshift—especially in disk-dominated LTGs—transitioning into a steep, linear trend at higher masses and redshifts. In the high-mass regime, the behavior resembles that of early-type galaxies.

The results suggest that late type galaxies of higher mass or located at relatively higher redshifts appear to be more baryon-dominated, with a greater proportion of baryonic matter compared to dark matter. This can be attributed to factors such as efficient baryon condensation in the compact, gas-rich galaxies of the relatively early Universe, less concentrated dark matter halos at higher redshifts, and dynamic gas flows that accumulate baryons at galaxy centers, reducing the influence of dark matter on their dynamics. 

Moreover, our results indicate that this evolution is not discrete but follows a continuous transition between morphological regimes. Dark matter within LTGs is at most equal to $\Delta \log \mathbf{M}$, depending on the impact of the IMF and SPS on stellar mass estimation. Although SPS-based stellar masses do not include the gas component, previous studies have shown that galaxies with log($\mathbf{M_{Stellar}/M_{Solar}}) > 10$ at $z \leq 0.3$ are predominantly stellar-mass dominated. Most galaxies in our sample fall within this regime, minimizing the impact of gas exclusion. Our findings go beyond the scope of individual galaxies, providing insight into the nearby Universe and highlighting the role of dark matter in determining galaxies' structure and evolution.

\end{abstract}






\section{Introduction}
\label{intro}

For the past five decades, the determination of galaxy masses has stood as a focal point in astrophysics, as evidenced by seminal works such as those by \citet[]{Burbidge1975,Fich1991}, and recently by \citet{Courteauetal2014}, and \citet{Sch24}, among numerous others. Over this period, the field has evolved significantly, and at present, three primary methods have emerged as the cornerstone for calculating galaxy masses. Each of these methods presents distinct technical challenges and advantages, requiring a multifaceted approach to the understanding of galactic mass determination. The following describes the specifics of each method.

\begin{itemize}

\item Dynamical Masses: Dynamical masses are calculated using dynamic techniques that rely on rotation curves or the velocity dispersion of gas or stars within galaxies \citep{Nigoche22}. It is important to note that these methods assume the validity of Newtonian gravity at galactic scales. Determinations of dynamical masses for late-type galaxies (LTGs) are primarily derived from their rotation curves. Early attempts to determine galaxy masses using this method were made by \citet{Scheiner1899}, \citet{Slipher1914}, \citet{Pease1918}, and \citet{Opik1922}, where they inferred the mass of M31. Later, there were efforts to determine the mass of the Milky Way by \citet{KapteynvanRhijn1922}, \citet{Oort1932a}, and \citet{Oort1932b}. Determinations of M31's total mass using velocities derived from absorption lines were conducted by \citet{Babcock1939}, \citet {Mayall1951}, and \citet{Lallemand1960}. The flatness of rotation curves in various types of galaxies is a well-established phenomenon, as demonstrated by \citet[]{FaberGallagher1979,Rubinetal1985,SofueRubin2001}. Mass determinations of LTGs have also been made using emission lines such as H$_{\alpha}$, CO, and HI, with reasonable agreement among them.

\item Stellar, Luminous, and Baryonic Masses: In this work, we distinguish between stellar mass, luminous mass, and baryonic mass. The stellar mass refers to the mass contained in stars, whereas the luminous mass denotes the portion of the baryonic content directly traced by electromagnetic emission, predominantly from stars but potentially also from luminous gas components. The baryonic mass encompasses all baryonic components, including both luminous matter and nonluminous or weakly luminous constituents such as cold atomic and molecular gas, ionized gas, and stellar remnants.

The luminous mass of a galaxy can be inferred from its integrated luminosity in combination with an appropriate mass-to-light ratio (M/L). The M/L ratio is not universal; it varies among galaxies and depends on factors such as mass assembly history \citep{deLuciaetal2007}, spectral energy distribution \citep[SED,][]{Walcheretal2011,Conroy2013}, stellar population mix \citep{Oort1926,Baade1944,Van08,Con09,Mar09,GreggioRenzini2011,Mar11,Che12}, star formation rate \citep[SFR,][]{Sandage1986a,MacArthuretal2004}, and initial mass function \citep[IMF,][]{Salpeter1955,Scalo1986,Kroupa2001,Chabrier2003,Van08,cap12,Mar23}. Although determining the M/L ratio is a nontrivial task, once it is constrained, the luminous mass can be computed directly from the galaxy’s total luminosity and its distance.

The stellar mass is generally derived using stellar population synthesis (SPS) techniques, which model the integrated light of galaxies as the sum of contributions from stars of different ages and metallicities. SPS incorporates stellar evolutionary tracks from the main sequence to the remnant stage, stellar spectral libraries, empirical or theoretical dust attenuation models, and an assumed IMF \citep[and references therein]{Conroy2013}. 


It is important to note that both the luminous and stellar masses generally underestimate the total baryonic mass, as they neglect the contribution from nonstellar baryons stored in the interstellar medium and in other nonluminous or weakly luminous components (see Section~\ref{sec:mBaryonicMass}).

\item Mass Determination Using Relativistic Gravitational Lensing: This method provides a rough estimate of a single galaxy's mass. However, it usually requires a massive object like a galaxy cluster to gravitationally lens light. The first detection of weak gravitational lensing was achieved by \citet{Tysonetal1984} and \citet{BrainerdBlandfordS1996}. Strong gravitational lensing has also been observed, with its byproducts including measurements of galaxy cluster masses, by  \citet{Men17}, and \citet{Gho23} among others.

\end{itemize}

To apply the first two methods (dynamics, stellar), knowing the galaxy's distance is essential, and greater distances introduce more uncertainty into mass determination.

In the case of LTGs, these contain different structures (bulge, disk, spiral arms) with different levels of importance that make the proper calculation of dynamical mass difficult because the tracers of these structures are different. The best way to obtain the dynamical mass of LTGs is by using rotation curves. At low redshift, surveys such as SAMI (Sydney-AAO Multi-object Integral Field Spectrograph) and MaNGA (Mapping Nearby Galaxies at Apache Point Observatory) have provided valuable datasets. SAMI covers a redshift range of $0.004<z<0.095$ and includes approximately 3400 galaxies across diverse environments \citep{Cro12}, while MaNGA, part of the Sloan Digital Sky Survey (SDSS-IV), spans a redshift range of $0.01<z<0.15$ and contains a sample of around 10,000 galaxies \citep{Bun15}. At higher redshift, integral field units (IFUs) such as MUSE, KMOS, and SINFONI have made it possible to derive rotation curves up to $z \sim 3$, providing samples of order $\sim 1000$ galaxies~\citep{For06,Bac15,Wis15}. These surveys have opened a new window into the kinematics of distant LTGs. However, despite their scientific importance, the relatively small number of galaxies in these samples limits their use for statistical studies such as the one presented here, where large and homogeneous samples are required to minimize biases and obtain robust statistical inferences. In contrast, the Sloan Digital Sky Survey (SDSS) main sample contains more than one hundred thousand late-type galaxies (LTGs) within $0.00<z<0.35$, with both photometric and spectroscopic information. Although rotation curves are not available for this large sample, their dynamical masses can be estimated through the velocity dispersion of gas and/or stars. \citet{Nigoche22} exploited this information by calibrating dynamical masses derived from SDSS velocity dispersions with rotation curves from a smaller set of LTGs in the literature, thereby accounting for the different structural components (bulge, disk, arms) that characterize LTGs.

An important feature of the rotation curves of many spiral galaxies is that in their outer part (beyond where the bulk of the galaxy's luminosity is located), they are flat. This led to the idea, using Newton's mechanics, that there was much more matter than what could be photometrically detected \citep{Babcock1939}. This nonobservable matter was called dark matter. Currently, there is still no direct evidence of dark matter, and many particles have been proposed theoretically to justify it, including theories in which dark matter is dispensed with by modifying Newton's dynamics \citep[MOND,][]{Mil83,Mil88}. However, these theories, like MOND, can only explain very specific phenomena, so they cannot yet be considered valid 
\citep[e.g.][]{Cha23}.

According to the above, the total mass of a galaxy can be considered as the sum of baryonic mass and dark matter. Assuming that both baryonic matter and dark matter obey Newton’s law of universal gravitation, the total mass inferred from the galaxy’s dynamics can be referred to as the dynamical mass. The difference between the dynamical mass and the baryonic mass provides an estimate of the dark matter content.

In this work, we adopt the stellar mass as a proxy for the baryonic mass. This assumption simplifies the analysis but carries important implications: it underestimates the true baryonic mass because it neglects nonstellar baryonic components such as cold atomic and molecular gas, ionized gas, and stellar remnants. As a result, the inferred dark matter content may be correspondingly overestimated, particularly in gas-rich galaxies (details are provided in Section~\ref{sec:mBaryonicMass}).


Previous studies \citep{Nig15, Nig16, Nig19} have investigated the dependence of dark matter content in early-type galaxies (ETGs) on various parameters, including dynamical mass, redshift, and environment, reporting significant correlations with each. In the present work, we extend this framework to late-type galaxies (LTGs), incorporating a broader set of variables and accounting for potential biases that may influence dark matter estimates. Identifying these dependencies offers valuable insights into the formation histories and evolutionary pathways of galaxies across morphological types.

In order to ensure consistency with our previous studies and to enable direct comparisons with published results, we characterize the offset between dynamical and stellar masses using the logarithmic quantity

\begin{equation}
\Delta \log \mathbf{M} \equiv \log\left(\frac{\mathbf{M}_{\mathbf{Dyn}}}{\mathbf{M}_{\mathbf{Solar}}}\right) - \log\left(\frac{\mathbf{M}_{\mathbf{Stellar}}}{\mathbf{M}_{\mathbf{Solar}}}\right).
\end{equation}

This logarithmic mass difference offers several methodological advantages over the commonly used dark matter fraction. It compresses the dynamical range of the measurements, improving the visibility of systematic trends and reducing the influence of outliers. As a scale-independent and symmetric estimator, it enables consistent comparisons across galaxies spanning a wide range of masses, redshifts, and environments, and integrates naturally into log--log frameworks frequently employed in galaxy scaling relations. Furthermore, it is less sensitive to absolute uncertainties in mass estimates, particularly in regimes where $\mathbf{M_{\mathbf{Dyn}}} \sim \mathbf{M_{\mathbf{Stellar}}}$, and has served as a standard diagnostic in our previous work, ensuring methodological continuity.


By contrast, the dark matter fraction $f_{\mathrm{DM}}$, defined as

\begin{equation}
f_{\mathrm{DM}} = \frac{\mathbf{M}_{\mathbf{Dyn}} - \mathbf{M}_{\mathbf{Stellar}}}{\mathbf{M}_{\mathbf{Dyn}}},
\end{equation}

is inherently scale-dependent and asymmetric, and can be more susceptible to measurement uncertainties, especially when the stellar and dynamical masses are comparable. Although $f_{\mathrm{DM}}$ is widely used to quantify the relative contribution of dark matter in galactic systems, its dependence on the absolute scale of the system can introduce biases when comparing galaxies of different sizes or evolutionary stages.

In this study, we adopt the logarithmic mass difference as a more robust alternative. This metric captures the proportional relationship between stellar and dynamical components without being affected by the physical scale of the system, thereby facilitating more consistent statistical analyses across heterogeneous samples. Importantly, $f_{\mathrm{DM}}$ can be expressed as a function of $\Delta \log \mathbf{M}$:

\begin{equation}\label{eq:flog}
f_{\mathrm{DM}} = 1 - 10^{-\Delta \log \mathbf{M}},
\end{equation}

allowing for straightforward transformations between both quantities. This formulation enables direct comparison with previous results in the literature that employ either metric, while preserving the analytical and interpretive advantages of the logarithmic approach. It should be noted, however, that the general form of Equation~\ref{eq:flog} must include the total baryonic mass rather than the stellar mass alone. Consequently, when stellar mass is used as a proxy, Equation~\ref{eq:flog} is strictly valid only for systems in which nonstellar baryonic components (such as the gas mass) are negligible.

This work is organized as follows: In Section \ref{sec:sample}, we present our sample of galaxies and the method by which it was selected. Section \ref{sec:masses} presents the formal definition of stellar and dynamical masses. In Section \ref{sec:comparison}, we calculate the difference between the dynamical and the stellar mass of LTGs in our sample. In Section \ref{sec:discussion}, we discuss our findings, and finally, in Section \ref{sec:conclusions}, we present our conclusions. 

\section{The sample of LTGs}
\label{sec:sample}

We selected a sample of galaxies with exponential brightness profiles from the SDSS DR16 \citep{yor00,bla03}. To achieve the goals of this work, we required essential parameters: size and velocity dispersion from stars and gas. We used the SDSS $fracDeV$ parameter, which corresponds to the S\'ersic (1968) index $n$, with $n$ = 1 equivalent to $fracDeV$ = 0 and $n$ = 4 equivalent to $fracDeV$ = 1. Galaxies with $fracDeV < 0.5$ are well fitted by an exponential profile. We also considered that the $g$ and $r$ filters tend to have lower photometric uncertainties; equally there are lower spectroscopic uncertainties for velocity dispersions greater than 60 km/s, so we used the photometric data from these filters and applied this velocity dispersion cut-off. The resulting sample, which met the criteria of $fracDeV_g < 0.5$, $fracDeV_r < 0.5$, and stellar $\sigma > 60$ km/s according to the SDSS nomenclature, comprised 126,815 galaxies. These galaxies are distributed in a redshift ($z$) range of approximately 0.00 $<$ $z$ $<$ 0.35 and cover a magnitude range of approximately $\Delta m \sim 10$ mag.

\begin{figure*}[!ht]
\begin{center}
   \includegraphics[angle=0,width=12cm]{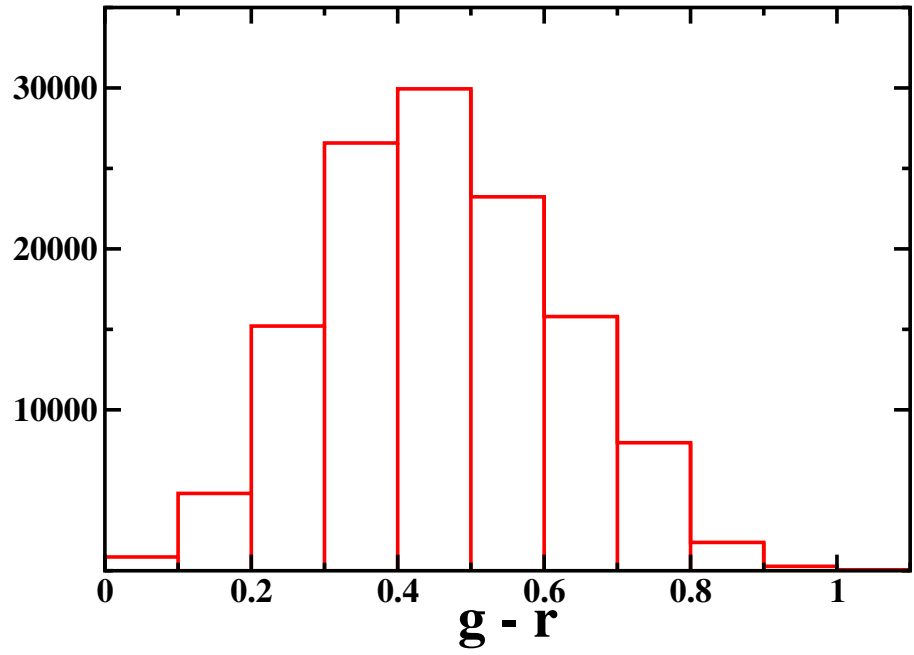}
   \caption{Color $g$-$r$ distribution of the sample of 126,815 LTGs used in this work.}
   \label{fig:g-r}
\end{center}
\end{figure*}

\begin{figure*}[!ht]
\begin{center}
   \includegraphics[angle=0,width=12cm]{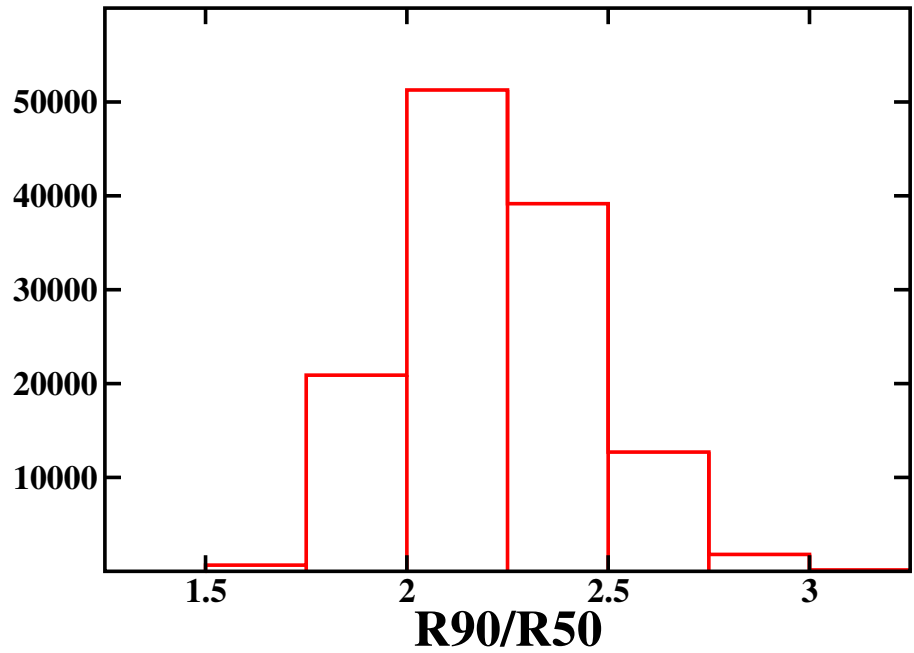}
   \caption{Concentration index (R90/R50) distribution of the sample of 126,815 LTGs used in this work.}
   \label{fig:r90r50}
\end{center}
\end{figure*}

\begin{figure*}[!ht]
\begin{center}
   \includegraphics[angle=0,width=12cm]{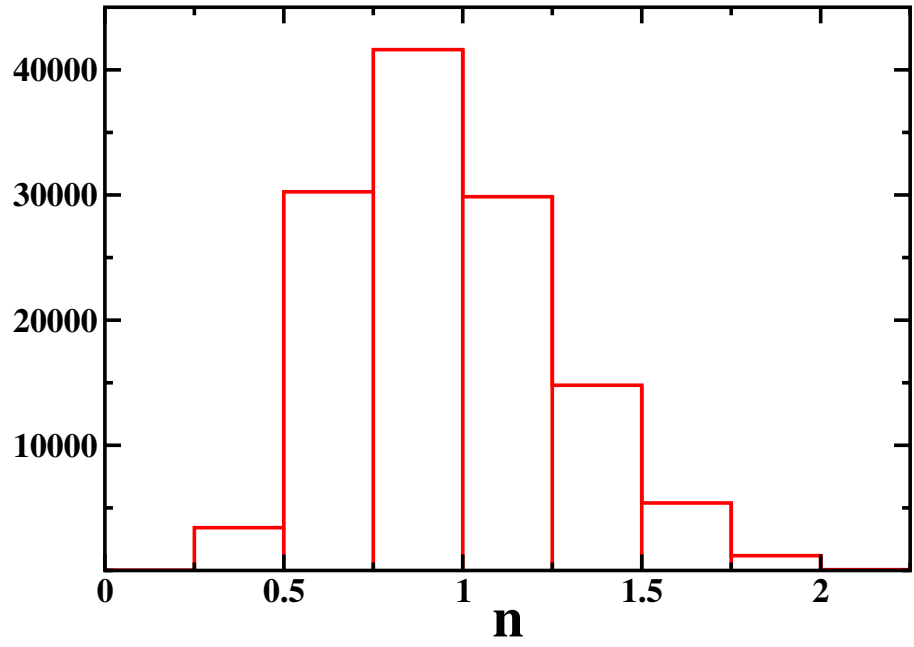}
   \caption{S\'ersic index ($n$) distribution of the sample of 126,815 LTGs used in this work.}
  \label{fig:n}
\end{center}
\end{figure*}

In Figures~\ref{fig:g-r}, \ref{fig:r90r50}, and 
\ref{fig:n}, we present the color $g-r$, the concentration index (R90/R50) and the S\'ersic index ($n$), respectively, for the selected sample.

\clearpage

The $g-r$ color was derived from the corrected $g$ and $r$ magnitudes in the SDSS (see Appendix \ref{sec:photospec}).

The concentration parameter was calculated by averaging the ratio of the Petrosian radii (R90/R50) in the $g$ and $r$ filters from the SDSS. R90 and R50 are the radii that contain 90\% and 50\% of the Petrosian flux, respectively.

The S\'ersic index ($n$) was determined by fitting the $n$ vs. R90/R50 data from Table 1 in \cite{gra05} using the corrected $g$ and $r$ filters from the SDSS. We then averaged these indices.

Figures \ref{fig:g-r}, \ref{fig:r90r50}, and 
\ref{fig:n} clearly show that the galaxy distribution is within the typical ranges for LTGs. Therefore, using the $fracDeV$ parameter to compile the LTGs sample adequately meets the goals of this research.

\section{The Mass Estimations of LTGs}
\label{sec:masses}

In this work, dynamical masses are those calculated from the velocity dispersion of gas or stars in galaxies (refer to section \ref{sec:virialmass}); these were calibrated using masses derived from rotation curves. It is crucial to emphasize that velocity dispersion and rotation curves are two different methods used for mass determination. Both mass values originate from the galaxy's gravitational potential and are calculated using Newtonian dynamics. Obtaining the dynamical mass of late-type galaxies (LTGs) is most effectively achieved through rotation curves. While integral field spectroscopy has enabled such measurements up to \( z \sim 3 \), the limited sample sizes and observational demands of IFU-based surveys constrain their applicability for large-scale statistical analyses. In contrast, photometric and spectroscopic data from wide-field surveys such as SDSS offer extensive coverage—over 100{,}000 LTGs—though they lack the spatial resolution required to derive full rotation curves. Nevertheless, dynamical masses can be estimated from gas and/or stellar velocity dispersions, providing a practical alternative for constructing homogeneous and statistically robust samples. This study leverages SDSS data to generate dynamical mass estimates for LTGs with broad coverage and relatively low associated uncertainties. By relying on spectroscopic measurements such as gas and stellar velocity dispersions, this approach enables the construction of statistically robust samples while minimizing systematic biases inherent to more spatially resolved methods. In particular, it avoids region-selection biases that arise when dynamical estimates are restricted to specific galactic zones—such as central bulges or star-forming disks—potentially misrepresenting the global mass distribution. Additionally, the use of homogeneous SDSS spectroscopic data helps mitigate instrumental and calibration biases, ensuring consistency across measurements and reducing uncertainties linked to aperture effects and cross-instrument variability. 

In the case of stellar mass, it is derived using SPS techniques (see section \ref{sec:luminousmass}). It is important to recognize that SPS models typically misjudge the quantities of neutral and ionized gas, dust, and stellar remnants, and other possible forms of baryonic matter in galaxies such as compact halo objects. These factors can result in an underestimation of a galaxy's baryonic mass. Additionally, the ratios of stars, gas, and stellar remnants can differ across galaxies, influencing their dynamic characteristics and evolution. In Section \ref{sec:discussion}, we will examine the possible implications of using this stellar mass as an estimate for total baryonic mass in our findings.

Below, we present the key variables, potential biases, and corrections that may affect the estimation of dynamical, stellar, and dark masses. A detailed description of these biases, along with the specific corrections applied to mitigate them, is provided in Appendix~\ref{sec:appendix_A}, where each factor is discussed in depth, and the corresponding correction formulae are outlined following \citet{Nig15} and related works. 



\begin{itemize}
     \item \textbf{Key variables.} 
   Velocity dispersion $\sigma$, effective radius $r_e$,  initial
mass functions IMF, color indices, redshift $z$, stellar population models, virial factor $K$ (see equation~\ref{eq:pov}), luminosity/magnitude, mass.


    \item \textbf{Main biases.}  
    \begin{itemize}
        \item \emph{Geometry/orientation:} inclination, aperture.  
        \item \emph{Observational:} seeing, extinction, cosmological dimming, Malmquist bias.  
        \item \emph{Model-dependent:} IMF choice, metallicity, dust, star formation history, stellar remnants.  
        \item \emph{Sample:} completeness, selection effects, wavelength dependence.  
    \end{itemize}

    \item \textbf{Corrections applied.}

\begin{itemize}
    \item Seeing and extinction corrections (for magnitudes and $r_e$).
    \item K-corrections and cosmological dimming adjustments.
    \item Evolutionary corrections to luminosity.
    \item Rest-frame corrections to effective radius.
    \item Aperture corrections to velocity dispersion.
\end{itemize}

    \end{itemize}

The uncertainties in the photometric and spectroscopic variables were obtained from the uncertainties provided by
the SDSS and propagated considering the mathematical expressions of each correction step.

\subsection{The dynamical mass from the stars or gas of LTGs}
\label{sec:virialmass}

For the dynamical mass, we adopt the principles of Newtonian dynamics, virial equilibrium, and attribute velocity dispersion to the size of the gravitational potential well. Our mass estimates for LTGs are derived from data pertaining to stars and gas, sourced from the SDSS DR16. These estimates have been fine-tuned through a dataset of galaxies for which we have conducted rotation curve measurements, as detailed in \citet{Nigoche22} (see section~\ref{sec:calibration} for more details). In SDSS DR16, velocity dispersion measurements for both stars and gas are derived from single-fiber spectroscopy obtained with the original SDSS spectrograph (used in SDSS-I/II) and the BOSS spectrograph (used in SDSS-III/IV). These instruments provide moderate-resolution spectra (R $\sim$ 1800–-2200) over the wavelength range 3600–-10400 Å. For stellar velocity dispersion, the SDSS DR16 pipeline employs full-spectrum fitting techniques on the observed galaxy spectra. The pipeline compares each spectrum to a grid of stellar population synthesis models~\citep{Bru03,Mar05}, which are broadened by a Gaussian line-of-sight velocity distribution (LOSVD). The best-fit model yields an estimate of the stellar velocity dispersion, after correcting for instrumental resolution. These measurements are performed using a chi-squared minimization over selected wavelength regions that avoid strong emission lines and sky residuals. The gas velocity dispersion is calculated by fitting Gaussian profiles to prominent nebular emission lines (Balmer lines). For each line, the width of the fitted Gaussian is corrected for instrumental broadening using the known line-spread function (LSF) of the spectrograph at the corresponding observed wavelength. Emission-line measurements are performed using the idlspec2d and SpecLine pipelines, which fit both the continuum and emission-line components simultaneously or iteratively. These measurements are made available in the DR16 SpecObjAll and SpecLineAll catalogs, which include velocity dispersions, line widths, and associated uncertainties for over 4 million unique spectra.

To address potential biases in the estimation of the dynamical mass (${\bf M_{Dyn}}$), we have calculated seven different dynamical masses using the following equation \citep{Poveda1958}:

\begin{equation} \label{eq:pov}
{\bf M_{Dyn}} \sim K\frac{ r_{e} \sigma_{e}^{2}}{G}.
\end{equation}

In this context, $r_{e}$ and $\sigma_{e}$ correspond to the effective radius and the velocity dispersion of stars (or gas) within $r_{e}$, respectively. $G$ is the gravitational constant, and $K$ is introduced as a proportion or scale factor. 

Equation \ref{eq:pov} represents an idealized scenario, overlooking potential influences on mass estimations for LTGs stemming from environmental factors, structural characteristics, and velocity dispersion anisotropies, among others. 

Traditionally, the scale factor $K$ has been held as a constant, as exemplified by the case of the de Vaucouleurs profile, where $K = 5.9$ \citep{cap06}. However, subsequent studies have indicated that this scale factor may exhibit variations dependent on factors such as the S\'ersic index and the inclination angle of the galactic plane \citep{cap06,moc12}. In our research, we will explore both constant and variable scale factors and take into account the importance of the different structures that make up the LTGs. 

Given that LTGs have distinct structural components (stars/gas, bulge/disk), it is important to see how these different components affect the estimation of dynamical masses. In the case of the disk, we have assumed that the global behavior of the gas is governed by gravitational interaction. This assumption is not far from reality, given that the force of gravity is the dominant force on large scales, keeping the gas within the galaxy and dictating its overall structure. The velocity dispersion used measures the gravitational effects on the gas. With this information, a mass estimation is made, which is calibrated using mass estimates obtained from rotation curves. We must remark that equation~\ref{eq:pov} provides an estimate of the dynamical mass under the assumption of virial equilibrium; however, this mass should not be confused with the virial mass associated with dark matter halos.

It is important to clarify that when estimating the dynamical mass of LTGs using the velocity dispersion derived from Balmer emission lines, we are tracing the kinematics of ionized hydrogen gas. These lines originate in HII regions, where hydrogen atoms have been ionized by ultraviolet radiation from young, massive stars. The subsequent recombination and electronic transitions produce the Balmer series in the optical spectrum. Therefore, the velocity dispersion measured from these lines reflects the motion of ionized gas within the galactic disk, which is typically associated with recent star formation. In the context of SDSS DR16, the Balmer lines provide a reliable tracer of the gravitational potential in star-forming regions, and their kinematics are used to infer dynamical mass under the assumption of virial equilibrium. 

Building on the previous discussion, we will offer definitions and a discussion of these mass estimates, summarized in Table~\ref{tab:mdyn_methods}. In Table~\ref{tab:mdyn_methods}, the ``Relevant Sections'' column identifies the parts of the manuscript where the variables and biases linked to each dynamical mass estimate are described, particularly those we seek to mitigate.

\begin{enumerate}

 \item $\mathbf{M_{KS}}$. This mass estimate is derived using a constant scale factor ($K$=5.9) in conjunction with the velocity dispersion of the stars. 
 

\item $\mathbf{M_{KA}}$. This estimate of mass is obtained using a constant scale factor ($K$=5.9) along with the velocity dispersion of gas ($H_{\alpha}$). 


\item $\mathbf{M_{KB}}$. For this estimate, we calculate the mass using a constant scale factor ($K$=5.9) and the velocity dispersion of gas (average velocity dispersion of the detected Balmer lines; $H_{\alpha},H_{\beta},H_{\gamma},H_{\delta},H_{\epsilon},H_{\zeta},H_{\eta}$). 


\item $\mathbf{M_{nS}}$. We determine this mass by employing a scale factor that varies with the S\'ersic index $n$ according to the formula $K=8.87-0.831n+0.0241n^2$ as proposed by  \citet{cap06}. The velocity dispersion of the stars is utilized in this calculation. 


For the remaining three estimates, we incorporate a scale factor ($K$) that depends on the galactic inclination angle ($i$) as follows:

\begin{equation}\label{eq:eq2}
    K=\left(\frac{2.3548}{sin(i)}\right)^2,
\end{equation}

\begin{equation}\label{eq:angle}
    sin(i)=\sqrt{\frac{1-(\frac{b}{a})^2}{1-(0.19)^2}},
\end{equation}

where $a$ and $b$ represent the apparent semi-major and semi-minor axes of the galactic disk. The value 2.3548 corresponds to the conversion factor between the standard deviation and the full width at half-maximum (FWHM) of a Gaussian profile, and is introduced to express the dispersion in terms of FWHM units. The inclination angle $i$ is estimated from the observed axial ratio ($b/a$), assuming an intrinsic flattening of 0.19 for edge-on late-type galaxies. This formulation follows the approach adopted by \citet[][]{moc12} where inclination corrections are essential for deriving intrinsic rotational velocities, surface brightness profiles, and dynamical parameters. The intrinsic axial ratio reflects the vertical thickness of galactic disks and is empirically calibrated from statistical samples of spiral galaxies. For more details on Equations~\ref{eq:eq2} and~\ref{eq:angle}, see \citet[][]{moc12}.

\item $\mathbf{M_{IS}}$. In this mass estimate, we apply a scale factor determined by the inclination angle (as defined in equation~\ref{eq:eq2}) in combination with the velocity dispersion of the stars. 


\item $\mathbf{M_{IA}}$. This mass is computed by employing a scale factor determined by the inclination angle (as specified in equation~\ref{eq:eq2}) along with the velocity dispersion of $H_{\alpha}$. 


\item $\mathbf{M_{IB}}$. We calculate this mass using a scale factor based on the inclination angle (as described in equation~\ref{eq:eq2}) and the average velocity dispersion of the Balmer lines. 

\end{enumerate}

These seven distinct mass estimates provide a comprehensive analysis of ${\bf M_{Dyn}}$ by considering different combinations of scale factors and sources of velocity dispersion.

\begin{table}[ht]
\centering

\begin{tabular}{llll}
\hline
Acronym & Scale Factor ($K$) & Velocity Dispersion Source & Relevant Sections \\
\hline
$\mathbf{M_{KS}}$ & Constant ($K = 5.9$) & Stars & 
\ref{sec:A1}(b), \ref{sec:A7}(a) \\
$\mathbf{M_{KA}}$ & Constant ($K = 5.9$) & H$_{\alpha}$ gas & \ref{sec:A1}(b–c), \ref{sec:A7}(a) \\
$\mathbf{M_{KB}}$ & Constant ($K = 5.9$) & Balmer lines (avg.) gas& \ref{sec:A1}(b–c), \ref{sec:A7}(a) \\
$\mathbf{M_{nS}}$ & Variable (Sérsic index) & Stars & 
\ref{sec:A1}(b), \ref{sec:A7}(b) \\
$\mathbf{M_{IS}}$ & Variable (inclination angle) & Stars & \ref{sec:A1}(b), \ref{sec:A7}(c) \\
$\mathbf{M_{IA}}$ & Variable (inclination angle) & H$_{\alpha}$ gas & \ref{sec:A1}(b–c), \ref{sec:A7}(c) \\
$\mathbf{M_{IB}}$ & Variable (inclination angle) & Balmer lines (avg.) gas & \ref{sec:A1}(b–c), \ref{sec:A7}(c) \\
\hline
\end{tabular}

\caption{Summary of the seven dynamical mass estimates ($\mathbf{M_{Dyn}}$) used in this study. Each method combines a specific scale factor ($K$)—either constant, Sérsic-dependent, or inclination-dependent—with velocity dispersion measurements from stars or ionized gas. The last column indicates the sections where the associated variables and biases are discussed.}

\label{tab:mdyn_methods}

\end{table}

\subsection{Calibration of the dynamical mass of LTGs.}
\label{sec:calibration}

\citet{Nigoche22} developed a method to calibrate the masses of LTGs derived from Equation \ref{eq:pov}. The calibration was conducted using a subsample of LTGs that had rotational velocity data \citep{cat05}. The calibration involved comparing the masses of the galaxies in the subsample obtained with Equation \ref{eq:pov} (dynamical mass) with those obtained through rotational velocity measurements, using linear regressions. Their analysis considered various variables and sources of bias relevant to dynamical mass estimation, including magnitude, the S\'ersic index $n$, concentration, color, and galactic inclination angle. The key outcomes, as reported by \citet{Nigoche22}, are summarized below and presented in Table~\ref{tab:mdyn_calibration_en}.

\begin{itemize}

\item When the magnitude is considered in the comparison of subsamples of LTGs, they find that the correlation coefficients attain the lowest values and the fit dispersions attain the highest values relative to the other variables considered in the analysis (S\'ersic index $n$, concentration, color, and angle of inclination). There are no significant differences found in the correlation fits or in the dispersion when comparing faint and bright galaxy samples (refer to their Figure 2 and Table 1). Therefore, the magnitude does not seem to influence the dynamical mass estimate. This paragraph addresses the variables and biases mentioned in Appendix~\ref{sec:A8}. 

\item When considering the S\'ersic index $n$, the correlation is relatively high in all fits (see their Table 2). They observe that for $n \leq 0.8$, the slopes are lower than those for $n > 0.8$. It is noticeable that for galaxies with $n \leq 0.8$, the correlation is lower when the dynamical mass is obtained using the stars compared to when it is obtained using the gas. For galaxies with $n > 0.8$, the opposite occurs. It appears that morphology impacts the estimation of the dynamical mass. This paragraph considers the variables and biases discussed
in Appendix~\ref{sec:A1}(b).

\item When considering the LTGs concentration parameter (R90/R50), they find that for R90/R50 $<$ 2.4, the slopes are lower than those for R90/R50 $>$ 2.4 (see their Table 3). They also find a relatively high correlation for all fits. In this case, they observe that the more concentrated LTGs (R90/R50 $<$ 2.4) show a larger correlation when using the gas to estimate the dynamical mass, as opposed to using the stars. For the less concentrated galaxies (R90/R50 $>$ 2.4), the opposite occurs. The structure seems to affect the calculation of the dynamical mass. This paragraph considers the variables and biases discussed
in Appendix~\ref{sec:A1}(b).

\item When considering the color of LTGs, they found that for blue galaxies ($g - r \leq 0.6$), the slopes are lower than those for red galaxies ($g - r > 0.6$; see their Tables 4 and 5). They also find a relatively high correlation in all cases. It can be noticed that for blue galaxies, the correlation is lower when the dynamical mass is obtained using the stars compared to when it is obtained using the gas. For red galaxies, the opposite occurs. The determination of the dynamical mass seems to be influenced by the color of the galaxy. This paragraph considers the variables and biases discussed in Appendix~\ref{sec:A1}(b).

\item When taking into account the inclination angle of the galaxy, they observe that for smaller inclination angles, the slopes are greater than for larger inclination angles (see their Table 6). They also found a clear difference in correlation coefficient and dispersion for samples with smaller inclination angles compared to samples with larger inclination angles. The correlation of the fits is higher (indicating lower dispersion) for galaxies with inclination angles $i > 66^{\circ}$, regardless of whether the velocity dispersion considered is that of the stars or the gas. The estimation of dynamical mass seems to be influenced by the galaxy's inclination angle. This paragraph considers the variables and biases discussed in Appendix~\ref{sec:A1}(a).

\item In general, the fits show a lower correlation (larger dispersion) when the galaxies are separated by luminosity values and a larger correlation (lower dispersion) when the galaxies are separated by inclination angle values (see their Tables 1-6 and Figures 2-8). In all cases, the best calibrations for the dynamic mass are those for which the inclination angle is relatively high ($i > 66^{\circ}$). On the other hand, for LTG samples exhibiting traits such as relatively low Sérsic index, low concentration, or blue color (indicating disk-dominated galaxies), the dynamical mass should be estimated using the gas velocity dispersion ($H_{\alpha}$ or the average velocity dispersion of the Balmer lines). In contrast, for cases with relatively high S\'ersic index, high concentration, or red color (suggesting bulge-dominated LTGs), it is more suitable to use the dynamical mass estimate derived from the velocity dispersion of the stars. The previous results are summarized in Table~\ref{tab:mdyn_calibration_en}.

\end{itemize}

\begin{deluxetable}{lll}[ht]
  \tablewidth{0pt}

  \tablecaption{Summary of calibration results for LTGs dynamical mass
    estimates based on \citet{Nigoche22}. Each row highlights the
    impact of a specific variable on the correlation between dynamical
    and rotational mass estimates, along with the preferred tracer
    (gas or stars) depending on structural and photometric
    properties.\label{tab:mdyn_calibration_en}}

  \tablehead{
    \colhead{Variable} &
    \colhead{Observed Trend} &
    \colhead{Recommended Tracer}
  }

  \startdata
  Magnitude & Weak correlation and high dispersion; no
  significant & Not applicable \\ \vspace{0.1cm}
  {}& difference between bright and faint subsamples.
  & {} \\ 
  S\'ersic index ($n$) &
  For $n \leq 0.8$ (disklike), gas yields better fits; for
  & Gas (low $n$), stars (high $n$)\\ \vspace{0.1cm}
  {}& $n > 0.8$ (bulge-like), stars perform better. &{} \\
  Concentration (R90/R50) & 
  R90/R50 $<$ 2.4 (disklike): better correlation with &
  Gas (low conc.), stars (high conc.)\\ \vspace{0.1cm}
  {}& gas; R90/R50 $>$ 2.4 (bulge-like): better with stars. & {}\\
  color ($g - r$) &
  Blue galaxies ($g - r$ $\leq 0.6$, disklike): gas preferred; &
  Gas (blue), stars (red)\\ \vspace{0.1cm}
  {}& red galaxies ($g - r$ $> 0.6$, bulge-like): stars preferred. & {}\\
  Inclination angle ($i$) &
  Galaxies with $i > 66^\circ$ show stronger correlations and &
  Either (if $i$ is high) \\ \vspace{0.1cm}
  {} & lower dispersion, regardless of tracer. & {}\\
  Overall trend &
  Inclination-based separation improves calibration; &
  Gas (disklike), stars (bulge-like) \\
  {} & tracer choice depends on morphology and color. & {}\\
  \enddata

\end{deluxetable}

\subsection{The stellar mass}
\label{sec:luminousmass}

Stellar population synthesis (SPS) methodologies are widely employed to determine the stellar masses of galaxies due to their more accurate and comprehensive estimations in comparison to alternative approaches. SPS relies on various components, including calculations of stellar evolution spanning from the main sequence to stellar death, spectral libraries of stars, empirical dust models, and stellar IMFs \citep[and references therein]{Conroy2013}. These elements work in concert to transform the evolution of a diverse population of stars with varying metallicities and ages into a prediction for the evolving integrated light emitted by that stellar population. In the SDSS DR16 there are eight estimations of the total stellar mass obtained considering the elements mentioned before. In this study, we use these eight estimates, summarized in Table~\ref{tab:stellar_mass_methods}, and briefly describe them below. In Table~\ref{tab:stellar_mass_methods}, the ``Relevant Sections'' column identifies the parts of the manuscript where the variables and biases linked to each stellar mass estimate are described, particularly those we seek to mitigate.

\begin{enumerate}

\item $\mathbf{M_{ED}}$. \textit{Stellar{\textbf{M}}assFSPSGran{\textbf{E}}arly{\textbf{D}}ust}. Stellar masses for SDSS and BOSS galaxies were estimated using the Granada method, which involves fitting SPS models by \citet{Con09} to SDSS photometry in the \textit{ugriz} bands. The fitting process utilizes extinction-corrected model magnitudes scaled to the $i$-band c-model magnitude. This particular implementation of the method, referred to as the ``early-star-formation'' version, imposes a constraint on the epoch of star formation, restricting it to within 2 billion years after the Big Bang. In addition, it incorporates a dust extinction fitting procedure and adopts an IMF of the form proposed by \citet{Van08}. This IMF is flexible with respect to a characteristic mass ($m_c$) and is continuous, thereby avoiding artificial discontinuities in the luminosity evolution of passive stellar populations. It closely resembles the widely used IMFs of \citet{Chabrier2003} and \citet{Kroupa2001} in the mass range around $1 \mathbf{M_{Solar}}$, but allows $m_c$ to vary in order to explore possible temporal evolution of the IMF. Unlike the segmented power-law form of the Kroupa IMF, this continuous functional form provides a smoother and more realistic treatment of stellar population synthesis.


    \item $\mathbf{M_{EN}}$. \textit{Stellar{\textbf{M}}assFSPSGran{\textbf{E}}arly{\textbf{N}}oDust}. Stellar masses for SDSS and BOSS galaxies were obtained using the Granada method, particularly the `early-star-formation with no dust' variant. This method involves fitting SPS models developed by \citet{Con09} to SDSS photometry across the \textit{ugriz} bands. The fitting process is conducted using extinction-corrected model magnitudes, which are then scaled to match the $i$-band c-model magnitude. In the `early-star-formation' version, the assumption is that star formation in the galaxy is constrained to a period within 2 billion years of the Big Bang. Furthermore, this version also includes an IMF using the form proposed by \citet{Van08}. 
    
    
    \item $\mathbf{M_{WD}}$. \textit{Stellar{\textbf{M}}assFSPSGran{\textbf{W}}ide{\textbf{D}}ust}. The estimated stellar masses for SDSS and BOSS galaxies, using the Granada method with wide-star-formation and accounting for dust, were calculated by fitting SPS models from \citet{Con09} to SDSS photometry data in the \textit{ugriz} filters. This fitting process involved extinction-corrected model magnitudes scaled to match the i-band c-model magnitude. This particular version, referred to as the `wide-star-formation' approach, encompasses an extended star-formation history, takes into account dust extinction, and utilizes an IMF from \citet{Van08}. 
    
    
    \item $\mathbf{M_{WN}}$. \textit{Stellar{\textbf{M}}assFSPSGran{\textbf{W}}ide{\textbf{N}}oDust}. The estimated stellar masses for SDSS and BOSS galaxies, employing the Granada method with wide-star-formation and no dust extinction, were derived by fitting SPS models developed by \citet{Con09} to SDSS photometry data in the \textit{ugriz} filters. This fitting process involved extinction-corrected model magnitudes scaled to match the $i$-band c-model magnitude. In particular, this version, referred to as the `wide-star-formation' approach, encompasses an extended star-formation history without considering dust extinction and utilizes an IMF from \citet{Van08}. 
    
\item $\mathbf{M_{PP}}$. \textit{Stellar{\textbf{M}}ass{\textbf{P}}assive{\textbf{P}}ort}. Stellar masses for SDSS and BOSS galaxies were determined using the Portsmouth method, specifically employing the `passive model' described by \citet{Mar09}. This approach fits passive stellar evolution models to SDSS photometry, incorporating known redshifts. The model assumes an instantaneous burst of star formation, with ages determined through the fitting process and a minimum age constraint of 3 billion years. The stellar population is composed of 97\% solar metallicity and 3\% metal-poor content by mass. A Kroupa IMF is adopted in this method. One of the main reasons for this choice is that the Kroupa IMF represents an intermediate option between the heavier Salpeter IMF and the lighter Chabrier IMF, making it a robust and widely accepted choice for population studies. 



    \item $\mathbf{M_{P03}}$. \textit{Stellar{\textbf{M}}ass{\textbf{P}}CAWiscBC{\textbf{03}}}. The stellar masses for SDSS and BOSS galaxies were calculated using the Wisconsin method and employing \citet{Bru03} stellar population synthesis models, as described in \citet{Che12}. A Kroupa IMF is assumed. 
    
    
    \item $\mathbf{M_{P11}}$. \textit{Stellar{\textbf{M}}ass{\textbf{P}}CAWiscM{\textbf{11}}}. The stellar masses for SDSS and BOSS galaxies were obtained using the Wisconsin method and Maraston stellar population synthesis models \citep{Mar11}, as described in \citet{Che12}. A Kroupa IMF is assumed. 
    
    
    \item $\mathbf{M_{SF}}$. \textit{Stellar{\textbf{M}}ass{\textbf{S}}tar{\textbf{F}}ormingPort}. The stellar masses for SDSS and BOSS galaxies were calculated using the Portsmouth method and a star-forming model. These calculations are based on the method developed by \citet{Mar09} and involve fitting stellar evolution models to the SDSS photometry, taking into account the known BOSS redshifts. These calculations assume a Kroupa IMF. 
    

\end{enumerate}

\begin{table}[ht]
\centering

\begin{tabular}{llllll}
\hline
Acronym & Method & SFH Type & Dust Treatment & IMF & Relevant Sections \\
\hline
$\mathbf{M_{ED}}$ & Granada & Early & Yes & Van Dokkum  & \ref{sec:A2}(a), \ref{sec:A3}(b), \ref{sec:A5}(a) \\
$\mathbf{M_{EN}}$ & Granada & Early & No & Van Dokkum  & 
\ref{sec:A2}(a), \ref{sec:A3}(a), \ref{sec:A5}(a) \\
$\mathbf{M_{WD}}$ & Granada & Wide & Yes & Van Dokkum  & 
\ref{sec:A2}(b), \ref{sec:A3}(b), \ref{sec:A5}(b) \\
$\mathbf{M_{WN}}$ & Granada & Wide & No & Van Dokkum  & 
\ref{sec:A2}(b), \ref{sec:A3}(a), \ref{sec:A5}(b) \\
$\mathbf{M_{PP}}$ & Portsmouth & Passive & No & Kroupa & 
\ref{sec:A2}(c), \ref{sec:A4}(a), \ref{sec:A5}(c) \\
$\mathbf{M_{P03}}$ & Wisconsin & PCA (BC03) & No & Kroupa &
\ref{sec:A2}(d), \ref{sec:A4}(c), \ref{sec:A5}(d) \\
$\mathbf{M_{P11}}$ & Wisconsin & PCA (M11) & No & Kroupa & \ref{sec:A2}(d), \ref{sec:A4}(c), \ref{sec:A5}(e) \\
$\mathbf{M_{SF}}$ & Portsmouth & Star-forming & No & Kroupa & \ref{sec:A2}(c), \ref{sec:A4}(b), \ref{sec:A5}(c) \\
\hline
\end{tabular}

\caption{Summary of the eight stellar mass ($\mathbf{M_{Stellar}}$) estimates available in SDSS DR16. Each method combines a specific SPS model, star formation history (SFH), dust treatment, and IMF assumption. The last column indicates the sections where the associated variables and biases are discussed.}

\label{tab:stellar_mass_methods}

\end{table}


\section{Comparison of dynamical and stellar mass}
\label{sec:comparison}

As discussed earlier, the estimates of dynamical and stellar masses are subject to various biases. Specifically for dynamical mass, \citet{Nigoche22} have observed that the most accurate calibrations for the dynamical masses of LTGs are linked to galaxies with relatively high inclination angles (see Section \ref{sec:calibration}). Their research highlights the importance of incorporating the morphology or color of these galaxies through the use of specific variables, such as the S\'ersic index ($n$), concentration index, or color. Their findings indicate that for LTGs dominated by the disk, gas should be employed, while for LTGs dominated by the bulge, stars should be utilized. Therefore, in this study, we consider these findings to obtain the most accurate dynamical mass estimations for the galaxies of our galaxy sample. That is, according to the parameters of \citet{Nigoche22}, the dynamical masses calculated through the velocity dispersion of stars were obtained from subsamples of relatively high $n$, high concentration, and red LTGs with high inclination angles, while the dynamical masses calculated through the velocity dispersion of gas were obtained from subsamples of relatively low $n$, low concentration, and blue LTGs with high inclination angles.

\citet{Nig16} have shown that comparing dynamical and stellar masses using heterogeneous samples of early-type galaxies (ETGs), i.e. samples covering wide ranges in mass and redshift, does not yield reliable results. This is because the difference between dynamical and stellar mass depends systematically on both mass and redshift. To disentangle these dependencies and minimize spurious correlations, they analyzed the relation between dynamical and stellar masses by constructing subsamples in quasi-constant mass and quasi-constant redshift intervals.

In this context, the term quasi-constant mass refers to logarithmic mass bins of width  0.1, while quasi-constant redshift corresponds to redshift bins of 0.01. This binning strategy significantly reduces the so-called geometric effect \citep{Nig15}, which arises when broad distributions in mass and redshift produce artificial trends in the dynamical-to-stellar mass relation due to projection effects in the mass–redshift plane. By working in narrow intervals of both variables, one isolates the intrinsic behavior of galaxies at fixed mass and redshift, allowing a cleaner investigation of how the difference between dynamical and stellar mass evolves with these parameters. Moreover, another strong reason to adopt narrow redshift ranges is that, within such intervals, sample completeness can be efficiently controlled, minimizing the impact of Malmquist bias (see section~\ref{sec:difference} for more details).

Applying this methodology, \citet{Nig16} demonstrated that the discrepancy between dynamical and stellar masses for ETGs increases with dynamical mass but decreases with redshift. Such trends would be partially obscured or even misinterpreted if wide heterogeneous samples were used without binning. For this reason, in the present work, we also adopt the approach of analyzing galaxy samples in quasi-constant mass and quasi-constant redshift intervals, ensuring that our conclusions about the relation between stellar and dynamical mass are not dominated by geometric biases but instead reflect genuine astrophysical dependencies.


In the present analysis, we use the procedures of \citet{Nig16} to analyze the relationship between the dynamical and stellar mass as a function of dynamical mass and redshift for the LTG samples (see Appendices \ref{sec:A10}(a), and \ref{sec:A10}(b)). We will compare the seven dynamical mass estimates with the eight stellar mass estimates, resulting in a total of 56 combinations. This examination will help us to determine whether the various variables and potential biases incorporated into these mass estimations have an impact on the outcomes related to dark matter.

\subsection{Difference between dynamical and stellar mass as function of dynamical mass and redshift for the LTGs samples}
\label{sec:difference}

It is important to emphasize that, in order to avoid Malmquist bias—--a selection effect whereby only the brightest objects are detected at larger distances—--the comparison between samples must be restricted to the same mass intervals, and these intervals should correspond to ranges where the samples are complete, i.e., where all galaxies present in the Universe are represented (see the discussion of Figures~\ref{fig:vVSr:MED} and~\ref{fig:sVSr:Mka} later in this same section). Completeness is evidenced by a well-populated distribution of galaxies across the considered mass and redshift ranges, a point that is crucial in the context of our analysis, since trends observed as a function of redshift or dynamical mass can otherwise be artificially driven by the sample's incompleteness. By ensuring that the comparisons are made within complete mass intervals, we minimize selection effects and strengthen the reliability of the inferred correlations between dynamical and stellar masses. For a more detailed discussion of this point, see Section \ref{sec:completeness_discussion}.

\begin{figure*}[!ht]
   \begin{center}

      \includegraphics[angle=0,width=9cm]{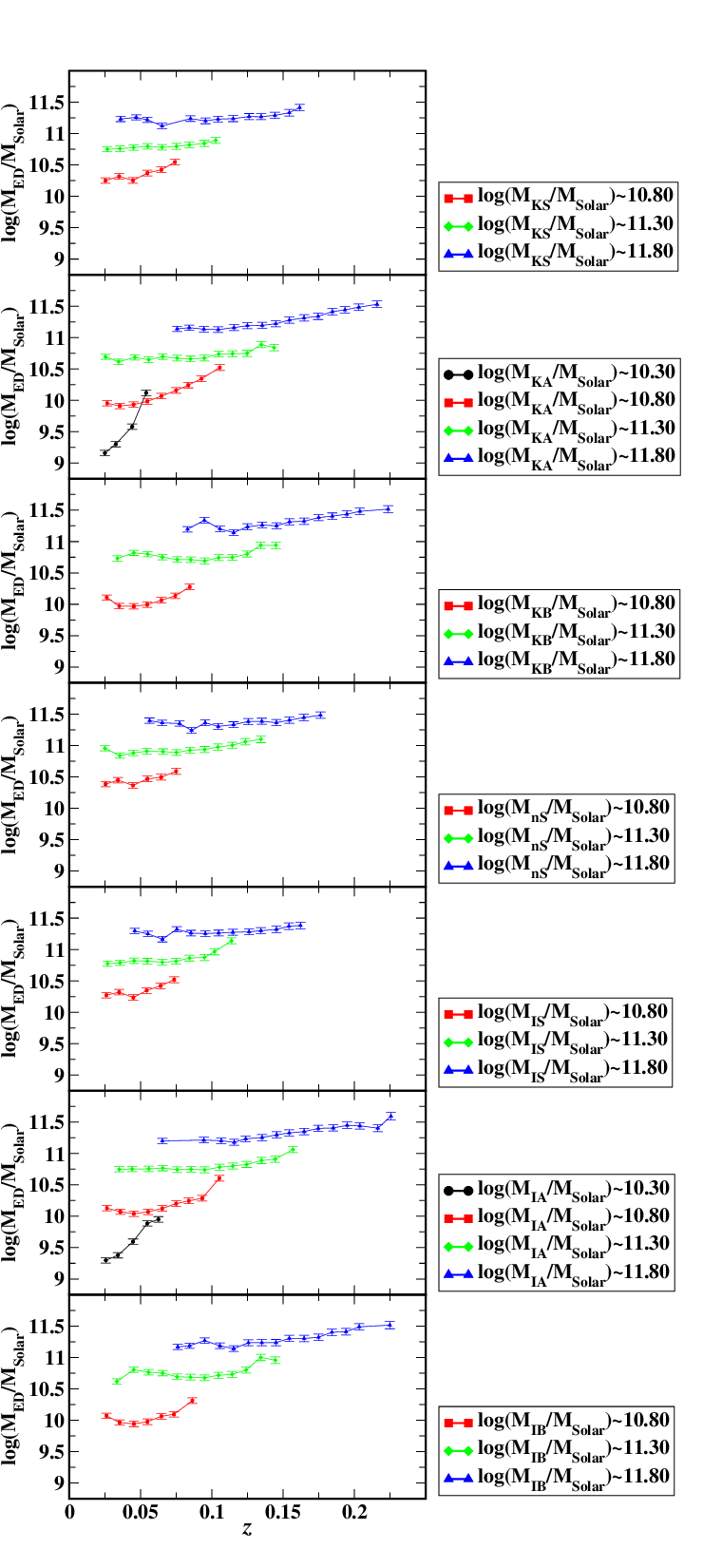}
               \caption{Behavior of stellar mass $({\bf M_{ED}})$ as a function of the redshift for quasi-constant dynamical mass. Each color and symbol represents quasi-constant dynamical mass. The lower-left part of the graph (black dots) corresponds to $\log({\bf M_{Dyn}/{\bf M_{Solar}}})$ $\sim$ 10.30, while the upper-right part of the graph (blue triangles) corresponds to $\log({\bf M_{Dyn}/{\bf M_{Solar}}})$ $\sim$ 11.80. The difference in $\log({\bf M_{Dyn}/{\bf M_{Solar}}})$ between consecutive symbols is approximately 0.5. The mean uncertainty of the $\log({\bf M_{ED}/{\bf M_{Solar}}})$ is approximately 0.062 dex.}

         \label{fig:eVSr:MED}
         \end{center}
   \end{figure*}

In Fig.~\ref{fig:eVSr:MED}, we present a mosaic showing the behavior of the stellar mass ($\mathbf{M_{ED}}$) as a function of the redshift, assuming quasi-constant dynamical mass across the seven dynamical mass estimates. The logarithm of the dynamical mass ranges from approximately 10.3 (lower-left part of the panels) to 11.8 (upper-right part). For a fixed dynamical mass, the stellar mass systematically increases with redshift. Moreover, the slope of this increase depends on the mass regime; in the high-mass range, the stellar mass grows with redshift more gently than in the low-mass range. We also find that red, high S\'ersic index $n$, and more concentrated LTGs—typically bulge-dominated—exhibit shallower slopes (see $\mathbf{M_{KS}}$, $\mathbf{M_{nS}}$, and $\mathbf{M_{IS}}$ in Figure~\ref{fig:eVSr:MED}), in contrast to blue, low $n$, and less concentrated LTGs—generally disk-dominated—which show steeper slopes (see $\mathbf{M_{KA}}$ and $\mathbf{M_{IA}}$ in Figure~\ref{fig:eVSr:MED}). The general observed pattern resembles that of ETGs reported by \citet[see their Figure~6]{Nig16}. In Appendix~\ref{sec:app:b}, we display mosaics for the other stellar mass estimates ($\mathbf{M_{ND}}$, $\mathbf{M_{WD}}$, $\mathbf{M_{WN}}$, $\mathbf{M_{PP}}$, $\mathbf{M_{P03}}$, $\mathbf{M_{P11}}$, $\mathbf{M_{SF}}$). These confirm the same global trend: the difference between dynamical and stellar mass decreases with redshift. Therefore, the variables and possible biases discussed in Appendix~\ref{sec:appendix_A}, which were considered in the different mass estimates, do not produce significant changes in the overall behavior. The only exceptions arise when dynamical masses are estimated using stars ($\mathbf{M_{KS}}$, $\mathbf{M_{nS}}$, $\mathbf{M_{IS}}$) or gas ($\mathbf{M_{KA}}$, $\mathbf{M_{KB}}$, $\mathbf{M_{IA}}$, $\mathbf{M_{IB}}$), indicating that structural components of galaxies are the main source of noticeable variations. Despite these differences, the overall trend remains robust; a higher redshift corresponds to a smaller difference between dynamical and stellar mass.

\clearpage

\begin{figure*}[!ht]
   \begin{center}

      \includegraphics[angle=0,width=11cm]{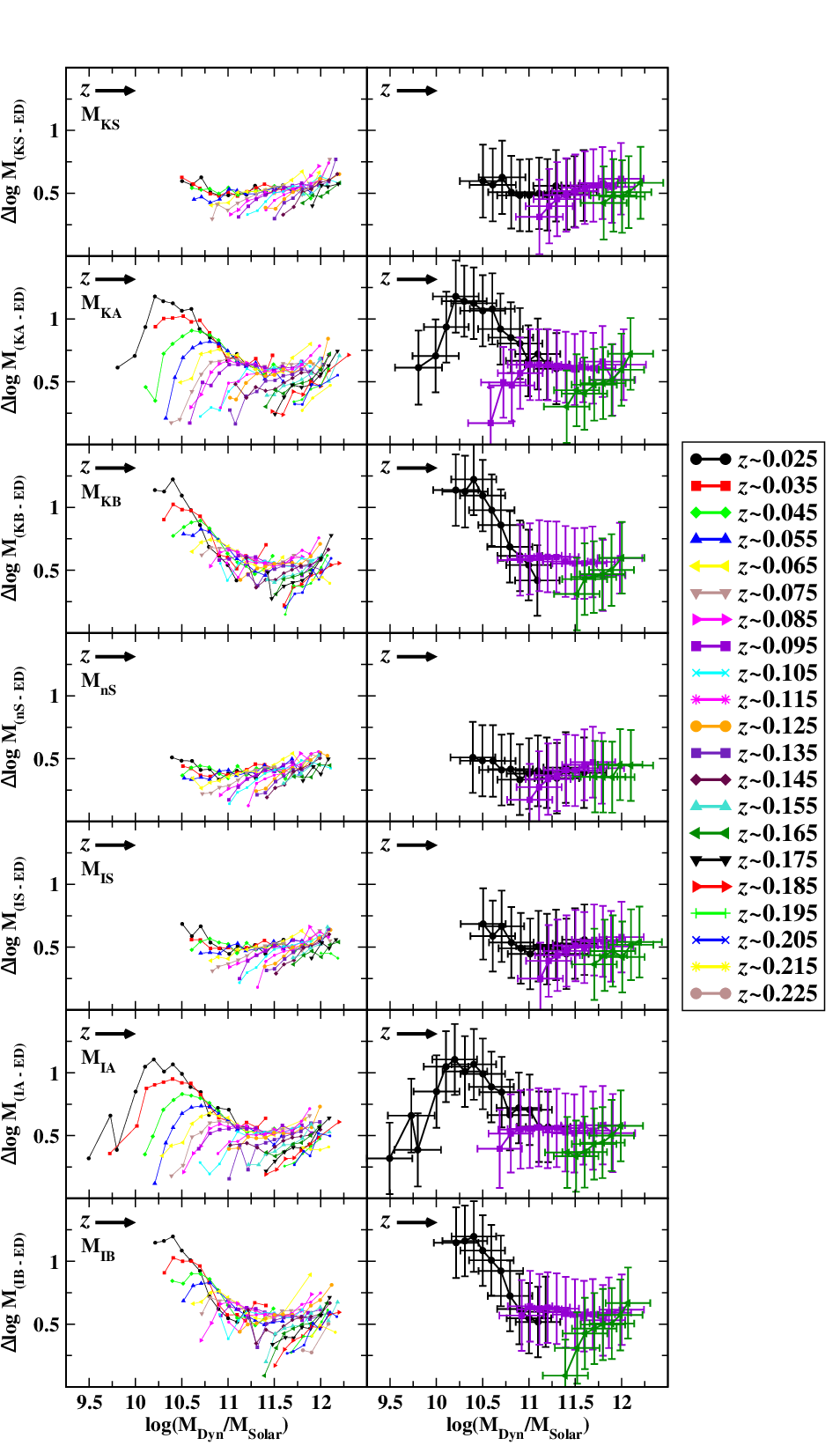}
      
         \caption{Difference between dynamical and stellar mass ($\Delta {\bf log M_{(Dyn - ED)}} $ = $\log({\bf M_{Dyn}/{\bf M_{Solar}}})$ - $\log({\bf M_{ED}/{\bf M_{Solar}}})$) as function of the dynamical mass for the LTGs samples. Each row corresponds to a specific estimation of dynamical mass ($\mathbf{M_{KS}}$, $\mathbf{M_{KA}}$, $\mathbf{M_{KB}}$, $\mathbf{M_{nS}}$, $\mathbf{M_{IS}}$, $\mathbf{M_{IA}}$, $\mathbf{M_{IB}}$). Each color and symbol represents a quasi-constant redshift. The difference in redshift between consecutive symbols is approximately 0.01. The mean uncertainty of the difference between $\log({\bf M_{Dyn}/{\bf M_{Solar}}})$ and $\log({\bf M_{ED}/{\bf M_{Solar}}})$ is approximately 0.280 dex. The left column of the mosaic contains the full range of redshift, while the right column only contains three specific redshifts: low (black dots, $z \sim 0.025$), intermediate (purple squares, $z \sim 0.095$), and high (dark green triangles, $z \sim 0.165$), the latter with the aim of appreciating more clearly the differences in dynamical and stellar masses due to redshift and also to display the uncertainties for each point.}
         \label{fig:oVSv:MED}
         \end{center}
   \end{figure*}
\begin{figure*}
   \begin{center}

    \includegraphics[angle=0,width=11cm]
    {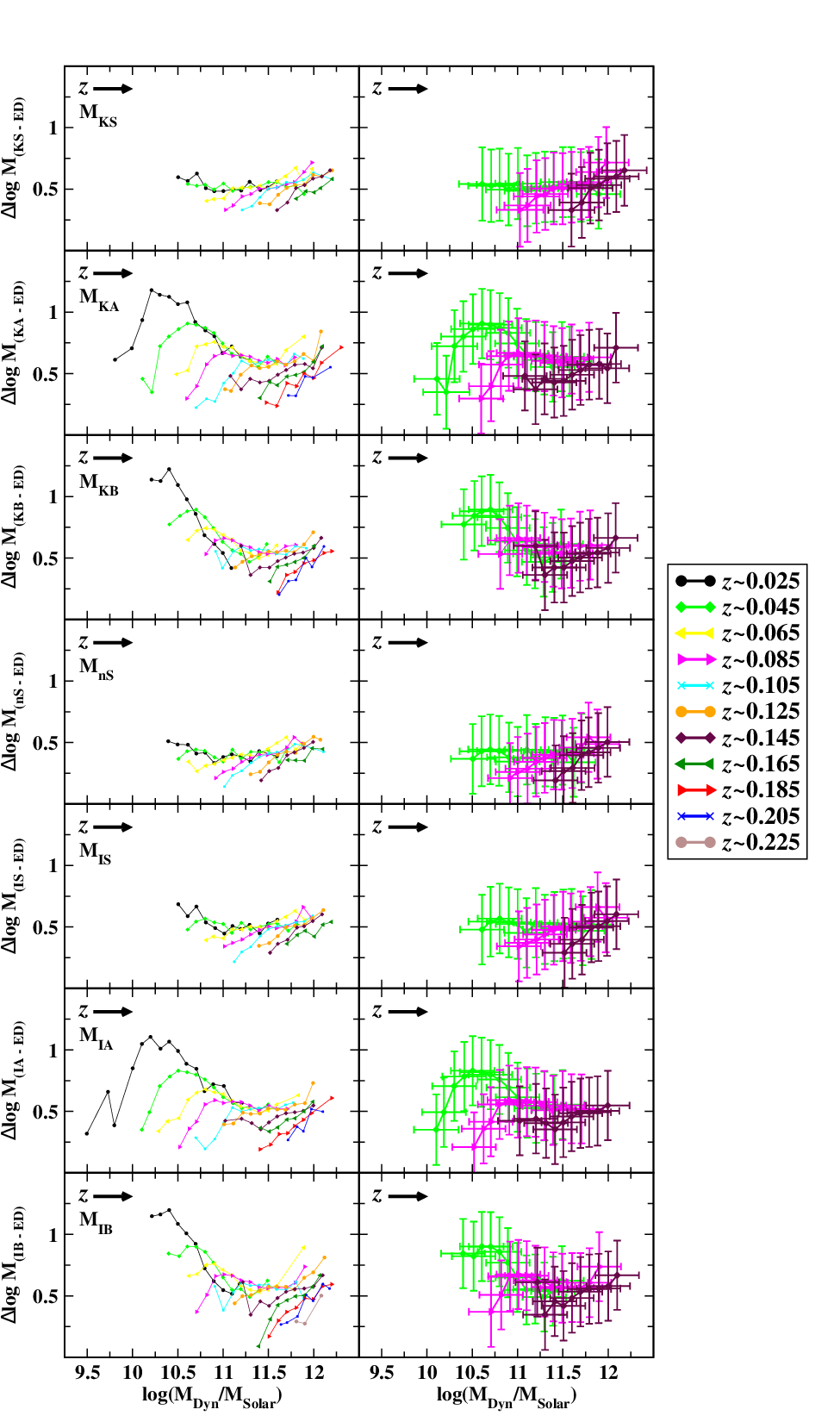}
         \caption{Mosaic of the difference between dynamical and stellar mass ($\Delta {\bf log M_{(Dyn - ED)}} $ = $\log({\bf M_{Dyn}/{\bf M_{Solar}}})$ - $\log({\bf M_{ED}/{\bf M_{Solar}}})$) as a function of the dynamical mass, analogous to Fig.~\ref{fig:oVSv:MED}. In the first column, the quasi-constant redshift trend lines are spaced every 0.02 instead of 0.01, which reduces crowding and makes the overall behavior easier to visualize. In the second column, three representative redshifts are shown: low (light green diamonds, $z \sim 0.045$), intermediate (magenta triangles, $z \sim 0.085$), and high (brown diamonds, $z \sim 0.145$). These examples complement those shown in Fig.~\ref{fig:oVSv:MED}, reinforcing the robustness of the behaviors discussed and facilitating the identification of completeness ranges shared by the different trends.}
\label{fig:oVSv:MED2}
\end{center}
\end{figure*}
\begin{figure*}[!ht]
   \begin{center}
   \includegraphics[angle=0,width=7.5cm]{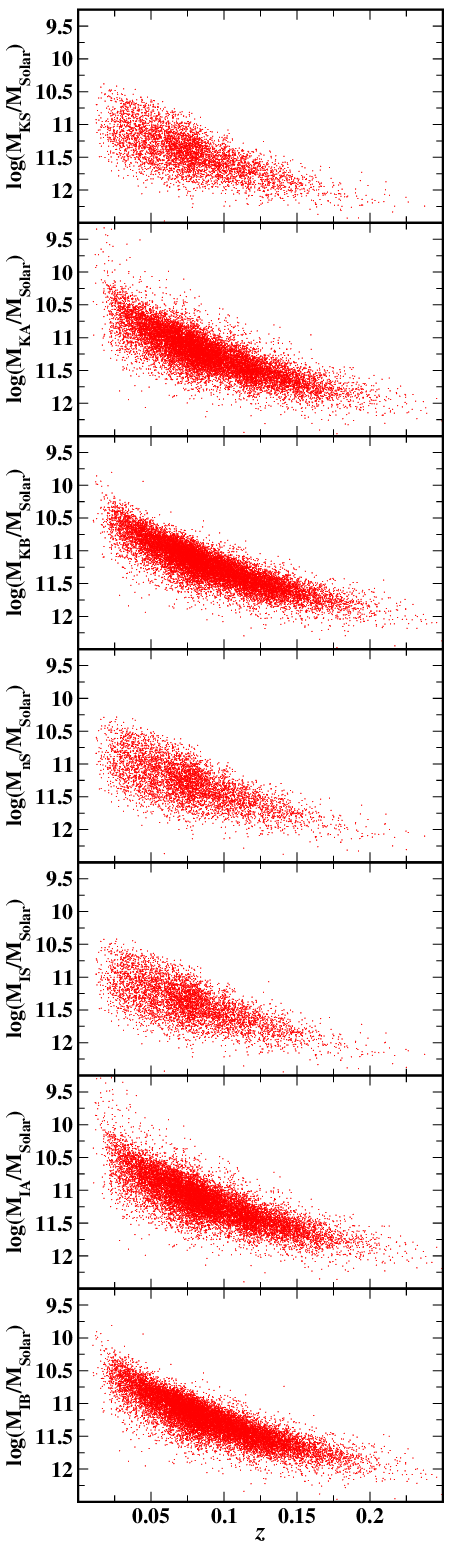}
   \caption{Dynamical mass as a function of the redshift. Each point represents a galaxy. Each graph corresponds to a specific estimation of dynamical mass ($\mathbf{M_{KS}}$, $\mathbf{M_{KA}}$, $\mathbf{M_{KB}}$, $\mathbf{M_{nS}}$, $\mathbf{M_{IS}}$, $\mathbf{M_{IA}}$, $\mathbf{M_{IB}}$).}
   \label{fig:vVSr:MED}
   \end{center}
\end{figure*}
\begin{figure*}
   \begin{center}
   \includegraphics[angle=0,width=7.5cm]{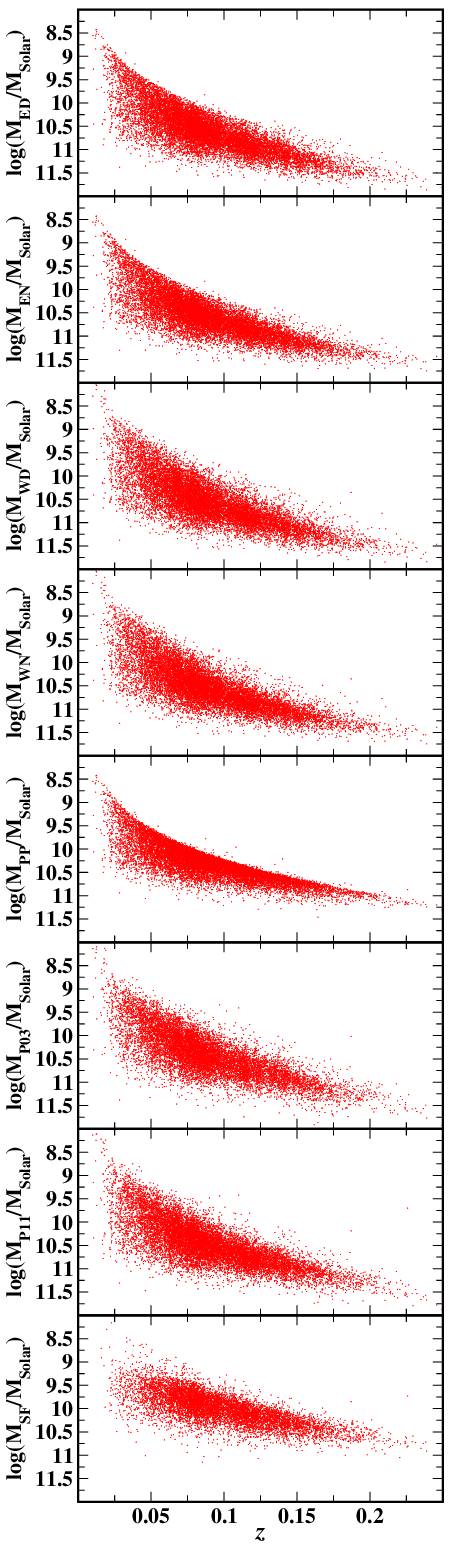}
   \caption{Stellar mass as a function of the redshift. Each point represents a galaxy. Each graph corresponds to a specific estimation of stellar mass ($\mathbf{M_{ED}}$, $\mathbf{M_{EN}}$, $\mathbf{M_{WD}}$, $\mathbf{M_{WN}}$, $\mathbf{M_{PP}}$, $\mathbf{M_{P03}}$, $\mathbf{M_{P11}}$, $\mathbf{M_{SF}}$).}
   \label{fig:sVSr:Mka}
   \end{center}
\end{figure*}

In Fig.~\ref{fig:oVSv:MED}, we show a mosaic of the difference between dynamical and stellar mass ($\mathbf{M_{ED}}$) as a function of the dynamical mass, this time assuming quasi-constant redshift across the seven dynamical mass estimates. The trend lines at quasi-constant redshift are spaced in intervals of 0.01 in redshift. Redshift spans from approximately 0.025 (upper-left panels) to 0.225 (lower-right panels). The first column of the figure displays the behavior of the mass difference as a function of the dynamical mass at quasi-constant redshift, covering the full redshift range available. To highlight the trends more clearly, the second column shows the same relation but restricted to three representative quasi-constant redshifts: low (black dots, $z \sim 0.025$), intermediate (purple squares, $z \sim 0.095$), and high (dark green triangles, $z \sim 0.165$). In this second column, we can see clearly that the trend lines are systematically offset, one below the other, indicating that at relatively low redshift the difference between dynamical and stellar mass is larger than at higher redshift (see also Figure~\ref{fig:oVSv:MED2}). It also becomes evident that the evolution of the mass difference with dynamical mass at quasi-constant redshift is not the same for less massive galaxies (black dots, $z \sim 0.025$) compared to more massive ones (dark green triangles, $z \sim 0.165$). Specifically, the relation displays a saddlelike shape at low redshift, which changes slope when moving toward more massive galaxies and higher redshifts. 


\clearpage

\clearpage


Overall, this mosaic in Fig.~\ref{fig:oVSv:MED} reveals a more complex behavior than that shown in Fig.~\ref{fig:eVSr:MED}. For bulge-dominated LTGs (see the case of $\mathbf{M_{KS}}$, $\mathbf{M_{nS}}$, and $\mathbf{M_{IS}}$) we can see that the vast majority of galaxies are in the relatively high dynamical mass regime (log($\mathbf{M_{Dyn}/M_{Solar}})>11$), we can also see that at low redshift, the mass difference declines marginally with dynamical mass, while at high redshift it increases with dynamical mass, with a gradual transition between these regimes. For disk-dominated LTGs (see the case of $\mathbf{M_{KA}}$, $\mathbf{M_{KB}}$, $\mathbf{M_{IA}}$, and $\mathbf{M_{IB}}$) we can see that  galaxies cover a wide dynamical mass interval. We also see that in the relatively low dynamical mass regime, the mass difference as a function of the dynamical mass exhibits a saddlelike shape at low redshift (see black dots, $z \sim 0.025$). This feature evolves with redshift, gradually flattening (see purple squares, $z \sim 0.095$) and eventually transforming into an almost straight line with a steadily increasing slope ({\bf dark} green triangles, $z \sim 0.165$). In all cases, within the high-mass regime (log($\mathbf{M_{Dyn}/M_{Solar}})>11$), the behavior closely resembles that reported for ETGs by \citet[see their Figure~5]{Nig16}; in particular, the mass difference increases as a function of the dynamical mass and decreases as a function of the redshift. Both disk- and bulge-dominated LTGs display the saddlelike feature, though it is more pronounced in the disk-dominated systems. In Appendix~\ref{sec:app:c}, we present mosaics for the remaining stellar mass estimates ($\mathbf{M_{EN}}$, $\mathbf{M_{WD}}$, $\mathbf{M_{WN}}$, $\mathbf{M_{PP}}$, $\mathbf{M_{P03}}$, $\mathbf{M_{P11}}$, $\mathbf{M_{SF}}$), which confirm the same general trends. As before, the variables and possible biases outlined in Appendix~\ref{sec:appendix_A} do not significantly alter the observed patterns. The only exceptions are when dynamical masses are derived from stars ($\mathbf{M_{KS}}$, $\mathbf{M_{nS}}$, $\mathbf{M_{IS}}$) or gas ($\mathbf{M_{KA}}$, $\mathbf{M_{KB}}$, $\mathbf{M_{IA}}$, $\mathbf{M_{IB}}$), reinforcing the conclusion that differences are mainly linked to the structural components of galaxies. In summary, while the detailed mass dependence varies between bulge- and disk-dominated systems, the overall evolutionary trend is consistent; at low redshift, the difference follows a saddlelike shape, which with increasing redshift evolves into a straight line with a progressively steeper slope.


In Figure~\ref{fig:vVSr:MED}, we present the distribution of dynamical mass as a function of the redshift, with each data point representing a galaxy. It can be observed that the different selection criteria in mass definitions imposed on the original sample of approximately 126,000 LTGs, as described in \ref{sec:virialmass} and \ref{sec:luminousmass}, reduce the redshift range to an approximate interval of 0--0.25. Samples with information on stellar velocity dispersion ($\mathbf{M_{KS}}$, $\mathbf{M_{nS}}$, and $\mathbf{M_{IS}}$) contain approximately 5000 LTGs, while samples with gas velocity dispersion information ($\mathbf{M_{KA}}$, $\mathbf{M_{KB}}$, $\mathbf{M_{IA}}$ and $\mathbf{M_{IB}}$) contain approximately 15,000 LTGs. The figure~\ref{fig:vVSr:MED} unveils the Malmquist bias (see Appendix~\ref{sec:A5}(c)) affecting our galaxy samples. This bias becomes evident as we observe that, at greater distances, only the most luminous and massive objects are sampled. At relatively low redshifts ($z \sim 0.025$), the samples demonstrate approximate completeness for $\log(\mathbf{M_{Dyn}/M_{Solar}}) >10.5$. Similarly, at intermediate redshifts ($z \sim 0.095$), the samples are approximately complete for $\log(\mathbf{M_{Dyn}/M_{Solar}})>11$, while at relatively high redshifts ($z \sim 0.165$), the samples are approximately complete for $\log(\mathbf{M_{Dyn}/M_{Solar}})>11.5$. Figure~\ref{fig:sVSr:Mka} shows the distribution of stellar mass as a function of the redshift, revealing a behavior similar to that observed in Figure~\ref{fig:vVSr:MED}. In fact, the completeness limit at each redshift for stellar masses is shifted toward lower masses by approximately 1.0 dex. This information is crucial in affirming the consistency of the trends shown in Figure~\ref{fig:oVSv:MED}. In particular, the observed difference in the dynamical and stellar mass as a function of dynamical mass and redshift does not appear to stem from biases in sample completeness, because the comparisons are made exclusively within complete mass ranges and within the ranges shared by the different trend lines. A comprehensive analysis of this point is provided in Section \ref{sec:completeness_discussion}.


\clearpage


\clearpage

Figures~\ref{fig:vVSr:MED} and~\ref{fig:sVSr:Mka}, which display the distribution of galaxies in the redshift–mass plane (both dynamical and stellar), provide additional support for the need to constrain the redshift range to relatively narrow intervals in order to ensure reliable control over sample completeness. In particular, when broad redshift ranges are used, relatively low-mass galaxies become increasingly underrepresented due to the Malmquist bias, potentially leading to systematic distortions in the results. This consideration further strengthens the argument presented in Section~\ref{sec:comparison}, emphasizing the importance of conducting comparisons between samples within quasi-constant intervals of mass and redshift to preserve statistical integrity and minimize selection effects.

In Figure~\ref{fig:oVSv:MED2}, we present a mosaic with the same structure as Fig.~\ref{fig:oVSv:MED}, with identical axes and panel organization. However, in the first column of this figure, the quasi-constant-redshift trend lines are spaced every 0.02 in redshift, instead of every 0.01 as in Fig.~\ref{fig:oVSv:MED}. This modification reduces the crowding of trend lines, leaving more separation between them and thus clarifying the overall behavior. In the second column of Fig.~\ref{fig:oVSv:MED2}, instead of the three redshifts, $z \sim 0.025$, $z \sim 0.095$, and $z \sim 0.165$, we display three alternative values that also represent relatively low, intermediate, and high redshifts: light green diamonds ($z \sim 0.045$), magenta triangles ($z \sim 0.085$), and brown diamonds ($z \sim 0.145$). These additional cases allow us to more clearly identify the completeness ranges shared by the different trends and further support the analysis made for Fig.~\ref{fig:oVSv:MED}. Taken together, Fig.~\ref{fig:oVSv:MED2} confirms the behaviors discussed for Fig.~\ref{fig:oVSv:MED}, but with enhanced clarity and with complementary examples of the evolution of the mass difference at representative redshifts.

\section{Discussion}
\label{sec:discussion}

In the preceding section, we found a dependence of the difference between the dynamical and stellar mass of LTGs on both the dynamical mass and the redshift. Notably, LTGs dominated by the disk behave differently from those dominated by the bulge, and the transition from one regime to another seems complex. Next, we will analyze possible biases due to mass estimations. Once the potential biases and uncertainties associated with mass estimations are characterized, we will discuss the statistical significance of the observed behaviors. Additionally, our discussion extends to the role that the SPS models and IMF might play in the observed behavior, considering that the SPS models might underestimate the baryonic mass, and the IMF utilized in estimating stellar masses might not be universally applicable.

The estimated mean uncertainties in the stellar masses for our datasets (derived from the uncertainties provided by the original authors) are as follows: 5\% in $\mathbf{M_{ED}}$, 3.5\% in $\mathbf{M_{EN}}$, 12\% in $\mathbf{M_{WD}}$, 7\% in $\mathbf{M_{WN}}$, 11\% in $\mathbf{M_{PP}}$, 38\% in $\mathbf{M_{P03}}$, 30\% in $\mathbf{M_{P11}}$, and 20\% in $\mathbf{M_{SF}}$. For a more comprehensive understanding of the uncertainties associated with each stellar mass, it is crucial to consult the references provided in Section \ref{sec:luminousmass}. In addition to these values, we also account for the systematic uncertainty introduced by neglecting the gas component, estimated to be at most $\sim 10\%$ \citep{Pop15}. This contribution, discussed in detail in Section~\ref{sec:mBaryonicMass}, has been added in quadrature to the stellar mass errors and propagated through all subsequent calculations.

The uncertainties in the dynamical mass were computed by us, taking into account the SDSS uncertainties in each of the parameters involved. Nevertheless, the primary source of uncertainty stems from the calibration process. As described in Section \ref{sec:calibration}, estimating the dynamical masses of LTGs is particularly challenging due to the significant influence of the galaxies' inclination angles. This section explains the uncertainties associated with dynamical mass estimations obtained through galaxy mass calibration. According to \cite{Nigoche22}, a stronger relationship is observed in the calibration of dynamical masses for galaxies with higher inclination angles ($i > 66^{o}$). This relationship is primarily attributed to reduced projection effects at higher inclinations, leading to more accurate mass estimates. \cite{Nigoche22} report that the dispersion in calibration fits for these galaxies is approximately 0.23-0.25 on a logarithmic scale (see their Table 6), indicating the level of uncertainty in dynamical mass estimates. In the primary analysis of this study, we focus mainly on LTGs with inclination angles greater than $66^{o}$, resulting in dynamical mass estimates that could vary within this dispersion range from the calculated value. The dispersion quantifies this uncertainty, which can be contextualized by evaluating both the absolute and relative uncertainties of the estimated dynamical mass. On a linear scale, this uncertainty is approximately 50\%.

Considering all the aforementioned factors, the typical uncertainty in the difference between dynamic and stellar mass ranges roughly from 50\% to 70\%.

It is worth noting that the individual uncertainties in dynamical and stellar mass are displayed in the various graphs of this study only when the figures have relatively few points and lines, and the inclusion of error bars did not impede the visualization of the depicted behaviors.

Based on the provided information about the uncertainties, an assessment of the differences between dynamical and stellar masses reveals some patterns:

\begin{enumerate}
\item
Stellar masses with significant differences.
In the case of stellar masses $\mathbf{M_{ED}}$, $\mathbf{M_{EN}}$, $\mathbf{M_{WD}}$, $\mathbf{M_{WN}}$, and $\mathbf{M_{PP}}$ considering all estimations of dynamical mass, the differences between dynamical and stellar mass as a function of the dynamical mass and redshift show clear underlying trends (see Figures \ref{fig:oVSv:MED}, \ref{fig:c1}, \ref{fig:c2}, \ref{fig:c3}, \ref{fig:c4}). 

\item
Stellar masses with mixed significance.
Regarding stellar masses $\mathbf{M_{P03}}$, $\mathbf{M_{P11}}$, and $\mathbf{M_{SF}}$ it is found that when considering dynamical masses obtained from stars, the difference between dynamical and stellar mass exhibits considerable dispersion and less distinct underlying trends. However, for the dynamical masses obtained from gas, the trends are more evident (see Figures \ref{fig:c5}, \ref{fig:c6}, \ref{fig:c7}). 

\item
Increased dispersion with higher uncertainty stellar masses.
Less accurate methods for calculating stellar masses ($\mathbf{M_{P03}}$, $\mathbf{M_{P11}}$, and $\mathbf{M_{SF}}$) yield plots with more fluctuations. This loss of clarity in trends is more evident in the case of $\mathbf{M_{SF}}$.

\item
Consistent general behavior.
It is important to emphasize that in all combinations of dynamical and stellar masses, the general behavior is similar. 

\end{enumerate}

\begin{figure*}
   \begin{center}
   \includegraphics[angle=0,width=12cm]{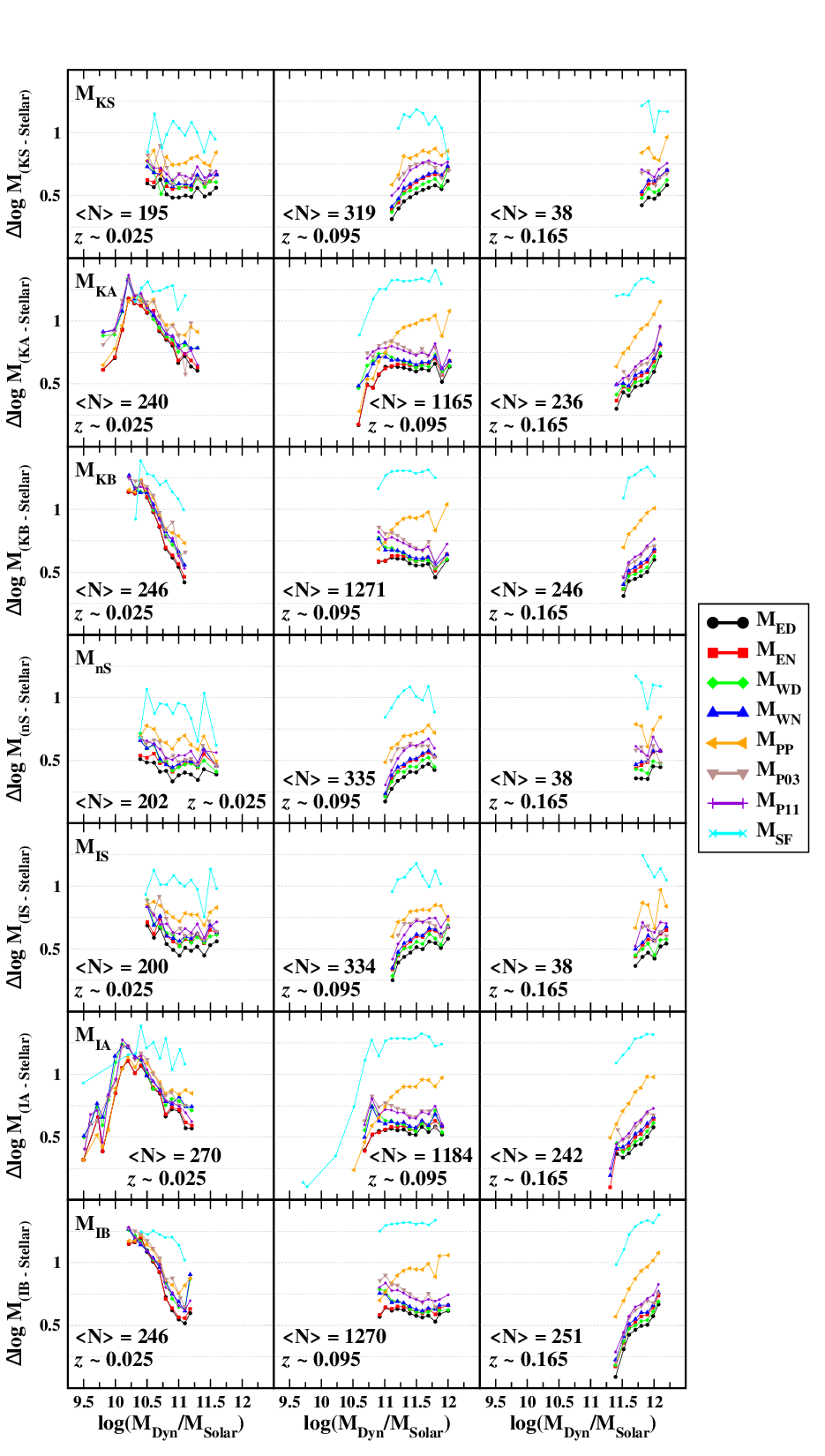}
   \caption{Difference between dynamical and stellar mass ($\Delta {\bf log M_{(Dyn - Stellar)}} $ = $\log({\bf M_{Dyn}/{\bf M_{Solar}}})$ - $\log({\bf M_{Stellar}/{\bf M_{Solar}}})$) as a function of the dynamical mass for the LTGs samples and three particular redshifts (low, medium, and high). Each row corresponds to a specific estimation of dynamical mass ($\mathbf{M_{KS}}$, $\mathbf{M_{KA}}$, $\mathbf{M_{KB}}$, $\mathbf{M_{nS}}$, $\mathbf{M_{IS}}$, $\mathbf{M_{IA}}$, $\mathbf{M_{IB}}$). Each color and symbol represents a specific stellar mass. The mean uncertainty of the difference between $\log({\bf M_{Dyn}/{\bf M_{Solar}}})$ and $\log({\bf M_{Stellar}/{\bf M_{Solar}}})$ is approximately 0.280, 0.279, 0.285, 0.281, 0.284, 0.331,
   0.312, and 0.294 dex for $\mathbf{M_{ED}}$, $\mathbf{M_{EN}}$, $\mathbf{M_{WD}}$, $\mathbf{M_{WN}}$, $\mathbf{M_{PP}}$, $\mathbf{M_{P03}}$, $\mathbf{M_{P11}}$, and $\mathbf{M_{SF}}$ respectively. The samples at low ($z \sim 0.025$), medium ($z \sim 0.095$), and high redshifts ($z \sim 0.165$) are approximately complete for $\log({\bf M_{Dyn}/M_{Solar}})$ greater than 10.5, 11, and 11.5, respectively. $<N>$ is the average number of galaxies in the different samples considered.}

   \label{fig:oVSv}
   \end{center}
\end{figure*}

\begin{figure*}
  \begin{center}

     \includegraphics[angle=0,width=11cm]
     {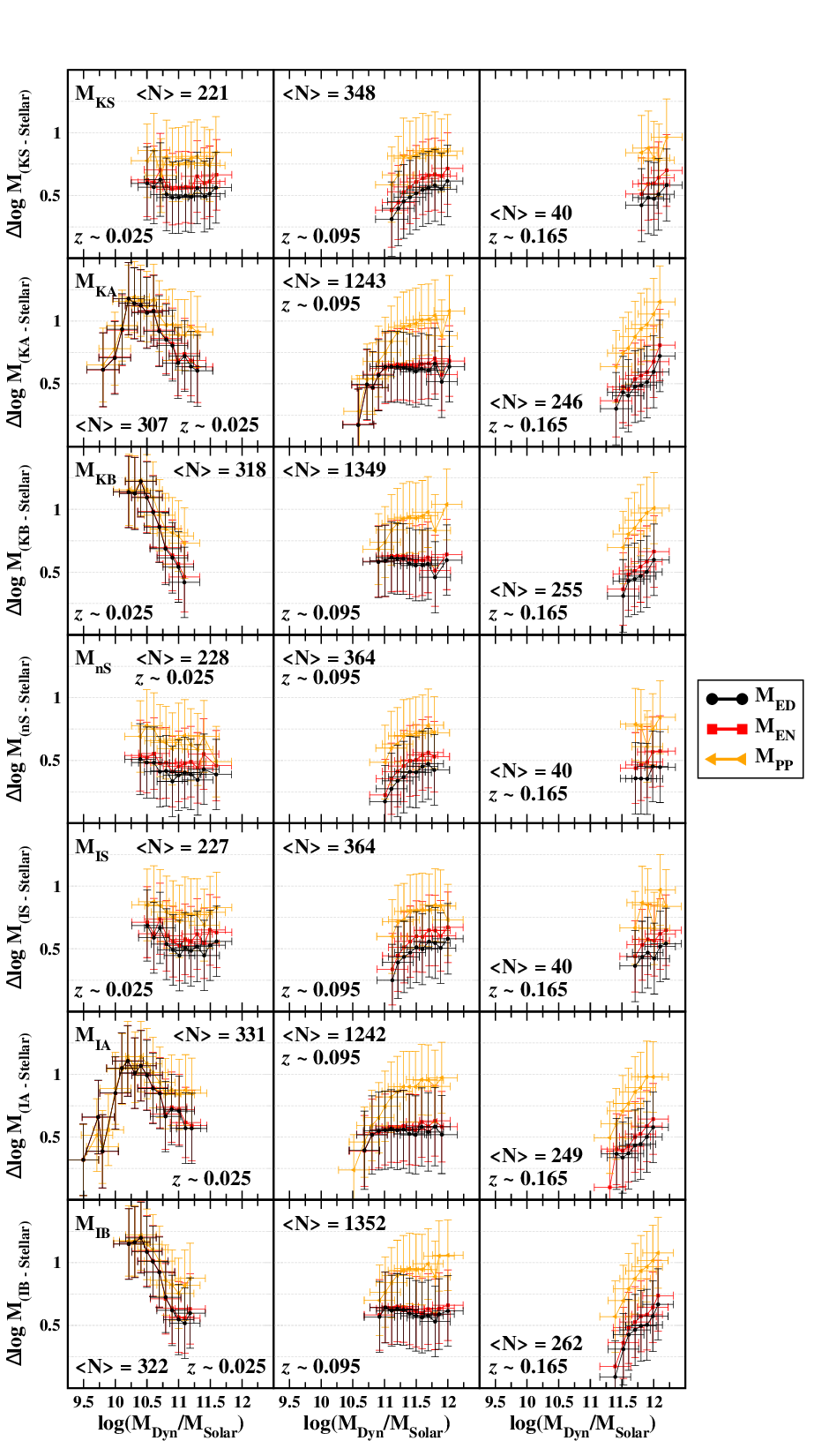}
 
           \caption{Difference between dynamical and stellar mass ($\Delta {\bf log M_{(Dyn - Stellar)}} $ = $\log({\bf M_{Dyn}/{\bf M_{Solar}}})$ - $\log({\bf M_{Stellar}/{\bf M_{Solar}}})$) as a function of the dynamical mass for the LTG samples at three representative redshifts (low, medium, and high), analogous to Fig.~\ref{fig:oVSv}. Each row corresponds to a specific estimation of dynamical mass ($\mathbf{M_{KS}}$, $\mathbf{M_{KA}}$, $\mathbf{M_{KB}}$, $\mathbf{M_{nS}}$, $\mathbf{M_{IS}}$, $\mathbf{M_{IA}}$, $\mathbf{M_{IB}}$). Each color and symbol corresponds to a specific stellar mass estimate. The mean uncertainties in the difference between $\log({\bf M_{Dyn}/M_{Solar}})$ and $\log({\bf M_{Stellar}/M_{Solar}})$ are
           0.280, 0.279, and 0.284 dex for $\mathbf{M_{ED}}$, $\mathbf{M_{ND}}$, and $\mathbf{M_{PP}}$, respectively. The samples are approximately complete for $\log({\bf M_{Dyn}/M_{Solar}}) > 10.5$, 11.0, and 11.5 at low ($z \sim 0.025$), medium ($z \sim 0.095$), and high ($z \sim 0.165$) redshift, respectively. $<N>$ is the average number of galaxies in the different samples considered.}
   \label{fig:oVSv2}
  \end{center}
\end{figure*}

In Figure~\ref{fig:oVSv}, graphs are presented showing the difference between dynamical and stellar mass as a function of the dynamical mass for the seven dynamical masses and the eight stellar masses using three different redshifts (low, medium, and high) with the aim of observing possible differences due to the methods used for the estimation of these masses. Here, it is important to emphasize that the samples at low ($z \sim 0.025$), medium ($z \sim 0.095$), and high redshifts ($z \sim 0.165$) are approximately complete for $\log({\bf M_{Dyn}/M_{Solar}})$ greater than 10.5, 11, and 11.5, respectively. In the said Figure~\ref{fig:oVSv}, it can be observed that the general behavior is similar in all cases, as discussed in the previous paragraph. It is noteworthy that when dynamic masses obtained from stars are used (see $\mathbf{M_{KS}}$, $\mathbf{M_{nS}}$, and $\mathbf{M_{IS}}$), there is a relatively high dispersion among the estimations of the different stellar masses, and this dispersion is similar at different redshifts. In the case of dynamical masses obtained using gas (see $\mathbf{M_{KA}}$, $\mathbf{M_{KB}}$, $\mathbf{M_{IA}}$, and $\mathbf{M_{IB}}$), it is observed that at low redshift, the dispersion is relatively lower than at medium and high redshifts. However, it is also important to mention that the dispersion in the difference between dynamical and stellar mass due to different estimations of stellar mass is of the order of or smaller than the associated uncertainties (except for $\mathbf{M_{SF}}$), so that these differences due to the methods used for the estimation of the stellar mass are not significant. It can be seen that the stellar mass $\mathbf{M_{SF}}$ significantly deviates from the rest of the stellar masses. Figure~\ref{fig:oVSv} does not include error bars, since the trend lines of the different stellar mass estimation methods (eight in total) are very close to each other, and adding error bars would increase the crowding of the plots. To address this, we include Figure~\ref{fig:oVSv2}, where only three particular trend lines are shown (the cases of $\mathbf{M_{ED}}$, $\mathbf{M_{EN}}$, and $\mathbf{M_{PP}}$), allowing error bars to be displayed more clearly without overlapping with a dense set of curves.


\clearpage

\clearpage

\subsection{The Impact of a Constant IMF Assumption on Dark Matter Estimates in LTGs}

As we have seen previously (see also Appendix~\ref{sec:A3}), the IMF is a fundamental concept that describes the distribution of stellar masses at birth. It is often assumed to be universal and constant. However, the implications of this assumption, especially for dark matter estimates in LTGs, warrant closer examination.

\subsection{Some effects of a constant IMF assumption:}

\begin{itemize}
\item Systematic uncertainties. A constant IMF assumption can introduce systematic uncertainties in dark matter estimates. If the actual IMF varies within a galaxy or between galaxies due to different variables, the derived dark matter and total masses may be biased.

\item Impact on baryonic Tully--Fisher relation. The baryonic Tully--Fisher relation, which relates the baryonic mass of a galaxy to its rotation velocity, is sensitive to the assumed IMF. Errors in the IMF can lead to scatter in this relation and affect our understanding of the coevolution of baryons and dark matter.

\item Implications for galaxy formation models. The IMF plays a crucial role in galaxy formation models. A constant IMF assumption can limit our ability to accurately model the processes of star formation and feedback, which are essential for understanding the distribution of dark matter in galaxies.

\end{itemize}

\subsection{Evidence for a variable IMF}

While a universal IMF is a convenient assumption, there is growing evidence to suggest that it may not be universally valid. Some studies have explored the impact of a variable IMF on galaxies using detailed dynamical modeling and simulations. For instance, \cite{Bau05} investigated the effects of varying the IMF on the Tully--Fisher relation and found that a top-heavy (higher proportion of massive stars) IMF could alleviate some of the tension between observations and theoretical models. \citet{cap12} and \citet{Dut13} claim that the IMF is not universal, but rather it depends on the stellar mass of the galaxy. In a more recent study, \cite{Bar18} delve into hydrodynamical simulations showing how the IMF changes in galaxies, taking into account elements like velocity dispersion and its effects on galaxy formation and evolution.

\subsection{Missing Baryonic Mass in LTGs Masses Derived from SPS}
\label{sec:mBaryonicMass}

SPS models provide valuable estimates of stellar masses in galaxies, but they could underestimate baryonic matter (see Appendix~\ref{sec:A6}(e)). While SPS models primarily account for the light emitted by stars, they typically do not directly consider:

\begin{itemize}
\item Neutral and ionized gas. These components could constitute a substantial fraction of a galaxy's baryonic mass, especially in star-forming regions.

\item Stellar remnants. White dwarfs, neutron stars, and black holes represent the end products of stellar evolution and contribute to the total baryonic mass.

\item Dust. Dust grains absorb and reemit stellar radiation, affecting the observed spectral energy distribution and potentially leading to underestimates of baryonic mass.

\item Other possible types of baryonic matter such as compact halo objects may count toward the total baryonic mass of a galaxy.
\end{itemize}

Consequently, the difference between dynamical and stellar mass likely arises from a combination of factors, including unaccounted baryonic components, possible IMF variations, and the presence of dark matter. In our estimates, gaseous components are not included; their addition would partially alleviate the mass discrepancy and lower the inferred amount of dark matter. However, obtaining reliable gas masses on a galaxy-by-galaxy basis is not feasible for a sample of over 120{,}000 objects. Although empirical scaling relations~\citep[e.g.][]{Pop15,Cat18} provide an alternative, they suffer from large intrinsic scatter (0.2–-0.4 dex or more) and depend on observables such as UV–-optical colors, morphology, disk scalelength, or star formation rates, which are not consistently available in our dataset. Applying such relations individually would therefore introduce systematic biases that compromise the statistical robustness of our analysis. On the other hand, previous works show that relatively massive and relatively low redshift galaxies are predominantly dominated by their stellar component, particularly \citealt{Pop15} show that galaxies with $\log(\mathbf{M_{Stellar}/M_{Solar}}) > 10$ at $z < 0.3$ typically contain only 5–10\% of gas mass compared to stellar mass  (see their Figure 6), a fraction far smaller than the scale of the mass discrepancies reported here. Since the vast majority of our galaxies fall within this mass and redshift range, neglecting the gas component is unlikely to significantly affect our conclusions. Within this context, the inferred amount of dark matter should be regarded as an upper limit, directly tied to the accuracy of the baryonic mass estimates. Moreover, the omission of the gas mass can be considered as an additional source of uncertainty in the stellar mass and mass difference. Based on \citealt{Pop15}, this uncertainty is at most $\sim 10\%$, which in logarithmic scale corresponds to $\sim 0.04$ dex; this systematic has been added in quadrature to our error estimates and is included in all calculations and figures where relevant.

 Given the absence of a definitive model for the IMF's variation with the previously mentioned variables and the challenge in measuring the quantity of weak and nonluminous baryonic matter, in this work, we have considered stellar mass obtained using SPS with constant IMFs (see Section~\ref{sec:luminousmass}) as an approximation of baryonic mass. A more in-depth exploration of IMF variations and the quantification of weak and nonluminous baryonic mass is beyond the scope of this work.

Finally, the environment, mainly represented by galaxy density and demographics, could impact the presence and quantity of dark matter within LTGs (see Section~\ref{sec:A10}(c)). Given the extensive analysis required to explore the potential influence of the environment on LTGs, a forthcoming article will be entirely dedicated to exploring this subject.


\subsection{Comparison with results from the literature}

Although directly comparable works are limited, several studies provide useful context for our analysis. Below, we summarize the most relevant findings.

\citet{Nig16} analyzed the relation between dynamical and stellar masses in early-type galaxies (ETGs) using SDSS DR9. Their sample of 19,000–98,000 ETGs was modeled under Newtonian dynamics with different light profiles and constant IMFs. They reported that the difference between dynamical and stellar mass increases with dynamical mass but decreases with redshift, a trend consistent with contributions from dark matter, a nonuniversal IMF, or both. These findings suggest that ETG evolution is tightly connected to the interplay between stellar and dark matter components.

\citet{Ghosh16} studied the role of dark matter halos in shaping global spiral modes in galaxies. They found that in 
low-surface-brightness (LSB) galaxies, dark matter accounts for up to $\sim 90\%$ of the total mass within the optical disk, dominating even the inner region. The ``inner region'' typically refers to radii of approximately a few kiloparsecs from the galactic center, generally within the range of 1 to 4 kpc, where the contributions of the bulge and the disk are most significant in the galaxy's dynamics. In contrast, in high-surface-brightness (HSB) systems such as the Milky Way, stellar mass plays a more prominent role in the center, while dark matter dominates at larger radii. These results indicate that LSBs, stabilized by their high dark matter content, tend to suppress small-scale structures, whereas HSBs preserve more complex features due to a more balanced mass distribution. 

\citet{Gen17} investigated the dynamics of baryon-dominated massive disk galaxies at $z\sim 2$, showing that baryons played a stronger role in galaxy assembly in the early Universe. Their results indicate that high gas fractions and large velocity dispersions led to decreasing rotation curves with radius. This suggests that baryons efficiently condensed at halo centers in the early Universe, triggering rapid star formation and structural evolution.

\citet{For20} present a comprehensive review of galaxy evolution, focusing on the physical processes that shape galaxies during the epoch of peak star formation, commonly referred to as ``cosmic noon'' ($z \sim 2$). The study examines star-forming galaxies (SFGs) in this era, highlighting the emergence of scaling relations among key properties such as kinematics, structure, star formation, and feedback mechanisms. Particular emphasis is placed on internal drivers of evolution—including gravitational instabilities, secular processes, and baryon cycling—which play a central role in regulating galaxy growth. The review also notes that massive galaxies at $z \sim 2$ frequently host dense cores and active galactic nuclei (AGNs), both of which are likely contributors to star formation quenching. Notably, the stellar mass function exhibits minimal evolution in its characteristic mass from $z \sim 2$ to the present, suggesting a degree of stability in the mass scale at which galaxies transition in their evolutionary pathways.

\citet{Sorini24} explored the concentration–mass relation of dark matter halos using both hydrodynamical and dark-matter-only simulations across the mass range $10^{9.5}$--$10^{15.5}~M_\odot$ and redshifts $0 < z < 7$. They found that baryonic effects significantly enhance halo concentrations at high redshift through adiabatic contraction, while feedback processes tend to reduce concentrations in lower-mass halos at later times. Although their analysis does not explicitly classify halos by environment or galaxy morphology, it spans a wide range of halo masses—from dwarf-galaxy scales to massive clusters—and focuses on internal structural properties rather than environmental interactions. Therefore, while a direct comparison with our galaxy sample is not straightforward, their results are referenced here to contextualize the general influence of baryons on halo evolution.

\citet{Sch24}, through dynamical modeling of CALIFA galaxies, demonstrated that galaxy properties such as age, metallicity, angular momentum, SFR, and morphology correlate with both stellar and halo mass. At fixed stellar mass, higher halo mass is associated with younger ages, lower metallicities, higher rotational support, and later morphologies, reinforcing the importance of halo mass in regulating baryonic evolution.

\citet{Sharma24} analyzed 263 star-forming disklike galaxies up to $z \sim 2.5$ using KMOS$^{\text{3D}}$, KGES, and KROSS data. They found that the dark matter fraction increases with radius and decreases with redshift, although it consistently remains above 50\% within the effective radius. Their results suggest that galaxies at high redshift are more baryon-dominated compared to local systems, in line with \citet{Gen17}.

Table~\ref{tab:sample_comparison} summarizes the key characteristics of the samples in the cited works and in this study, including the number of galaxies, redshift range, stellar mass range, and morphological type. The following points highlight similarities and differences with our results:

\begin{itemize}

    \item \textbf{Sample size and galaxy type.}
    \citet{Ghosh16}, \citet{Gen17}, \citet{For20}, \citet{Sharma24}, and \citet{Sch24} conducted analyses on relatively small samples of galaxies, each targeting distinct populations such as LSBs and HSBs, massive high-redshift disks, star-forming galaxies (SFGs), and late-type systems. In contrast, \citet{Nig16} examined a substantially larger sample of ETGs, ranging from 19,000 to 98,000 objects drawn from SDSS DR9. Our study builds on this scale by focusing on a robust sample of $\sim$126,000 LTGs from SDSS DR16, with subsamples of 5000–15,000 galaxies used for specific analyses.

    \item \textbf{Mass discrepancies.}  
    Both \citet{Nig16} and the present study identify systematic differences between dynamical and stellar mass estimates, which may reflect the influence of dark matter and/or variations in the IMF. Notably, our findings indicate that the dynamical mass dependence of these differences varies between bulge- and disk-dominated systems, exhibiting a smooth transition across morphological regimes. In a complementary context, \citet{Sharma24} found that dark matter fractions tend to increase with galactocentric radius but decrease with redshift. Similarly, \citet{Ghosh16} reported that the degree of dark matter dominance varies significantly between LSB and HSB galaxies, highlighting the role of structural properties in shaping the dark matter distribution.

    \item \textbf{Redshift dependence.}  
    Our finding that mass discrepancies decrease with redshift is consistent with \citet{Nig16} and \citet{Sharma24}, as well as with \citet{Gen17}, who argue that early-Universe galaxies were more baryon-dominated. \citet{Sorini24} further show that baryonic effects significantly influence halo concentrations at high redshift.  

    \item \textbf{Evolutionary Implications.}
    \citet{Nig16} and the present work suggest that the evolution of galaxies is closely linked to the interplay between dark matter and stellar mass, with implications for galaxy formation and evolution. This aligns with the broader evolutionary perspectives found in \citet{Gen17}, \citet{Sharma24}, and \citet{Sorini24}.

\end{itemize}

These results suggest that the most distant galaxies, that is, those with a relatively higher redshift, tend to be more baryon-dominated, meaning they have a greater proportion of baryonic matter compared to dark matter. This phenomenon could be explained by several factors:

1. Baryon Condensation. In the early Universe, galaxies were richer in gas and more compact. This allowed baryons to efficiently condense at the center of dark matter halos, resulting in a greater dominance of baryonic matter in the inner regions of galaxies.

2. Lower Dark Matter Concentration. At higher redshifts, dark matter halos may not be as concentrated as in the local Universe. This means that dark matter could be distributed differently, contributing to the fact that forming galaxies in the past are less influenced by dark matter in their dynamics.

3. Effects of Gas Dynamics. Galaxies in the early Universe experience a high gas flow, which can lead to dynamic interactions that allow gas to accumulate at the center of the galaxy, reducing the influence of dark matter on the observed dynamics.

 It is important to note that our findings go beyond the results discussed in the previous paragraphs, as we found that the difference between total (dynamical) and stellar mass depends complexly on both total mass and redshift. This nuanced dependence underscores that the evolution of galaxies within the nearby Universe differs for galaxies dominated by the disk and those dominated by the bulge. Moreover, our results indicate that the evolution is not separated, but rather there is a continuous transition from one regime to another. These results imply that the role of dark matter in galaxy formation and evolution is more complex than previously thought and has implications for the formation and evolution models of galaxies, suggesting that they may need to account for a more fluid transition between different morphological types.

\begin{deluxetable}{lllll}[ht]
\tabletypesize{\scriptsize}
\tablewidth{0pt}

\tablecaption{Summary of the main properties of the galaxy
samples analyzed in the referenced studies and in this work. The
columns list the sample size, redshift range ($z$), stellar mass
range ($\log(\mathbf{M_{Stellar}/M_{Solar}})$), and morphological type.} \label{tab:sample_comparison}

\tablehead{
    \colhead{Author} & \colhead{Sample Size} & \colhead{$z$} &
    \colhead{$\log(\mathbf{M_{Stellar}/M_{Solar}})$} &
    \colhead{Morphological Type}
}

\startdata
   \citet{Nig16}\textsuperscript{O} 
   & 19,000--98,000 galaxies & 0.00--0.35
   & 9.00--12.00 & Early-type galaxies\\
   \citet{Ghosh16}\textsuperscript{A} & Two different models & 0.00--0.01 & 
   9.25--10.80 & LSB and HSB galaxies\\
   \citet{Gen17}\textsuperscript{O} & 6 galaxies  &  0.9--2.4 &
   10.60--11.10 & Massive star-forming galaxies\\
   \citet{For20}\textsuperscript{O/S} & \nodata  & 1--3 & 9.00--11.50 &
   Star-forming galaxies \\
    \citet{Sorini24}\textsuperscript{S} & $\sim100,000$ halos & 0--7 & \nodata
    & Dwarf galaxies \\
    & & & & to superclusters\\
    \citet{Sharma24}\textsuperscript{O} & 263 galaxies &  0.6--2.5 &
    8.30--11.70 & Disklike galaxies\\
    \citet{Sch24}\textsuperscript{O} & 260 galaxies   & 0.005--0.029 &
    9.00--11.75 & Ellipticals to late-type\\
    & & & & spiral galaxies\\ 
    This work\textsuperscript{O} & $\sim126,000$ galaxies   & 0.00--0.35 &
    9.00--12.00 & Late-type galaxies\\
 \enddata
\tablecomments{Type of Study: O: Observations; A: Analytical/Theoretical; S: Simulations.}
\end{deluxetable}

\subsection{{\bf Completeness of quasi-constant mass and redshift bins, and comparison with two local group galaxies.}}

\label{sec:completeness_discussion}

To ensure that comparisons across masses and redshift are meaningful, all trends presented in this work are computed using quasi-constant dynamical mass bins ($\Delta$ $\log(\mathbf{M_{Dyn}})$ = 0.1 dex) and quasi-constant redshift bins ($\Delta z = 0.01$). It is very important to emphasize that a range is a relatively broad interval of data, while a bin is a smaller, more specific subdivision within that range.

Two primary methods exist for assessing sample completeness. The first option would involve plotting the distribution (histograms) of stellar masses within each of the $21$ redshift bins ($z$) analyzed in Figure~\ref{fig:oVSv:MED} and its equivalents. However, this approach is unfeasible due to the massive volume of plots it would generate; considering the seven dynamical mass estimates and eight stellar mass estimates, the result totals $1,176$ histograms. This volume of figures is neither practical nor efficient for a scientific paper. The alternative method, adopted in this study and presented in Figures~\ref{fig:vVSr:MED} and~\ref{fig:sVSr:Mka}, focuses on showing the distribution of galaxies in the plane dynamical mass vs. redshift. This visualization offers a compact and exceptionally clear representation of the trend: low-mass galaxies are progressively absent at higher redshifts. This tendency, which defines the Malmquist bias, is captured efficiently with just two figures, making it a superior method to the excessive use of histograms. Nevertheless, for the purpose of direct visualization and as a reference for the internal distribution, a specific set of histograms is included in the Appendix~\ref{sec:appendix_A} (Figure \ref{fig:histo}) for the case of dynamical mass $\mathbf{M_{KA}}$ combined with stellar mass $\mathbf{M_{ED}}$.

In Figures~\ref{fig:vVSr:MED} and~\ref{fig:sVSr:Mka} we identify regions in this plane that are well-populated, meaning that the survey includes all galaxies that exist in that simultaneous interval of mass and redshift. A quasi-constant bin is considered complete when the survey contains all galaxies that can exist in that mass--redshift window. Importantly, completeness does not imply equal point density across bins; differences in point density {could} reflect intrinsic physical properties of the Universe (e.g., comoving volume, large-scale structure).

To further clarify how we controlled the Malmquist bias, we have added three figures to Appendix~\ref{sec:appendix_A} (Figures~\ref{fig:smal-1}-\ref{fig:smal-3}). These figures were extracted from the last row of Figure~\ref{fig:vVSr:MED} and visually outline our binning strategy in dynamical mass and redshift. They show the distribution of dynamical mass as a function of the redshift, and illustrate: the quasi-constant redshift bin for three different redshifts, including the full available mass range for each redshift bin (Figure~\ref{fig:smal-1}); the overlapping mass range for these three quasi-constant redshifts (Figure~\ref{fig:smal-2}); and, finally, a specific quasi-constant bin of dynamical mass at three quasi-constant redshifts (Figure~\ref{fig:smal-3}). The color code is the same as in Figures~\ref{fig:oVSv:MED2} and~\ref{fig:oVSv:MED2extra}, namely light green for $z = 0.045$, magenta for $z = 0.085$, and brown for $z = 0.145$. These figures illustrate two key zones:\bigskip

Zone 1: The region dominated by Malmquist bias, where lower-mass galaxies become progressively undetected at higher redshifts.\bigskip


Zone 2: The region where the survey can detect galaxies.\bigskip

Our comparative analysis of galaxy properties is intentionally restricted to Zone 2, and only in the overlapping mass range where the survey can detect galaxies across the different redshift bins. This ensures a ``fair comparison'' by using only the data that is robustly available at all considered redshifts.


When comparing bins of the same dynamical mass (e.g., bin centered at $\log(\mathbf{M_{Dyn}})=11.5$, width $0.1$) at different redshifts (e.g., bin centered at $z=0.045$ width $0.01$, bin centered at $z=0.085$ width $0.01$ and bin centered at $z=0.145$ width $0.01$), only the overlapping mass interval is viable for analysis (see Figures~\ref{fig:smal-2}-\ref{fig:smal-3}). This procedure can also be seen in Figures~\ref{fig:oVSv:MEDextra} and~\ref{fig:oVSv:MED2extra} (using the same convention as in Figures~\ref{fig:oVSv:MED} and~\ref{fig:oVSv:MED2}) where the overlapping  mass ranges for some specific redshifts are limited with orange lines. In these ranges, the higher-$z$ trend lies below the lower-$z$ trend, implying that galaxies of the same dynamical mass present less difference in dynamical and stellar mass at higher redshifts. This same behavior can be observed systematically in consecutive redshift trends implying that galaxies of the same dynamical mass present less difference in dynamical and stellar mass at higher redshifts. 

We must to note that the distribution within the overlapping mass bins is not uniform (e.g., see Figure~\ref{fig:histo3}). Figure~\ref{fig:histo3} presents the frequency distribution of galaxies across quasi-constant dynamical mass bins at three different quasi-constant redshifts; the orange bars represent the limits of the overlapping dynamical mass ranges. These distributions are derived from the same data as in the second column of Figures~\ref{fig:oVSv:MED2} and~\ref{fig:oVSv:MED2extra}, and they share the same color coding for consistency. If the observed differences in galaxy density were totally due to the Malquist effect, we would expect the number of galaxies to decrease with redshift. However, this is not the case. For example, if we observe in Figure~\ref{fig:histo3} the number of galaxies for a dynamical mass log(M$_{Dyn}$)=11.5, the quantity is small at $z=0.045$, increases at $z$=0.085, and then decreases again at $z=0.145$. Our analysis of specific, quasi-constant mass (e.g., log(M$_{Dyn}$)=11.5) and quasi-constant redshift bins ($z=0.045$, $z=0.085$, $z=0.145$), as shown in Figure~\ref{fig:smal-3}, confirms a nonuniform distribution that cannot be totally explained by Malmquist bias. The fundamental reasoning is: if a galaxy of a given mass is detectable at a relatively higher redshift ($z=0.145$), it is unquestionably detectable at a lower redshift ($z=0.045$). Therefore, a deficiency of such galaxies at lower redshifts is unlikely to be a selection effect and is more plausibly due to intrinsic galaxy evolution or another, yet unidentified, bias.

While we have thoroughly controlled for all \textit{known} biases, as discussed in Section~\ref{sec:masses} and Appendix~\ref{sec:appendix_A}, we acknowledge the possible presence of an \textit{unknown} systematic effect that could subtly influence our results. We emphasize that all biases currently recognized in the literature and within our analysis framework have been carefully addressed. Nevertheless, we remain open to the possibility of additional, as yet unidentified, sources of systematic uncertainty. We are committed to investigating this further in future work by analyzing surveys with a broader range of completeness, noting that incorporating and analyzing such additional datasets lies beyond the scope of the present paper.

On the other hand, the large uncertainties affecting our data do not erase this systematic behavior; this affirmation can be supported following~\citet{Nig+09}. In Appendix~\ref{sec:appendix_A} of that paper, they showed, using bootstrap resampling and hypothesis testing, that even when uncertainties are comparable to a particular trend amplitude, the monotonic trend remains statistically significant. 

Finally, we calculated the logarithmic dynamical-to-baryonic mass difference for the Milky Way and Andromeda (M31) inside their optical radii, obtaining $\Delta$ $\log(\mathbf{M_{MW}})=0.54\pm0.15$ dex and $\Delta$ $\log(\mathbf{M_{M31}})=0.53\pm0.15$ dex. The detailed calculation of these parameters is presented in Appendix~\ref{sec:milky}. The positions of these two Local Group galaxies in Figures~\ref{fig:oVSv:MEDextra} and~\ref{fig:oVSv:MED2extra} (blue dot for the Milky Way and red dot for Andromeda) fall within the trends obtained in this work for galaxies of comparable dynamical mass and redshift. The results show that the dark matter inside the optical radius of both galaxies is greater than that observed for galaxies of similar mass at high redshift. This finding reinforces that our methodology and conclusions regarding the mass difference at higher redshifts are consistent with well-characterized nearby galaxies.

\section{Conclusions}
\label{sec:conclusions}

In this study, we utilize a sample of approximately 126,000 LTGs from the SDSS DR16, spanning the redshift range of approximately $0.00 < z < 0.35$ and dynamical mass range of $9.5 < \log(\mathbf{M_{ Dyn}/M_{Solar}}) < 12.5$. We conduct a comparison between the stellar mass and the dynamical mass, considering eight stellar mass estimates obtained through different population synthesis models and different IMFs. Additionally, we consider seven dynamical mass estimates based on Newtonian dynamics, virial equilibrium, different components (gas, stars), and several variables that may impact the estimation of dynamical mass (galactic inclination angle, color, concentration index, scale factor $K$, Sérsic index). The calibration of dynamical masses was performed following the methodology outlined by \citet{Nigoche22}. The analysis has yielded the following results: 

\begin{itemize}


\item The difference between dynamical and stellar mass appears to depend on both mass and redshift.


\item The difference between dynamical and stellar mass, in general, appears to decrease as redshift rises (see Fig.~\ref{fig:eVSr:MED} and the figures in Appendix~\ref{sec:app:b}). This behavior is similar when considering different variables in the dynamical mass (LTGs dominated by the disk or by the bulge) and stellar mass (different population synthesis models, different IMF) estimates. However, this behavior appears to be more pronounced for those LTGs dominated by the disk. 


\item The difference between dynamical and stellar mass can range from nearly zero to approximately 95\% of the dynamical mass, depending on dynamical mass and redshift (see Fig.~\ref{fig:oVSv:MED}).

\item The behavior of the difference between dynamical and stellar mass as a function of the dynamical mass and redshift reveals a complex pattern (see Fig.~\ref{fig:oVSv:MED} and figures in appendix~\ref{sec:app:c}). At low dynamical mass and low redshift, the difference is relatively large and tends to decrease as dynamical mass increases, producing a saddlelike shape. This saddle is more pronounced in disk-dominated LTGs, though it is also present in bulge-dominated systems. As redshift increases, this saddlelike behavior gradually transitions into a steeper, more linear trend, where the mass difference increases with dynamical mass. In the high-mass and high-redshift regime, the relation approximates a straight line with positive slope, resembling the trend reported for ETGs by \citet{Nig16}. In this regime, the mass difference grows with dynamical mass, while overall decreasing with redshift.



\item Since the current study estimates stellar masses using SPS models with fixed IMFs, any discrepancy between dynamical and stellar mass may arise from the assumption of a constant IMF, potential underestimations of the baryonic content, and/or the contribution of dark matter. This implies that the inferred amount of dark matter is sensitive to how the IMF and SPS models influence the stellar mass calculation. Although SPS-based stellar mass estimates do not explicitly account for the gas component, previous studies—such as \citet{Pop15}—have shown that galaxies with log($\mathbf{M_{Stellar}/M_{Solar}}) > 10$ at \( z \leq 0.3 \) are predominantly dominated by their stellar component. Given that the majority of galaxies in our sample fall within this mass and redshift regime, the exclusion of the gas component is not expected to significantly impact our conclusions.

\end{itemize}

Our general results suggest that galaxies at relatively higher redshifts appear to be more baryon-dominated, with a greater proportion of baryonic matter compared to dark matter.  This can be attributed to factors such as efficient baryon condensation in the compact, gas-rich galaxies of the relatively early Universe, less concentrated dark matter halos at higher redshifts, and dynamic gas flows that accumulate baryons at galaxy centers, reducing the influence of dark matter on their dynamics. The good agreement between the results for the Milky Way and Andromeda and the general trends validates our approach, demonstrating the coherence of our methodology and conclusions when compared with these well-characterized nearby galaxies. 

Moreover, our results suggest that the evolution of galaxies within the nearby Universe depends on whether the galaxy is dominated by the disk or dominated by the bulge. Nonetheless, rather than being distinct, the evolution exhibits a continuous transition from one regime to another.

In conclusion, the implications of this research extend beyond individual galaxies, offering a broader perspective on the nearby Universe and the role of dark matter in shaping the Universe's structure and evolution.

These findings indicate that the influence of dark matter in galaxy formation and evolution is more complex than previously understood. They suggest that models of galaxy formation and evolution may need to incorporate a more seamless transition between different morphological types.

However, while our methodology utilizes narrow dynamical mass bins ($\Delta$ log M$_{Dyn}=0.1$ dex) and narrow redshift bins ($\Delta$$z=0.01$) to compare galaxies of similar mass across different epochs, Figures~\ref{fig:vVSr:MED}, and~\ref{fig:sVSr:Mka} show that the sample at higher redshifts lacks sufficient representation in the lower-mass bins. Consequently, at the highest redshifts, our comparisons effectively involve higher-mass systems. This limitation should be borne in mind when interpreting any evolution, or lack thereof, in the derived quantities. We identify the use of surveys with more extensive completeness ranges as a crucial next step, and we commit to this line of inquiry in future work, noting that such an analysis exceeds the boundaries of the current study. 




\begin{acknowledgements}
We thank the Instituto de Astronom\'{\i}a y Meteorolog\'{\i}a (UdG, M\'exico) for all the facilities provided for the 
realization of this project. A.N.-N. acknowledges support from CONAHCyT and PRODEP (M\'exico). E.D.L.F. thanks Colegio departamental de F\'isica, authorities of CUCEI, and Coordinaci\'on General Acad\'emica y de Innovaci\'on (CGAI-UDG). P.L. gratefully acknowledges support by the GEMINI ANID
project No. 32240002. The work of R.J.D. is supported by the International Gemini Observatory, a program of NSF NOIRLab, which is managed by the Association of Universities for Research in Astronomy (AURA) under a cooperative agreement with the U.S. National Science Foundation, on behalf of the Gemini partnership of Argentina, Brazil, Canada, Chile, the Republic of Korea, and the United States of America.
\end{acknowledgements}


\bibliography{referenciasDM_en_Espirales}

@INCOLLECTION{Burbidge1975,
       author = {{Burbidge}, E.~M. and {Burbidge}, G.~R.},
        title = "{The Masses of Galaxies}",
    booktitle = {Galaxies and the Universe},
         year = 1975,
       editor = {{Sandage}, Allan and {Sandage}, Mary and {Kristian}, Jerome},
       publisher = {University of Chicago Press, Stars and Stellar Systems, Volume 9},
        pages = {81},
       adsurl = {https://ui.adsabs.harvard.edu/abs/1975gaun.book...81B}
}

@ARTICLE{Slipher1914,
       author = {{Slipher}, V.~M.},
        title = "{The Radial Velocity of the Andromeda Nebula}",
      journal = {Popular Astronomy},
         year = 1914,
        month = jan,
       volume = {22},
        pages = {19-21},
       adsurl = {https://ui.adsabs.harvard.edu/abs/1914PA.....22...19S}
}

@ARTICLE{Pease1918,
       author = {{Pease}, F.~G.},
        title = "{The Rotation and Radial Velocity of the Central Part of the Andromeda Nebula}",
      journal = {Proceedings of the National Academy of Science},
         year = 1918,
        month = jan,
       volume = {4},
       number = {1},
        pages = {21-24},
          doi = {10.1073/pnas.4.1.21},
       adsurl = {https://ui.adsabs.harvard.edu/abs/1918PNAS....4...21P}
}

@ARTICLE{Opik1922,
       author = {{Opik}, E.},
        title = "{An estimate of the distance of the Andromeda Nebula.}",
      journal = {\apj},
         year = 1922,
        month = jun,
       volume = {55},
        pages = {406-410},
          doi = {10.1086/142680},
       adsurl = {https://ui.adsabs.harvard.edu/abs/1922ApJ....55..406O}
}

@ARTICLE{KapteynvanRhijn1922,
       author = {{Kapteyn}, J.~C. and {van Rhijn}, P.~J.},
        title = "{The proper motions of {\ensuremath{\delta}} Cephei stars and the distances of the globular clusters}",
      journal = {\bain},
         year = 1922,
        month = mar,
       volume = {1},
        pages = {37},
       adsurl = {https://ui.adsabs.harvard.edu/abs/1922BAN.....1...37K}
}

@ARTICLE{Oort1932a,
       author = {{Oort}, J.~H.},
        title = "{Note on the distribution of luminosities of K and M giants}",
      journal = {\bain},
         year = 1932,
        month = aug,
       volume = {6},
        pages = {289},
       adsurl = {https://ui.adsabs.harvard.edu/abs/1932BAN.....6..289O}
}

@ARTICLE{Oort1932b,
       author = {{Oort}, J.~H.},
        title = "{The force exerted by the stellar system in the direction perpendicular to the galactic plane and some related problems}",
      journal = {\bain},
         year = 1932,
        month = aug,
       volume = {6},
        pages = {249},
       adsurl = {https://ui.adsabs.harvard.edu/abs/1932BAN.....6..249O}
}

@ARTICLE{Babcock1939,
       author = {{Babcock}, Horace W.},
        title = "{The rotation of the Andromeda Nebula}",
      journal = {Lick Observatory Bulletin},
         year = 1939,
        month = jan,
       volume = {498},
        pages = {41-51},
          doi = {10.5479/ADS/bib/1939LicOB.19.41B},
       adsurl = {https://ui.adsabs.harvard.edu/abs/1939LicOB..19...41B}
}

@ARTICLE{Lallemand1960,
       author = {{Lallemand}, A. and {Duchesne}, M. and {Walker}, M.~F.},
      journal = {\pasp},
         year = 1960,
        month = aug,
       volume = {72},
        pages = {283},
       adsurl = {https://ui.adsabs.harvard.edu/abs/1960PASP...72..283L}
}

@ARTICLE{FaberGallagher1979,
       author = {{Faber}, S.~M. and {Gallagher}, J.~S.},
        title = "{Masses and mass-to-light ratios of galaxies.}",
      journal = {\araa},
         year = 1979,
        month = jan,
       volume = {17},
        pages = {135-187},
          doi = {10.1146/annurev.aa.17.090179.001031},
       adsurl = {https://ui.adsabs.harvard.edu/abs/1979ARA&A..17..135F}
}

@ARTICLE{Rubinetal1985,
       author = {{Rubin}, V.~C. and {Burstein}, D. and {Ford}, W.~K., Jr. and {Thonnard}, N.},
        title = "{Rotation velocities of 16 SA galaxies and a comparison of Sa, SB and SC rotation properties.}",
      journal = {\apj},
         year = 1985,
        month = feb,
       volume = {289},
        pages = {81-104},
          doi = {10.1086/162866},
       adsurl = {https://ui.adsabs.harvard.edu/abs/1985ApJ...289...81R}
}

@ARTICLE{SofueRubin2001,
       author = {{Sofue}, Yoshiaki and {Rubin}, Vera},
        title = "{Rotation Curves of Spiral Galaxies}",
      journal = {\araa},
         year = 2001,
        month = jan,
       volume = {39},
        pages = {137-174},
          doi = {10.1146/annurev.astro.39.1.137},
archivePrefix = {arXiv},
       eprint = {astro-ph/0010594},
 primaryClass = {astro-ph},
       adsurl = {https://ui.adsabs.harvard.edu/abs/2001ARA&A..39..137S}
}

@ARTICLE{Nigoche22,
       author = {{Nigoche-Netro}, A. and {de la Fuente}, E. and {Diaz}, R. J. and {Agüero}, M. P. and {Kemp}, S. N. and {Marquez-Lugo}, R. A. and {Lagos}, P. and {Ruelas-Mayorga}, A. and {López-Contreras}, N. L.},
        title = "{Virial masses of late-type galaxies from the SDSS DR16}",
      journal = {\mnras},
         year = 2022,
        month = sep,
       volume = {515},
        pages = {27},
          doi = {10.1093/mnras/stac1872},
       adsurl = {https://ui.adsabs.harvard.edu/abs/2022MNRAS.515.2351N/abstract}
}

@ARTICLE{Poveda1958,
       author = {{Poveda}, A.},
        title = "{The Masses of Spherical Galaxies M32. A likely application}",
      journal = {Boletin de los Observatorios Tonantzintla y Tacubaya},
         year = 1958,
        month = apr,
       volume = {2},
        pages = {3-7},
       adsurl = {https://ui.adsabs.harvard.edu/abs/1958BOTT....2q...3P}
}

@ARTICLE{deLuciaetal2007,
       author = {{De Lucia}, Gabriella and {Poggianti}, Bianca M. and {Arag{\'o}n-Salamanca}, Alfonso and {White}, Simon D.~M. and {Zaritsky}, Dennis and {Clowe}, Douglas and {Halliday}, Claire and {Jablonka}, Pascale and {von der Linden}, Anja and {Milvang-Jensen}, Bo and {Pell{\'o}}, Roser and {Rudnick}, Gregory and {Saglia}, Roberto P. and {Simard}, Luc},
        title = "{The build-up of the colour-magnitude relation in galaxy clusters since z \raisebox{-0.5ex}\textasciitilde 0.8}",
      journal = {\mnras},
         year = 2007,
        month = jan,
       volume = {374},
       number = {3},
        pages = {809-822},
          doi = {10.1111/j.1365-2966.2006.11199.x},
archivePrefix = {arXiv},
       eprint = {astro-ph/0610373},
 primaryClass = {astro-ph},
       adsurl = {https://ui.adsabs.harvard.edu/abs/2007MNRAS.374..809D}
}

@ARTICLE{Walcheretal2011,
       author = {{Walcher}, Jakob and {Groves}, Brent and {Budav{\'a}ri}, Tam{\'a}s and {Dale}, Daniel},
        title = "{Fitting the integrated spectral energy distributions of galaxies}",
      journal = {\apss},
         year = 2011,
        month = jan,
       volume = {331},
        pages = {1-52},
          doi = {10.1007/s10509-010-0458-z},
archivePrefix = {arXiv},
       eprint = {1008.0395},
 primaryClass = {astro-ph.CO},
       adsurl = {https://ui.adsabs.harvard.edu/abs/2011Ap&SS.331....1W}
}

@ARTICLE{Conroy2013,
       author = {{Conroy}, Charlie},
        title = "{Modeling the Panchromatic Spectral Energy Distributions of Galaxies}",
      journal = {\araa},
         year = 2013,
        month = aug,
       volume = {51},
       number = {1},
        pages = {393-455},
          doi = {10.1146/annurev-astro-082812-141017},
archivePrefix = {arXiv},
       eprint = {1301.7095},
 primaryClass = {astro-ph.CO},
       adsurl = {https://ui.adsabs.harvard.edu/abs/2013ARA&A..51..393C}
}

@ARTICLE{Oort1926,
       author = {{Oort}, J.~H.},
        title = "{Asymmetry in the distribution of stellar velocities}",
      journal = {The Observatory},
         year = 1926,
        month = oct,
       volume = {49},
        pages = {302-304},
       adsurl = {https://ui.adsabs.harvard.edu/abs/1926Obs....49..302O}
}

@ARTICLE{Baade1944,
       author = {{Baade}, W.},
        title = "{The Resolution of Messier 32, NGC 205, and the Central Region of the Andromeda Nebula.}",
      journal = {\apj},
         year = 1944,
        month = sep,
       volume = {100},
        pages = {137},
          doi = {10.1086/144650},
       adsurl = {https://ui.adsabs.harvard.edu/abs/1944ApJ...100..137B}
}

@BOOK{GreggioRenzini2011,
       author = {{Greggio}, Laura and {Renzini}, Alvio},
        title = "{Stellar Populations. A User Guide from Low to High Redshift}",
         year = 2011,
       publisher = {Wiley VCH Verlag, Series in Cosmology, 265 pages},
       adsurl = {https://ui.adsabs.harvard.edu/abs/2011spug.book.....G}
}

@ARTICLE{Sandage1986a,
       author = {{Sandage}, A.},
        title = "{Star formation rates, galaxy morphology and the Hubble sequence.}",
      journal = {\aap},
         year = 1986,
        month = jun,
       volume = {161},
        pages = {89-101},
       adsurl = {https://ui.adsabs.harvard.edu/abs/1986A&A...161...89S}
}

@ARTICLE{MacArthuretal2004,
       author = {{MacArthur}, Lauren A. and {Courteau}, St{\'e}phane and {Bell}, Eric and {Holtzman}, Jon A.},
        title = "{Structure of Disk-dominated Galaxies. II. Color Gradients and Stellar Population Models}",
      journal = {\apjs},
         year = 2004,
        month = jun,
       volume = {152},
       number = {2},
        pages = {175-199},
          doi = {10.1086/383525},
archivePrefix = {arXiv},
       eprint = {astro-ph/0401437},
 primaryClass = {astro-ph},
       adsurl = {https://ui.adsabs.harvard.edu/abs/2004ApJS..152..175M}
}

@ARTICLE{Salpeter1955,
       author = {{Salpeter}, Edwin E.},
        title = "{The Luminosity Function and Stellar Evolution.}",
      journal = {\apj},
         year = 1955,
        month = jan,
       volume = {121},
        pages = {161},
          doi = {10.1086/145971},
       adsurl = {https://ui.adsabs.harvard.edu/abs/1955ApJ...121..161S}
}

@ARTICLE{Scalo1986,
       author = {{Scalo}, J.~M.},
        title = "{The Stellar Initial Mass Function}",
      journal = {\fcp},
         year = 1986,
        month = may,
       volume = {11},
        pages = {1-278},
       adsurl = {https://ui.adsabs.harvard.edu/abs/1986FCPh...11....1S}
}

@ARTICLE{Kroupa2001,
       author = {{Kroupa}, Pavel},
        title = "{On the variation of the initial mass function}",
      journal = {\mnras},
         year = 2001,
        month = apr,
       volume = {322},
       number = {2},
        pages = {231-246},
          doi = {10.1046/j.1365-8711.2001.04022.x},
archivePrefix = {arXiv},
       eprint = {astro-ph/0009005},
 primaryClass = {astro-ph},
       adsurl = {https://ui.adsabs.harvard.edu/abs/2001MNRAS.322..231K}
}

@ARTICLE{Chabrier2003,
       author = {{Chabrier}, Gilles},
        title = "{Galactic Stellar and Substellar Initial Mass Function}",
      journal = {\pasp},
         year = 2003,
        month = jul,
       volume = {115},
       number = {809},
        pages = {763-795},
          doi = {10.1086/376392},
archivePrefix = {arXiv},
       eprint = {astro-ph/0304382},
 primaryClass = {astro-ph},
       adsurl = {https://ui.adsabs.harvard.edu/abs/2003PASP..115..763C}
}

@ARTICLE{Tysonetal1984,
       author = {{Tyson}, J.~A. and {Valdes}, F. and {Jarvis}, J.~F. and {Mills}, A.~P., Jr.},
        title = "{Galaxy mass distribution from gravitational light deflection}",
      journal = {\apjl},
         year = 1984,
        month = jun,
       volume = {281},
        pages = {L59-L62},
          doi = {10.1086/184285},
       adsurl = {https://ui.adsabs.harvard.edu/abs/1984ApJ...281L..59T}
}

@INPROCEEDINGS{BrainerdBlandfordS1996,
       author = {{Brainerd}, T.~G. and {Blandford}, R.~D. and {Smail}, I.},
        title = "{Weak Gravitational Lensing by Galaxies - Implications for Dark Matter Halos}",
    booktitle = {American Astronomical Society Meeting Abstracts \#188},
         year = 1996,
       series = {American Astronomical Society Meeting Abstracts},
       volume = {188},
        month = may,
          eid = {13.02},
        pages = {13.02},
       adsurl = {https://ui.adsabs.harvard.edu/abs/1996AAS...188.1302B}
}

@ARTICLE{yor00,
       author = {{York}, Donald G. and {Adelman}, J. and {Anderson}, John E., Jr. and {Anderson}, Scott F. and {Annis}, James and {Bahcall}, Neta A. and {Bakken}, J.~A. and {Barkhouser}, Robert and {Bastian}, Steven and {Berman}, Eileen and {Boroski}, William N. and {Bracker}, Steve and {Briegel}, Charlie and {Briggs}, John W. and {Brinkmann}, J. and {Brunner}, Robert and {Burles}, Scott and {Carey}, Larry and {Carr}, Michael A. and {Castander}, Francisco J. and {Chen}, Bing and {Colestock}, Patrick L. and {Connolly}, A.~J. and {Crocker}, J.~H. and {Csabai}, Istv{\'a}n and {Czarapata}, Paul C. and {Davis}, John Eric and {Doi}, Mamoru and {Dombeck}, Tom and {Eisenstein}, Daniel and {Ellman}, Nancy and {Elms}, Brian R. and {Evans}, Michael L. and {Fan}, Xiaohui and {Federwitz}, Glenn R. and {Fiscelli}, Larry and {Friedman}, Scott and {Frieman}, Joshua A. and {Fukugita}, Masataka and {Gillespie}, Bruce and {Gunn}, James E. and {Gurbani}, Vijay K. and {de Haas}, Ernst and {Haldeman}, Merle and {Harris}, Frederick H. and {Hayes}, J. and {Heckman}, Timothy M. and {Hennessy}, G.~S. and {Hindsley}, Robert B. and {Holm}, Scott and {Holmgren}, Donald J. and {Huang}, Chi-hao and {Hull}, Charles and {Husby}, Don and {Ichikawa}, Shin-Ichi and {Ichikawa}, Takashi and {Ivezi{\'c}}, {\v{Z}}eljko and {Kent}, Stephen and {Kim}, Rita S.~J. and {Kinney}, E. and {Klaene}, Mark and {Kleinman}, A.~N. and {Kleinman}, S. and {Knapp}, G.~R. and {Korienek}, John and {Kron}, Richard G. and {Kunszt}, Peter Z. and {Lamb}, D.~Q. and {Lee}, B. and {Leger}, R. French and {Limmongkol}, Siriluk and {Lindenmeyer}, Carl and {Long}, Daniel C. and {Loomis}, Craig and {Loveday}, Jon and {Lucinio}, Rich and {Lupton}, Robert H. and {MacKinnon}, Bryan and {Mannery}, Edward J. and {Mantsch}, P.~M. and {Margon}, Bruce and {McGehee}, Peregrine and {McKay}, Timothy A. and {Meiksin}, Avery and {Merelli}, Aronne and {Monet}, David G. and {Munn}, Jeffrey A. and {Narayanan}, Vijay K. and {Nash}, Thomas and {Neilsen}, Eric and {Neswold}, Rich and {Newberg}, Heidi Jo and {Nichol}, R.~C. and {Nicinski}, Tom and {Nonino}, Mario and {Okada}, Norio and {Okamura}, Sadanori and {Ostriker}, Jeremiah P. and {Owen}, Russell and {Pauls}, A. George and {Peoples}, John and {Peterson}, R.~L. and {Petravick}, Donald and {Pier}, Jeffrey R. and {Pope}, Adrian and {Pordes}, Ruth and {Prosapio}, Angela and {Rechenmacher}, Ron and {Quinn}, Thomas R. and {Richards}, Gordon T. and {Richmond}, Michael W. and {Rivetta}, Claudio H. and {Rockosi}, Constance M. and {Ruthmansdorfer}, Kurt and {Sandford}, Dale and {Schlegel}, David J. and {Schneider}, Donald P. and {Sekiguchi}, Maki and {Sergey}, Gary and {Shimasaku}, Kazuhiro and {Siegmund}, Walter A. and {Smee}, Stephen and {Smith}, J. Allyn and {Snedden}, S. and {Stone}, R. and {Stoughton}, Chris and {Strauss}, Michael A. and {Stubbs}, Christopher and {SubbaRao}, Mark and {Szalay}, Alexander S. and {Szapudi}, Istvan and {Szokoly}, Gyula P. and {Thakar}, Anirudda R. and {Tremonti}, Christy and {Tucker}, Douglas L. and {Uomoto}, Alan and {Vanden Berk}, Dan and {Vogeley}, Michael S. and {Waddell}, Patrick and {Wang}, Shu-i. and {Watanabe}, Masaru and {Weinberg}, David H. and {Yanny}, Brian and {Yasuda}, Naoki and {SDSS Collaboration}},
        title = "{The Sloan Digital Sky Survey: Technical Summary}",
      journal = {\aj},
         year = 2000,
        month = sep,
       volume = {120},
       number = {3},
        pages = {1579-1587},
          doi = {10.1086/301513},
archivePrefix = {arXiv},
       eprint = {astro-ph/0006396},
 primaryClass = {astro-ph},
       adsurl = {https://ui.adsabs.harvard.edu/abs/2000AJ....120.1579Y}
}

@ARTICLE{bla03,
       author = {{Blanton}, Michael R. and {Hogg}, David W. and {Bahcall}, Neta A. and {Baldry}, Ivan K. and {Brinkmann}, J. and {Csabai}, Istv{\'a}n and {Eisenstein}, Daniel and {Fukugita}, Masataka and {Gunn}, James E. and {Ivezi{\'c}}, {\v{Z}}eljko and {Lamb}, D.~Q. and {Lupton}, Robert H. and {Loveday}, Jon and {Munn}, Jeffrey A. and {Nichol}, R.~C. and {Okamura}, Sadanori and {Schlegel}, David J. and {Shimasaku}, Kazuhiro and {Strauss}, Michael A. and {Vogeley}, Michael S. and {Weinberg}, David H.},
        title = "{The Broadband Optical Properties of Galaxies with Redshifts 0.02<z<0.22}",
      journal = {\apj},
         year = 2003,
        month = sep,
       volume = {594},
       number = {1},
        pages = {186-207},
          doi = {10.1086/375528},
archivePrefix = {arXiv},
       eprint = {astro-ph/0209479},
 primaryClass = {astro-ph},
       adsurl = {https://ui.adsabs.harvard.edu/abs/2003ApJ...594..186B}
}

@ARTICLE{nig08,
       author = {{Nigoche-Netro}, A. and {Ruelas-Mayorga}, A. and {Franco-Balderas}, A.},
        title = "{The Kormendy relation for early-type galaxies. Dependence on the magnitude range}",
      journal = {\aap},
         year = 2008,
        month = dec,
       volume = {491},
       number = {3},
        pages = {731-738},
          doi = {10.1051/0004-6361:200810211},
archivePrefix = {arXiv},
       eprint = {0805.0961},
 primaryClass = {astro-ph},
       adsurl = {https://ui.adsabs.harvard.edu/abs/2008A&A...491..731N}
}

@ARTICLE{jor95a,
       author = {{J\o rgensen}, Inger and {Franx}, Marijn and {Kjaergaard}, Per},
        title = "{Multicolour CCD surface photometry for E and S0 galaxies in 10 clusters}",
      journal = {\mnras},
         year = 1995,
        month = apr,
       volume = {273},
       number = {4},
        pages = {1097-1128},
          doi = {10.1093/mnras/273.4.1097},
       adsurl = {https://ui.adsabs.harvard.edu/abs/1995MNRAS.273.1097J}
}

@ARTICLE{hyd09,
       author = {{Hyde}, Joseph B. and {Bernardi}, Mariangela},
        title = "{Curvature in the scaling relations of early-type galaxies}",
      journal = {\mnras},
         year = 2009,
        month = apr,
       volume = {394},
       number = {4},
        pages = {1978-1990},
          doi = {10.1111/j.1365-2966.2009.14445.x},
archivePrefix = {arXiv},
       eprint = {0810.4922},
 primaryClass = {astro-ph},
       adsurl = {https://ui.adsabs.harvard.edu/abs/2009MNRAS.394.1978H}
}

@ARTICLE{jor95b,
       author = {{J\o rgensen}, Inger and {Franx}, Marijn and {Kjaergaard}, Per},
        title = "{Spectroscopy for E and S0 galaxies in nine clusters}",
      journal = {\mnras},
         year = 1995,
        month = oct,
       volume = {276},
       number = {4},
        pages = {1341-1364},
          doi = {10.1093/mnras/276.4.1341},
       adsurl = {https://ui.adsabs.harvard.edu/abs/1995MNRAS.276.1341J}
}

@ARTICLE{cap06,
       author = {{Cappellari}, Michele and {Bacon}, R. and {Bureau}, M. and {Damen}, M.~C. and {Davies}, Roger L. and {de Zeeuw}, P.~T. and {Emsellem}, Eric and {Falc{\'o}n-Barroso}, Jes{\'u}s and {Krajnovi{\'c}}, Davor and {Kuntschner}, Harald and {McDermid}, Richard M. and {Peletier}, Reynier F. and {Sarzi}, Marc and {van den Bosch}, Remco C.~E. and {van de Ven}, Glenn},
        title = "{The SAURON project - IV. The mass-to-light ratio, the virial mass estimator and the Fundamental Plane of elliptical and lenticular galaxies}",
      journal = {\mnras},
         year = 2006,
        month = mar,
       volume = {366},
       number = {4},
        pages = {1126-1150},
          doi = {10.1111/j.1365-2966.2005.09981.x},
archivePrefix = {arXiv},
       eprint = {astro-ph/0505042},
 primaryClass = {astro-ph},
       adsurl = {https://ui.adsabs.harvard.edu/abs/2006MNRAS.366.1126C}
}

@ARTICLE{moc12,
       author = {{Mocz}, P. and {Green}, A. and {Malacari}, M. and {Glazebrook}, K.},
        title = "{The Tully-Fisher relation for 25 000 Sloan Digital Sky Survey galaxies as a function of environment}",
      journal = {\mnras},
         year = 2012,
        month = sep,
       volume = {425},
       number = {1},
        pages = {296-310},
          doi = {10.1111/j.1365-2966.2012.21458.x},
archivePrefix = {arXiv},
       eprint = {1206.1662},
 primaryClass = {astro-ph.CO},
       adsurl = {https://ui.adsabs.harvard.edu/abs/2012MNRAS.425..296M}
}

@ARTICLE{cat05,
       author = {{Catinella}, Barbara and {Haynes}, Martha P. and {Giovanelli}, Riccardo},
        title = "{Rotational Widths for Use in the Tully-Fisher Relation. I. Long-Slit Spectroscopic Data}",
      journal = {\aj},
         year = 2005,
        month = sep,
       volume = {130},
       number = {3},
        pages = {1037-1048},
          doi = {10.1086/432543},
archivePrefix = {arXiv},
       eprint = {astro-ph/0506148},
 primaryClass = {astro-ph},
       adsurl = {https://ui.adsabs.harvard.edu/abs/2005AJ....130.1037C}
}

@ARTICLE{gra05,
       author = {{Graham}, Alister W. and {Driver}, Simon P. and {Petrosian}, Vah{\'e} and {Conselice}, Christopher J. and {Bershady}, Matthew A. and {Crawford}, Steven M. and {Goto}, Tomotsugu},
        title = "{Total Galaxy Magnitudes and Effective Radii from Petrosian Magnitudes and Radii}",
      journal = {\aj},
         year = 2005,
        month = oct,
       volume = {130},
       number = {4},
        pages = {1535-1544},
          doi = {10.1086/444475},
archivePrefix = {arXiv},
       eprint = {astro-ph/0504287},
 primaryClass = {astro-ph},
       adsurl = {https://ui.adsabs.harvard.edu/abs/2005AJ....130.1535G}
}

@ARTICLE{nig10,
       author = {{Nigoche-Netro}, A. and {Aguerri}, J.~A.~L. and {Lagos}, P. and {Ruelas-Mayorga}, A. and {S{\'a}nchez}, L.~J. and {Machado}, A.},
        title = "{The Faber-Jackson relation for early-type galaxies: dependence on the magnitude range}",
      journal = {\aap},
         year = 2010,
        month = jun,
       volume = {516},
          eid = {A96},
        pages = {A96},
          doi = {10.1051/0004-6361/200912719},
archivePrefix = {arXiv},
       eprint = {1004.5496},
 primaryClass = {astro-ph.CO},
       adsurl = {https://ui.adsabs.harvard.edu/abs/2010A&A...516A..96N}
}

@ARTICLE{Nig15,
       author = {{Nigoche-Netro}, A. and {Ruelas-Mayorga}, ~A. and {Lagos}, P. and {Ramos-Larios}, G. and {Kehrig}, C. and {Kemp}, S. N. and {Montero-Dorta}, A. D. and {Gonz\'alez-Cervantes}, J.},
        title = "{How much dark matter is there inside early-type galaxies?}",
      journal = {\mnras},
         year = 2015,
        month = jan,
       volume = {446},
          eid = {85},
        pages = {18},
          doi = {10.1093/mnras/stu2045},
archivePrefix = {arXiv},
       eprint = {1410.1469},
 primaryClass = {astro-ph.CO},
       adsurl = {https://ui.adsabs.harvard.edu/abs/2015MNRAS.446...85N/abstract}
}

@ARTICLE{Nig16,
       author = {{Nigoche-Netro}, A. and {Ramos-Larios}, G. and {Lagos}, P. and {Ruelas-Mayorga}, A. and {de la Fuente}, E. and {Kemp}, S. N. and {Navarro}, S. G. and {Corral}, L. J. and {Hidalgo-G\'amez} A. M. },
        title = "{Dark matter inside early-type galaxies as function of mass and redshift}",
      journal = {\mnras},
         year = 2016,
        month = oct,
       volume = {462},
        pages = {11},
          doi = {10.1093/mnras/stw1661},
       adsurl = {https://ui.adsabs.harvard.edu/abs/2016MNRAS.462..951N/abstract}
}

@ARTICLE{Nig19,
       author = {{Nigoche-Netro}, A. and {Ramos-Larios}, G. and {Lagos}, P. and {de la Fuente}, E.  and {Ruelas-Mayorga}, A. and {Mendez-Abreu}, J. and {Kemp}, S. N.  and {Diaz}, R. J. },
        title = "{The quantity of dark matter in early-type galaxies and its relation to the environment}",
      journal = {\mnras},
         year = 2019,
        month = sep,
       volume = {488},
        pages = {15},
          doi = {10.1093/mnras/stz1786 },
       adsurl = {https://ui.adsabs.harvard.edu/abs/2019MNRAS.488.1320N/abstract}
}

@article{Courteauetal2014,
  title = {Galaxy masses},
  author = {Courteau, St\'ephane and Cappellari, Michele and de Jong, Roelof S. and Dutton, Aaron A. and Emsellem, Eric and Hoekstra, Henk and Koopmans, L. V. E. and Mamon, Gary A. and Maraston, Claudia and Treu, Tommaso and Widrow, Lawrence M.},
  journal = {Rev. Mod. Phys.},
  volume = {86},
  issue = {1},
  pages = {47--119},
  numpages = {0},
  year = {2014},
  month = {Jan},
  publisher = {American Physical Society},
  doi = {10.1103/RevModPhys.86.47},
  url = {https://link.aps.org/doi/10.1103/RevModPhys.86.47}
}

@ARTICLE{Scheiner1899,
       author = {{Scheiner}, J.},
        title = "{On the spectrum of the great nebula in Andromeda.}",
      journal = {\apj},
         year = 1899,
        month = mar,
       volume = {9},
        pages = {149-150},
          doi = {10.1086/140564},
       adsurl = {https://ui.adsabs.harvard.edu/abs/1899ApJ.....9..149S}
}

@ARTICLE{Mayall1951,
       author = {{Mayall}, N.~U.},
        title = "{Comparison of Rotational Motions Observed in the Spirals M31 and M33 and in The Galaxy}",
      journal = {Publ. Obs. Univ. Michigan},
         year = 1951,
        month = jan,
       volume = {10},
        pages = {19},
       adsurl = {https://ui.adsabs.harvard.edu/abs/1951POMic..10...19M}
}

@ARTICLE{Fich1991,
       author = {{Fich}, Michel and {Tremaine}, Scott},
        title = "{The mass of the Galaxy.}",
      journal = {\araa},
         year = 1991,
        month = jan,
       volume = {29},
        pages = {409-445},
          doi = {10.1146/annurev.aa.29.090191.002205},
       adsurl = {https://ui.adsabs.harvard.edu/abs/1991ARA&A..29..409F}
}

@ARTICLE{Van08,
       author = {{Van Dokkum}, P. G.},
        title = "{Evidence of Cosmic Evolution of the Stellar Initial Mass Function}",
      journal = {\apj},
         year = 2008,
        month = feb,
       volume = {674},
        pages = {16},
          doi = {10.1086/525014 },
       adsurl = {https://ui.adsabs.harvard.edu/abs/2008ApJ...674...29V/abstract}
}

@ARTICLE{Con09,
       author = {{Conroy}, C. and {Gunn}, J. E. and {White}, M.},
        title = "{The propagation of uncertainties in stellar population synthesis modeling I: The relevance of uncertain aspects of stellar evolution and the IMF to the derived physical properties of galaxies}",
      journal = {\apj},
         year = 2009,
        month = jun,
       volume = {699},
        pages = {486-506},
          doi = {10.1088/0004-637X/699/1/486},
       adsurl = {https://ui.adsabs.harvard.edu/abs/2009ApJ...699..486C/abstract}
}

@ARTICLE{Mar09,
       author = {{Maraston}, C. and {Strömbäck}, T. and {Thomas}, D. and {Wake}, D. A. and {Nichol}, R. C},
        title = "{Modelling the colour evolution of luminous red galaxies - improvements with empirical stellar spectra}",
      journal = {\mnras},
         year = 2009,
        month = mar,
       volume = {394},
        pages = {5},
          doi = {10.1111/j.1745-3933.2009.00621.x},
       adsurl = {https://ui.adsabs.harvard.edu/abs/2009MNRAS.394L.107M/abstract}
}

@ARTICLE{Che12,
       author = {{Chen}, Y. and {Kauffmann}, G. and {Tremonti}, C. A. and {White}, S. and {Heckman}, T. M. and {Kovač}, K. and {Bundy}, K. and {Chisholm}, J. and {Maraston}, C. and {Schneider}, D. P. and {Bolton}, A. S. and {Weaver}, B. A. and {Brinkmann}, J. },           
        title = "{Evolution of the most massive galaxies to z= 0.6 - I. A new method for physical parameter estimation}",
      journal = {\mnras},
         year = 2012,
        month = mar,
       volume = {421},
        pages = {20},
          doi = {10.1111/j.1365-2966.2011.20306.x },
       adsurl = {https://ui.adsabs.harvard.edu/abs/2012MNRAS.421..314C/abstract}
}

@ARTICLE{Bru03,
       author = {{Bruzual}, G. and {Charlot}, S.},
        title = "{Stellar population synthesis at the resolution of 2003}",
      journal = {\mnras},
         year = 2003,
        month = oct,
       volume = {344},
        pages = {35},
          doi = {10.1046/j.1365-8711.2003.06897.x},
       adsurl = {https://ui.adsabs.harvard.edu/abs/2003MNRAS.344.1000B/abstract}
}

@ARTICLE{Mar11,
       author = {{Maraston}, C. and {Strömbäck}, G.},
        title = "{Stellar population models at high spectral resolution}",
      journal = {\mnras},
         year = 2011,
        month = dec,
       volume = {418},
        pages = {30},
          doi = {10.1111/j.1365-2966.2011.19738.x},
       adsurl = {https://ui.adsabs.harvard.edu/abs/2011MNRAS.418.2785M/abstract}
}

@ARTICLE{cap12,
       author = {{Cappellari}, M. and {McDermid}, R. M. and {Alatalo}, K. and {Alatalo}, K. and {Blitz}, L. and {Bois}, M. and {Bournaud}, F. and {Frédéric}, B.},
        title = "{Systematic variation of the stellar initial mass function in early-type galaxies}",
      journal = {Nature},
         year = 2012,
        month = apr,
       volume = {484},
        pages = {4},
          doi = {10.1038/nature10972},
       adsurl = {https://ui.adsabs.harvard.edu/abs/2012Natur.484..485C/abstract}
}

@ARTICLE{Mar23,
       author = {{Martín-Navarro}, I. and {de Lorenzo-Cáceres}, A. and {Gadotti}, D. and {Méndez-Abreu}, J. and {Falcón-Barroso}, J. and {Sánchez-Blázquez}, P. and {Coelho}, P. and {Neumann}, J. and {van de Ven}, G. and {Pérez}, I.},
        title = "{The universal variability of the stellar initial mass function probed by the TIMER survey}",
      journal = {Astronomy \& Astrophysics},
         year = 2023,
        month = dec,
       volume = {arXiv e-prints},
        pages = {16},
          doi = {10.48550/arXiv.2312.13355 },
       adsurl = {https://ui.adsabs.harvard.edu/abs/2023arXiv231213355M/abstract}
}

@ARTICLE{Sch24,
       author = {{Scholz-Díaz}, L. and {Martín-Navarro}, I. and {Falcón-Barroso}, J. and {Lyubenova}, M. and {van de Ven}, G. },
        title = "{Baryonic properties of nearby galaxies across the stellar-to-total dynamical mass relation}",
      journal = {Nature Astronomy},
         year = 2024,
        month = feb,
       volume = {8},
        pages = {648-656},
          doi = {10.1038/s41550-024-02209-8},
       adsurl = {https://ui.adsabs.harvard.edu/abs/2024NatAs.tmp...42S/abstract}
}

@ARTICLE{Gho23,
       author = {{Ghosh}, A. and {Adams}, D. and {Williams}, L. L. R. and {Liesenborgs}, J. and {Alavi}. A. and {Scarlata}, C.},
        title = "{An excursion into the core of the cluster lens Abell 1689}",
      journal = {MNRAS},
         year = 2023,
        month = oct,
       volume = {525},
        pages = {14},
          doi = {10.1093/mnras/stad2418},
       adsurl = {https://ui.adsabs.harvard.edu/abs/2023MNRAS.525.2519G/abstract}
}

@ARTICLE{Men17,
       author = {{Meneghetti}, M. and {Natarajan}, P. and {Coe}, D. and {Contini}, E. and {De Lucia}, G. and {Giocoli}, C. and {Acebron}, A. and {Borgani}, S. and {Bradac}, M. and {Diego}, J. M. and {Hoag}, A. and {Ishigaki}, M. and {Johnson}, T. L. and {Jullo}, E. and {Kawamata}, R. and {Lam}, D. and {Limousin}, M. and {Liesenborgs}, J. and {Oguri}, M. and {Sebesta}, K. and {Sharon}, K. and {Williams}, L. L. R. and {Zitrin}, A.},
        title = "{The Frontier Fields lens modelling comparison project}",
      journal = {MNRAS},
         year = 2017,
        month = dec,
       volume = {472},
        pages = {38},
          doi = {10.1093/mnras/stx2064 },
       adsurl = {https://ui.adsabs.harvard.edu/abs/2017MNRAS.472.3177M/abstract}
}

@ARTICLE{Mil83,
       author = {{Milgrom}, M.},
        title = "{A modification of the newtonian dynamics : implications for galaxy systems.}",
      journal = {ApJ},
         year = 1983,
        month = jul,
       volume = {270},
        pages = {6},
          doi = {10.1086/161132},
       adsurl = {https://ui.adsabs.harvard.edu/abs/1983ApJ...270..384M/abstract}
}

@ARTICLE{Mil88,
       author = {{Milgrom}, M.},
        title = "{On the Use of Galaxy Rotation Curves to Test the Modified Dynamics}",
      journal = {ApJ},
         year = 1988,
        month = oct,
       volume = {333},
        pages = {6},
          doi = {10.1086/166777},
       adsurl ={https://ui.adsabs.harvard.edu/abs/1988ApJ...333..689M/abstract}
}

@ARTICLE{Cha23,
       author = {{Chan}, M. H. and {Law}, K. C.},
        title = "{A Severe Challenge to the Modified Newtonian Dynamics Phenomenology in Our Galaxy}",
      journal = {ApJ},
         year = 2023,
        month = nov,
       volume = {957},
        pages = {6},
          doi = {10.3847/1538-4357/acf8c0},
       adsurl = {https://ui.adsabs.harvard.edu/abs/2023ApJ...957...24C/abstract}
}

@ARTICLE{Sorini24,
       author = {{Sorini}, D. and {Bose}, S and {Pakmor}, R. and {Hernquist}, L. and {Springel}, V and {Hadzhiyska}, B. and {Hernández-Aguayo}, C and {Kannan}, R},
        title = "{The impact of baryons on the internal structure of dark matter haloes
from dwarf galaxies to superclusters in the redshift range 0 < z < 7}",
      journal = {\mnras},
         year = 2025,
        month = Jan,
       volume = {536, 1},
        pages = {728-751},
          doi = {10.1093/mnras/stae2613},
       adsurl = {https://ui.adsabs.harvard.edu/abs/2025MNRAS.536..728S/abstract}
}

@ARTICLE{Ghosh16,
       author = {{Ghosh}, S. and {Saini}, T. D. and {Jog}, C. J.},
        title = "{Effect of dark matter halo on global spiral modes in galaxies}",
      journal = {MNRAS},
         year = 2016,
        month = Nov,
       volume = {456},
        pages = {943–950},
          doi = {https://doi.org/10.1093/mnras/stv2652},
       adsurl = {https://academic.oup.com/mnras/article/456/1/943/1067406?login=false}
}

@ARTICLE{Sharma24,
       author = {{Sharma}, G. and {van de Ven}, G. and {Salucci,}, P. and {Martorano}, M},
        title = "{Dark Matter Fraction in Disk-Like Galaxies Over the Past 10 Gyr}",
      journal = {\aap},
         year = 2025,
        month = Sep,
       volume = {699, id.A164},
        pages = {1-28},
          doi = {10.1051/0004-6361/202347922},
       adsurl = {https://ui.adsabs.harvard.edu/abs/2025A%26A...699A.164S/abstract}
}

@ARTICLE{Dut13,
       author = {{Dutton}, Aaron A. and {Macci{\`o}}, Andrea V. and {Mendel}, J. Trevor and {Simard}, Luc},
        title = "{Universal IMF versus dark halo response in early-type galaxies: breaking the degeneracy with the Fundamental Plane}",
      journal = {\mnras},
         year = 2013,
        month = jul,
       volume = {432},
       number = {3},
        pages = {2496-2511},
          doi = {10.1093/mnras/stt608},
archivePrefix = {arXiv},
       eprint = {1204.2825},
 primaryClass = {astro-ph.CO},
       adsurl = {https://ui.adsabs.harvard.edu/abs/2013MNRAS.432.2496D}
}

@ARTICLE{Bau05,
       author = {{Baugh}, C.~M. and {Lacey}, C.~G. and {Frenk}, C.~S. and {Granato}, G.~L. and {Silva}, L. and {Bressan}, A. and {Benson}, A.~J. and {Cole}, S.},
        title = "{Can the faint submillimetre galaxies be explained in the {\ensuremath{\Lambda}} cold dark matter model?}",
      journal = {\mnras},
         year = 2005,
        month = jan,
       volume = {356},
       number = {3},
        pages = {1191-1200},
          doi = {10.1111/j.1365-2966.2004.08553.x},
archivePrefix = {arXiv},
       eprint = {astro-ph/0406069},
 primaryClass = {astro-ph},
       adsurl = {https://ui.adsabs.harvard.edu/abs/2005MNRAS.356.1191B}
}

@ARTICLE{Bar18,
       author = {{Barber}, Christopher and {Crain}, Robert A. and {Schaye}, Joop},
        title = "{Calibrated, cosmological hydrodynamical simulations with variable IMFs I: method and effect on global galaxy scaling relations}",
      journal = {\mnras},
         year = 2018,
        month = oct,
       volume = {479},
       number = {4},
        pages = {5448-5473},
          doi = {10.1093/mnras/sty1826},
archivePrefix = {arXiv},
       eprint = {1804.09079},
 primaryClass = {astro-ph.GA},
       adsurl = {https://ui.adsabs.harvard.edu/abs/2018MNRAS.479.5448B}
}

@ARTICLE{For20,
       author = {{F{\"o}rster Schreiber}, Natascha M. and {Wuyts}, Stijn},
        title = "{Star-Forming Galaxies at Cosmic Noon}",
      journal = {\araa},
         year = 2020,
        month = aug,
       volume = {58},
        pages = {661-725},
          doi = {10.1146/annurev-astro-032620-021910},
archivePrefix = {arXiv},
       eprint = {2010.10171},
 primaryClass = {astro-ph.GA},
       adsurl = {https://ui.adsabs.harvard.edu/abs/2020ARA&A..58..661F}
}

@ARTICLE{Gen17,
       author = {{Genzel}, R. and {F{\"o}rster Schreiber}, N.~M. and {{\"U}bler}, H. and {Lang}, P. and {Naab}, T. and {Bender}, R. and {Tacconi}, L.~J. and {Wisnioski}, E. and {Wuyts}, S. and {Alexander}, T. and {Beifiori}, A. and {Belli}, S. and {Brammer}, G. and {Burkert}, A. and {Carollo}, C.~M. and {Chan}, J. and {Davies}, R. and {Fossati}, M. and {Galametz}, A. and {Genel}, S. and {Gerhard}, O. and {Lutz}, D. and {Mendel}, J.~T. and {Momcheva}, I. and {Nelson}, E.~J. and {Renzini}, A. and {Saglia}, R. and {Sternberg}, A. and {Tacchella}, S. and {Tadaki}, K. and {Wilman}, D.},
        title = "{Strongly baryon-dominated disk galaxies at the peak of galaxy formation ten billion years ago}",
      journal = {\nat},
         year = 2017,
        month = mar,
       volume = {543},
       number = {7645},
        pages = {397-401},
          doi = {10.1038/nature21685},
archivePrefix = {arXiv},
       eprint = {1703.04310},
 primaryClass = {astro-ph.GA},
       adsurl = {https://ui.adsabs.harvard.edu/abs/2017Natur.543..397G}
}

@ARTICLE{Bun15,
       author = {{Bundy}, Kevin and {Bershady}, Matthew A. and {Law}, David R. and {Yan}, Renbin and {Drory}, Niv and {MacDonald}, Nicholas and {Wake}, David A. and {Cherinka}, Brian and {S{\'a}nchez-Gallego}, Jos{\'e} R. and {Weijmans}, Anne-Marie and {Thomas}, Daniel and {Tremonti}, Christy and {Masters}, Karen and {Coccato}, Lodovico and {Diamond-Stanic}, Aleksandar M. and {Arag{\'o}n-Salamanca}, Alfonso and {Avila-Reese}, Vladimir and {Badenes}, Carles and {Falc{\'o}n-Barroso}, J{\'e}sus and {Belfiore}, Francesco and {Bizyaev}, Dmitry and {Blanc}, Guillermo A. and {Bland-Hawthorn}, Joss and {Blanton}, Michael R. and {Brownstein}, Joel R. and {Byler}, Nell and {Cappellari}, Michele and {Conroy}, Charlie and {Dutton}, Aaron A. and {Emsellem}, Eric and {Etherington}, James and {Frinchaboy}, Peter M. and {Fu}, Hai and {Gunn}, James E. and {Harding}, Paul and {Johnston}, Evelyn J. and {Kauffmann}, Guinevere and {Kinemuchi}, Karen and {Klaene}, Mark A. and {Knapen}, Johan H. and {Leauthaud}, Alexie and {Li}, Cheng and {Lin}, Lihwai and {Maiolino}, Roberto and {Malanushenko}, Viktor and {Malanushenko}, Elena and {Mao}, Shude and {Maraston}, Claudia and {McDermid}, Richard M. and {Merrifield}, Michael R. and {Nichol}, Robert C. and {Oravetz}, Daniel and {Pan}, Kaike and {Parejko}, John K. and {Sanchez}, Sebastian F. and {Schlegel}, David and {Simmons}, Audrey and {Steele}, Oliver and {Steinmetz}, Matthias and {Thanjavur}, Karun and {Thompson}, Benjamin A. and {Tinker}, Jeremy L. and {van den Bosch}, Remco C.~E. and {Westfall}, Kyle B. and {Wilkinson}, David and {Wright}, Shelley and {Xiao}, Ting and {Zhang}, Kai},
        title = "{Overview of the SDSS-IV MaNGA Survey: Mapping nearby Galaxies at Apache Point Observatory}",
      journal = {\apj},
         year = 2015,
        month = jan,
       volume = {798},
       number = {1},
          eid = {7},
        pages = {7},
          doi = {10.1088/0004-637X/798/1/7},
archivePrefix = {arXiv},
       eprint = {1412.1482},
 primaryClass = {astro-ph.GA},
       adsurl = {https://ui.adsabs.harvard.edu/abs/2015ApJ...798....7B}
}

@ARTICLE{Cro12,
       author = {{Croom}, Scott M. and {Lawrence}, Jon S. and {Bland-Hawthorn}, Joss and {Bryant}, Julia J. and {Fogarty}, Lisa and {Richards}, Samuel and {Goodwin}, Michael and {Farrell}, Tony and {Miziarski}, Stan and {Heald}, Ron and {Jones}, D. Heath and {Lee}, Steve and {Colless}, Matthew and {Brough}, Sarah and {Hopkins}, Andrew M. and {Bauer}, Amanda E. and {Birchall}, Michael N. and {Ellis}, Simon and {Horton}, Anthony and {Leon-Saval}, Sergio and {Lewis}, Geraint and {L{\'o}pez-S{\'a}nchez}, {\'A}. R. and {Min}, Seong-Sik and {Trinh}, Christopher and {Trowland}, Holly},
        title = "{The Sydney-AAO Multi-object Integral field spectrograph}",
      journal = {\mnras},
         year = 2012,
        month = mar,
       volume = {421},
       number = {1},
        pages = {872-893},
          doi = {10.1111/j.1365-2966.2011.20365.x},
archivePrefix = {arXiv},
       eprint = {1112.3367},
 primaryClass = {astro-ph.CO},
       adsurl = {https://ui.adsabs.harvard.edu/abs/2012MNRAS.421..872C}
}

@ARTICLE{Mar05,
       author = {{Maraston}, Claudia},
        title = "{Evolutionary population synthesis: models, analysis of the ingredients and application to high-z galaxies}",
      journal = {\mnras},
         year = 2005,
        month = sep,
       volume = {362},
       number = {3},
        pages = {799-825},
          doi = {10.1111/j.1365-2966.2005.09270.x},
archivePrefix = {arXiv},
       eprint = {astro-ph/0410207},
 primaryClass = {astro-ph},
       adsurl = {https://ui.adsabs.harvard.edu/abs/2005MNRAS.362..799M}
}

@ARTICLE{Pop15,
       author = {{Popping}, Gerg{\"o} and {Behroozi}, Peter S. and {Peeples}, Molly S.},
        title = "{Evolution of the atomic and molecular gas content of galaxies in dark matter haloes}",
      journal = {\mnras},
         year = 2015,
        month = may,
       volume = {449},
       number = {1},
        pages = {477-493},
          doi = {10.1093/mnras/stv318},
archivePrefix = {arXiv},
       eprint = {1409.1574},
 primaryClass = {astro-ph.GA},
       adsurl = {https://ui.adsabs.harvard.edu/abs/2015MNRAS.449..477P}
}

@ARTICLE{Cat18,
       author = {{Catinella}, Barbara and {Saintonge}, Am{\'e}lie and {Janowiecki}, Steven and {Cortese}, Luca and {Dav{\'e}}, Romeel and {Lemonias}, Jenna J. and {Cooper}, Andrew P. and {Schiminovich}, David and {Hummels}, Cameron B. and {Fabello}, Silvia and {Ger{\'e}b}, Katinka and {Kilborn}, Virginia and {Wang}, Jing},
        title = "{xGASS: total cold gas scaling relations and molecular-to-atomic gas ratios of galaxies in the local Universe}",
      journal = {\mnras},
         year = 2018,
        month = may,
       volume = {476},
       number = {1},
        pages = {875-895},
          doi = {10.1093/mnras/sty089},
archivePrefix = {arXiv},
       eprint = {1802.02373},
 primaryClass = {astro-ph.GA},
       adsurl = {https://ui.adsabs.harvard.edu/abs/2018MNRAS.476..875C}
}

@ARTICLE{Wis15,
       author = {{Wisnioski}, E. and {F{\"o}rster Schreiber}, N.~M. and {Wuyts}, S. and {Wuyts}, E. and {Bandara}, K. and {Wilman}, D. and {Genzel}, R. and {Bender}, R. and {Davies}, R. and {Fossati}, M. and {Lang}, P. and {Mendel}, J.~T. and {Beifiori}, A. and {Brammer}, G. and {Chan}, J. and {Fabricius}, M. and {Fudamoto}, Y. and {Kulkarni}, S. and {Kurk}, J. and {Lutz}, D. and {Nelson}, E.~J. and {Momcheva}, I. and {Rosario}, D. and {Saglia}, R. and {Seitz}, S. and {Tacconi}, L.~J. and {van Dokkum}, P.~G.},
        title = "{The KMOS$^{3D}$ Survey: Design, First Results, and the Evolution of Galaxy Kinematics from 0.7 <= z <= 2.7}",
      journal = {\apj},
         year = 2015,
        month = feb,
       volume = {799},
       number = {2},
          eid = {209},
        pages = {209},
          doi = {10.1088/0004-637X/799/2/209},
archivePrefix = {arXiv},
       eprint = {1409.6791},
 primaryClass = {astro-ph.GA},
       adsurl = {https://ui.adsabs.harvard.edu/abs/2015ApJ...799..209W}
}

@ARTICLE{Bac15,
       author = {{Bacon}, R. and {Brinchmann}, J. and {Richard}, J. and {Contini}, T. and {Drake}, A. and {Franx}, M. and {Tacchella}, S. and {Vernet}, J. and {Wisotzki}, L. and {Blaizot}, J. and {Bouch{\'e}}, N. and {Bouwens}, R. and {Cantalupo}, S. and {Carollo}, C.~M. and {Carton}, D. and {Caruana}, J. and {Cl{\'e}ment}, B. and {Dreizler}, S. and {Epinat}, B. and {Guiderdoni}, B. and {Herenz}, C. and {Husser}, T. -O. and {Kamann}, S. and {Kerutt}, J. and {Kollatschny}, W. and {Krajnovic}, D. and {Lilly}, S. and {Martinsson}, T. and {Michel-Dansac}, L. and {Patricio}, V. and {Schaye}, J. and {Shirazi}, M. and {Soto}, K. and {Soucail}, G. and {Steinmetz}, M. and {Urrutia}, T. and {Weilbacher}, P. and {de Zeeuw}, T.},
        title = "{The MUSE 3D view of the Hubble Deep Field South}",
      journal = {\aap},
         year = 2015,
        month = mar,
       volume = {575},
          eid = {A75},
        pages = {A75},
          doi = {10.1051/0004-6361/201425419},
archivePrefix = {arXiv},
       eprint = {1411.7667},
 primaryClass = {astro-ph.GA},
       adsurl = {https://ui.adsabs.harvard.edu/abs/2015A&A...575A..75B}
}

@ARTICLE{For06,
       author = {{F{\"o}rster Schreiber}, N.~M. and {Genzel}, R. and {Lehnert}, M.~D. and {Bouch{\'e}}, N. and {Verma}, A. and {Erb}, D.~K. and {Shapley}, A.~E. and {Steidel}, C.~C. and {Davies}, R. and {Lutz}, D. and {Nesvadba}, N. and {Tacconi}, L.~J. and {Eisenhauer}, F. and {Abuter}, R. and {Gilbert}, A. and {Gillessen}, S. and {Sternberg}, A.},
        title = "{SINFONI Integral Field Spectroscopy of z \raisebox{-0.5ex}\textasciitilde 2 UV-selected Galaxies: Rotation Curves and Dynamical Evolution}",
      journal = {\apj},
         year = 2006,
        month = jul,
       volume = {645},
       number = {2},
        pages = {1062-1075},
          doi = {10.1086/504403},
archivePrefix = {arXiv},
       eprint = {astro-ph/0603559},
 primaryClass = {astro-ph},
       adsurl = {https://ui.adsabs.harvard.edu/abs/2006ApJ...645.1062F}
}

@ARTICLE{Nig+09,
       author = {{Nigoche-Netro}, A. and {Ruelas-Mayorga}, A. and {Franco-Balderas}, A.},
        title = "{The Fundamental Plane for early-type galaxies: dependence on the magnitude range}",
      journal = {\mnras},
         year = 2009,
        month = jan,
       volume = {392},
       number = {3},
        pages = {1060-1069},
          doi = {10.1111/j.1365-2966.2008.14145.x},
archivePrefix = {arXiv},
       eprint = {0805.1142},
 primaryClass = {astro-ph},
       adsurl = {https://ui.adsabs.harvard.edu/abs/2009MNRAS.392.1060N}
}

@ARTICLE{BlandHawthorn2016,
       author = {{Bland-Hawthorn}, Joss and {Gerhard}, Ortwin},
        title = "{The Galaxy in Context: Structural, Kinematic, and Integrated Properties}",
      journal = {\araa},
         year = 2016,
        month = sep,
       volume = {54},
        pages = {529-596},
          doi = {10.1146/annurev-astro-081915-023441},
archivePrefix = {arXiv},
       eprint = {1602.07702},
 primaryClass = {astro-ph.GA},
       adsurl = {https://ui.adsabs.harvard.edu/abs/2016ARA&A..54..529B}
}

@ARTICLE{Navarro1997,
       author = {{Navarro}, Julio F. and {Frenk}, Carlos S. and {White}, Simon D.~M.},
        title = "{A Universal Density Profile from Hierarchical Clustering}",
      journal = {\apj},
         year = 1997,
        month = dec,
       volume = {490},
       number = {2},
        pages = {493-508},
          doi = {10.1086/304888},
archivePrefix = {arXiv},
       eprint = {astro-ph/9611107},
 primaryClass = {astro-ph},
       adsurl = {https://ui.adsabs.harvard.edu/abs/1997ApJ...490..493N}
}

@ARTICLE{Tamm2012,
       author = {{Tamm}, A. and {Tempel}, E. and {Tenjes}, P. and {Tihhonova}, O. and {Tuvikene}, T.},
        title = "{Stellar mass map and dark matter distribution in M 31}",
      journal = {\aap},
         year = 2012,
        month = oct,
       volume = {546},
          eid = {A4},
        pages = {A4},
          doi = {10.1051/0004-6361/201220065},
archivePrefix = {arXiv},
       eprint = {1208.5712},
 primaryClass = {astro-ph.CO},
       adsurl = {https://ui.adsabs.harvard.edu/abs/2012A&A...546A...4T}
}

@ARTICLE{Sofue2012,
       author = {{Sofue}, Yoshiaki},
        title = "{Grand Rotation Curve and Dark Matter Halo in the Milky Way Galaxy}",
      journal = {\pasj},
         year = 2012,
        month = aug,
       volume = {64},
          eid = {75},
        pages = {75},
          doi = {10.1093/pasj/64.4.75},
archivePrefix = {arXiv},
       eprint = {1110.4431},
 primaryClass = {astro-ph.GA},
       adsurl = {https://ui.adsabs.harvard.edu/abs/2012PASJ...64...75S}
}

@ARTICLE{Petac2020,
       author = {{Peta{\v{c}}}, Mihael},
        title = "{Equilibrium axisymmetric halo model for the Milky Way and its implications for direct and indirect dark matter searches}",
      journal = {\prd},
         year = 2020,
        month = dec,
       volume = {102},
       number = {12},
          eid = {123028},
        pages = {123028},
          doi = {10.1103/PhysRevD.102.123028},
archivePrefix = {arXiv},
       eprint = {2008.11172},
 primaryClass = {astro-ph.GA},
       adsurl = {https://ui.adsabs.harvard.edu/abs/2020PhRvD.102l3028P}
}
\bibliographystyle{aasjournal}

\appendix
\setcounter{figure}{0}
\renewcommand{\thefigure}{A\arabic{figure}}
\setcounter{table}{0}
\renewcommand{\thetable}{A\arabic{table}}
  
\section{Variables, biases, and corrections in the estimation of dynamical, stellar, and dark matter in galaxies}
\label{sec:appendix_A}

The estimation of the dynamical and stellar masses of LTGs is a fundamental aspect of our analysis. In this appendix, we provide a comprehensive compilation of the variables, corrections, and potential biases that affecting the estimation of dynamical, stellar, and dark matter. Here, we also include figures and tables that provide additional context for the results discussed in the main body of the article (see Figures \ref{fig:histo}--\ref{fig:smal-3} and Table \ref{TableA1}).

\subsection{Stellar or Gas Velocity Dispersion ($\sigma$): Used to estimate the dynamical mass through the virial relation (see equation \ref{eq:pov}).}
\label{sec:A1}
Biases:
\begin{itemize}
\item[(a)] Inclination angle. The orientation of the galaxy relative to the observer can affect the measurement of $\sigma$. 
\item[(b)] Stellar and nonstellar (gas) components. $\sigma$ may be influenced by the presence of stellar and nonstellar components. This bias is linked to the galaxies' structure, color, and morphological characteristics. 
\item[(c)] Spectral lines. It is possible to measure $\sigma$ using different spectral lines, and the choice of lines could affect the mass estimate.
\item[(d)] Aperture correction. In spectroscopy, observations are often made through limited apertures, which may not represent the entire galaxy, potentially biasing $\sigma$ measurement and, consequently, the dynamical mass.
\end{itemize}
        
\subsection{Effective Radius ($r_e$): Represents the radius within which half of the galaxy's total light is contained. Utilized for calculating the dynamic mass via the virial relation (see equation \ref{eq:pov}).}
\label{sec:A2}

Biases: 

\begin{itemize}
\item[(a)] Visual conditions (seeing) during observations. This factor refers to the blurring and distortion of astronomical objects caused by turbulence in Earth's atmosphere. Any inaccuracies in the correction of visual conditions during observations could skew the measured effective radius.
\item[(b)] Cosmological correction. This factor ensures that the sizes of galaxies are measured in a way that accurately reflects their intrinsic properties, free from observational biases introduced by redshift. Any inaccuracies in the cosmological correction may bias the measured effective radius. 

\item[(c)] Extinction correction. This factor refers to the process of adjusting observed astronomical data to account for the absorption and scattering of light by dust and gas, both within the observed galaxy and in the Milky Way. Inaccuracies in the extinction correction can lead to a skewed measurement of the effective radius.
\item[(d)] Inclination angle. The orientation of the galaxy relative to the observer can affect measurements of the effective radius.
\end{itemize}

\subsection{Initial Mass Function: Determines how stellar masses are distributed at the time of their formation.}
\label{sec:A3}

Biases:

\begin{itemize}        
\item[(a)] Different choices of IMF (e.g., Chabrier, Salpeter) can lead to significant differences in the estimated stellar mass.
\item[(b)] Dependencies of the IMF on several variables (e.g. velocity dispersion,
mass) may result in substantial variations in the calculated stellar mass.
\end{itemize}

\subsection{color Indices: Used to infer the stellar population and the age of the galaxy.}
\label{sec:A4}

Biases:

\begin{itemize}
\item[(a)] Dust extinction affects color indices, which can lead to errors in the estimation of stellar mass. 
\item[(b)] Metallicity influences color indices, potentially causing inaccuracies in the stellar mass estimation.
\end{itemize}

\subsection{Redshift ($z$): Affects both photometric and spectroscopic observations.}
\label{sec:A5}

Biases:

\begin{itemize}       
\item[(a)] K-correction. When observing galaxies at varying distances, the universe's expansion causes a shift of their light to longer wavelengths, affecting the observed colors and magnitudes. Accurate implementation of the K-correction is essential.
\item[(b)] Cosmological dimming. This factor is a result of the universe's expansion, making distant objects seem dimmer than if the Universe remained static. It is essential to correctly account for cosmological dimming.
\item[(c)] Malmquist bias. This bias occurs because, at greater distances, we are more likely to observe intrinsically brighter objects, while fainter objects may fall below the detection limit of surveys. It is essential to accurately consider Malmquist bias.
\end{itemize}

\subsection{Stellar Population Models: Used to convert observed luminosity of the stars into stellar mass.}
\label{sec:A6}

Biases: 

\begin{itemize}        
\item[(a)] Models using different IMFs can produce variations in stellar mass. This bias is related to ~\ref{sec:A3}(a).
\item[(b)] Models using different metallicity can lead to differences in the stellar mass. 
\item[(c)] Models with or without dust are capable of inducing changes in stellar mass.
\item[(d)] Models using different star formation histories can result in differences in stellar mass.
\item[(e)] Models underestimate the baryonic mass of galaxies because they do not completely account for the weak and 
nonluminous baryonic matter.
\end{itemize}

\subsection{The factor $K$ (refer to equation~\ref{eq:pov}): This factor indicates whether the proportion between mass and velocity dispersion is constant or varies among different systems or settings, and influences the calculation of the dynamical mass.}
\label{sec:A7}

Biases: 

\begin{itemize}        
\item[(a)] $K$ could be constant \citep{cap06}.
\item[(b)] $K$ could depend on the S\'ersic index \citep[morphology;][]{cap06,moc12}.
\item[(c)] $K$ could depend on the inclination angle of the galactic plane \citep{cap06,moc12}. This bias is related to~\ref{sec:A1}(a) and~\ref{sec:A2}(d).
\end{itemize}

\subsection{Magnitude (surface brightness): Used in Mass-to-light ratio and SPS models to estimate the luminous and stellar mass.}
\label{sec:A8}        

Biases: 

\begin{itemize}        
\item[(a)] Visual conditions (seeing) during observations (see~\ref{sec:A2}(a)).
\item[(b)] Galactic interstellar extinction (see~\ref{sec:A2}(c)).
\item[(c)] Cosmological dimming (see~\ref{sec:A2}(b)).
\item[(d)] Evolution correction. This correction in galaxies accounts for changes in their luminosity over time due to factors such as star formation and aging of stellar populations, allowing for more accurate comparisons across different redshifts.
\end{itemize}
    
\subsection{Other Biases:}
\label{sec:A9}    

\begin{itemize}  
\item[(a)] Sample completeness: This factor refers to the extent to which the surveys include all galaxies within their target range of brightness (Malmquist bias), distance, and other selection criteria, ensuring that the sample accurately represents the overall galaxy population. This bias is related to~\ref{sec:A5}(c).
\item[(b)] Selection bias: Galaxy samples may be biased toward more luminous galaxies (\ref{sec:A5}(c)) or those with certain spectral features (\ref{sec:A1}(c)), affecting both dynamical and luminous (or stellar) mass estimations.
\item[(c)] Wavelength: The use of one filter or another for the estimation of mass-related parameters can affect the results.
\end{itemize}
    
\subsection{Possible dark matter dependencies:}
\label{sec:A10}

\begin{itemize}
\item[(a)] Dynamical or luminous (stellar) mass: The amount of dynamical or luminous (stellar) mass could be related to the content of dark matter inside galaxies.
\item[(b)] Redshift: The evolution of galaxies over the cosmic time could be related to the content of dark matter inside galaxies.
\item[(c)] Density of galaxies: The environment could be related to the content of dark matter inside galaxies.
\end{itemize}

\noindent The previously listed factors are crucial for obtaining accurate estimates of the dynamical, stellar, and dark matter of galaxies. Proper corrections and awareness of these biases are essential for the interpretation of astrophysical data. In the following sections, we will address the corrections and biases that can affect mass estimations.

\subsection{Correction of Photometric and Spectroscopic Data}
\label{sec:photospec}

As we saw before, it is essential to correct both the photometric and spectroscopic data. Below, we outline the corrections applied following \citet{Nig15}. At the end of each subparagraph, we indicate the section where the variables and biases we intend to address were discussed:

\begin{itemize}

\item Seeing and Extinction Corrections: We utilize seeing-corrected parameters for the total magnitude and effective radius, along with extinction corrections, by applying the corresponding SDSS pipelines to the data \citep[see][and references therein]{yor00,bla03}. This paragraph considers the biases discussed in Appendices~\ref{sec:A2}(a), \ref{sec:A2}(c), \ref{sec:A8}(a), and \ref{sec:A8}(b).

\item K-correction: The values are derived using the following formulae~\citep{nig08}:

\begin{equation} \label{eq:kg}
k_g (z) = -5.261z^{1.197}, \end{equation}

\begin{equation} \label{eq:kr}
k_r (z) = -1.271z^{1.023}, \end{equation}

This paragraph considers the bias discussed in Appendix~\ref{sec:A5}(a).

\item Cosmological Dimming Correction. According to \citet{jor95a}, the effective surface brightness ($<\mu_e>$) or magnitude are adjusted by subtracting 10 $\log(1+z)$. This paragraph considers the bias discussed in Appendix~\ref{sec:A8}(c). 

\item Evolution Correction. We apply the evolution correction from~\citet{nig10}, which is given by:

\begin{equation} \label{eq:evg}
ev_g (z) = +1.15z, \end{equation}

\begin{equation} \label{eq:evr}
ev_r (z) = +0.85z, \end{equation}

This paragraph considers the bias discussed in Appendix~\ref{sec:A8}(d). 

\item Effective Radius Correction to the Rest Reference Frame. Following~\citet{hyd09}, we correct color gradients where the mean effective radius is smaller at longer wavelengths using:

\begin{equation} r_{e,g,rest} = \left[\frac{(1+z)\lambda_g-\lambda_r}{\lambda_g-\lambda_r}\right] \left(r_{e,g,obs} - r_{e,r,obs}\right) + r_{e,r,obs}, \end{equation}

with $\lambda_g$ = 4686\AA{} and $\lambda_r$ = 6165\AA{}.

This paragraph considers the bias discussed in Appendix~\ref{sec:A2}(b).

\item Aperture Correction to Velocity Dispersion. Following \citet{jor95b}, we compute the ratio between the SDSS velocity dispersion ($\sigma_{SDSS}$) and the corrected velocity dispersion ($\sigma_e$),  which corresponds to the velocity dispersion within the effective radius $r_e$, as:

\begin{equation} \log\left(\frac{\sigma_{SDSS}}{\sigma_e}\right)= -0.065 \log\left(\frac{r_{ap}}{r_e}\right) - 0.013\left[\log\left(\frac{r_{ap}}{r_e}\right)\right]^2, \end{equation}

where $r_{ap}$ = 1\farcs5 for SDSS data \citep[][]{yor00,bla03}.

This paragraph considers the bias discussed in Appendix~\ref{sec:A1}(d). 
\end{itemize}


\subsection{Logarithmic Mass Difference within the Optical Radius of the Milky Way and Andromeda}

\label{sec:milky}

\subsubsection{Methodology}
To quantify the relative contribution of baryonic and dark matter components within the optical extent of spiral galaxies, we computed the logarithmic difference between the total (dynamical) and baryonic (luminous) masses as

\begin{equation}
    \Delta \log M = \log_{10}\left(\frac{M_{\text{total}}}{M_{\text{bar}}}\right),
    \label{eq:dlogm}
\end{equation}
where \(M_{\text{total}}\) is the dynamical mass enclosed within the optical radius \(R_{\text{opt}}\), and \(M_{\text{bar}}\) is the corresponding baryonic (stellar + gaseous) mass within the same radius. 

The optical radius was defined as the radius where the $B$-band surface brightness falls to 25~mag~arcsec\(^{-2}\): \(R_{\text{opt}} = 15~\mathrm{kpc}\) for the Milky Way and \(R_{\text{opt}} = 25~\mathrm{kpc}\) for Andromeda (see \citealt{Sofue2012, BlandHawthorn2016,Tamm2012}).

\subsubsection{Mass modeling}

The total mass inside a radius \(r\) is given by the sum of the baryonic and dark matter (DM) components:
\begin{equation}
    M_{\text{total}}(r) = M_{\text{bar}}(r) + M_{\text{DM}}(r).
\end{equation}


The dark matter mass was computed assuming an Navarro–Frenk–White (NFW) density profile \citep{Navarro1997}:
\begin{equation}
    \rho(r) = \frac{\rho_s}{(r/r_s)\,(1+r/r_s)^2},
\end{equation}
where \(\rho_s\) and \(r_s\) are the scale density and scale radius, respectively. The enclosed DM mass is then obtained by integrating the profile:
\begin{equation}
    M_{\text{DM}}(<r) = 4\pi \rho_s r_s^3 \left[\ln(1 + x) - \frac{x}{1+x}\right],
    \quad x = \frac{r}{r_s}.
    \label{eq:mdm}
\end{equation}

\subsubsection*{Milky Way parameters}

For the Milky Way, we adopted the parameters derived by \citet{Petac2020} and \citet{Sofue2012}.

\begin{itemize}
    \item Scale density: \(\rho_s = 0.012~M_\odot~\mathrm{pc}^{-3}\);
    \item Scale radius: \(r_s = 18~\mathrm{kpc}\);
    \item Total baryonic mass: \(M_{\text{bar,tot}} = (6.0 \pm 1.2)\times10^{10}~M_\odot\).
\end{itemize}

Assuming that approximately 90\% of the baryonic mass lies within the optical radius, we take
\[
M_{\text{bar}}(R_{\text{opt}} = 15~\mathrm{kpc}) = (5.4 \pm 1.1)\times10^{10}~M_\odot.
\]
Using Eq.~\ref{eq:mdm}, the enclosed DM mass within \(15~\mathrm{kpc}\) is
\[
M_{\text{DM}}(R_{\text{opt}}) = 4\pi(1.2\times10^7)(18)^3
\left[\ln(1.833) - \frac{0.833}{1.833}\right]
= (1.33 \pm 0.40)\times10^{11}~M_\odot.
\]
The total dynamical mass inside this radius is
\[
M_{\text{total}}(R_{\text{opt}}) = (1.87 \pm 0.55)\times10^{11}~M_\odot.
\]
Hence, from Eq.~\ref{eq:dlogm}:
\[
\Delta\log M_{\text{MW}}(R_{\text{opt}}) =
\log_{10}\!\left(\frac{1.87\times10^{11}}{5.4\times10^{10}}\right)
= 0.54 \pm 0.15~\text{dex.}
\]


\subsubsection*{Andromeda (M31) parameters}

For Andromeda, we adopt dynamical and baryonic masses consistent with the modeling of~\citet{Tamm2012}. \citet{Tamm2012} report a total stellar mass in the range $(1.0\textrm{--}1.5)\times10^{11}\,M_\odot$ (see their Stellar mass distribution in Table 2 and Discussion). The observed rotation curve and enclosed mass table (see their Table~3 and Fig.~5) imply cumulative masses of order a few $10^{11}\,M_\odot$ within radii $\sim$ 20--30 kpc. Adopting
\[
R_{\rm opt}=25\ \mathrm{kpc},\quad M_{\text{total}}(R_{\rm opt})=(3.4\pm1.0)\times10^{11}\,M_\odot,\quad
M_{\text{bar}}(R_{\rm opt})=(1.0\pm0.2)\times10^{11}\,M_\odot,
\]


we obtain
\[
\Delta\log M_{\text{M31}}(R_{\rm opt})=\log_{10}\!\left(\frac{3.4\times10^{11}}{1.0\times10^{11}}\right)=0.53\pm0.15\ \mathrm{dex}.
\]





\begin{deluxetable}{lccccc}[ht]
  \tablecaption{Logarithmic mass difference within the optical
      radius for the Milky Way and Andromeda.}
      \label{TableA1}
    \tablewidth{0pt}

    \tablehead{\colhead{Galaxy} & \colhead{$R_{\text{opt}}$ (kpc)}
      & \colhead{$M_{\text{total}}$ ($M_\odot$)}
      & \colhead{$M_{\text{bar}}$ ($M_\odot$)}
      & \colhead{$\Delta \log M$ (dex)} & \colhead{$\sigma$ (dex)}
    }

    \startdata
    Milky Way & 15 & $1.9\times10^{11}$ & $5.4\times10^{10}$ & 0.54 & 0.15 \\
    Andromeda (M31) & 25 & $3.4\times10^{11}$ & $1.0\times10^{11}$ & 0.53 & 0.15 \\
    \enddata
\end{deluxetable}

\begin{figure*}
  \begin{center}
     \includegraphics[angle=0,width=11cm]{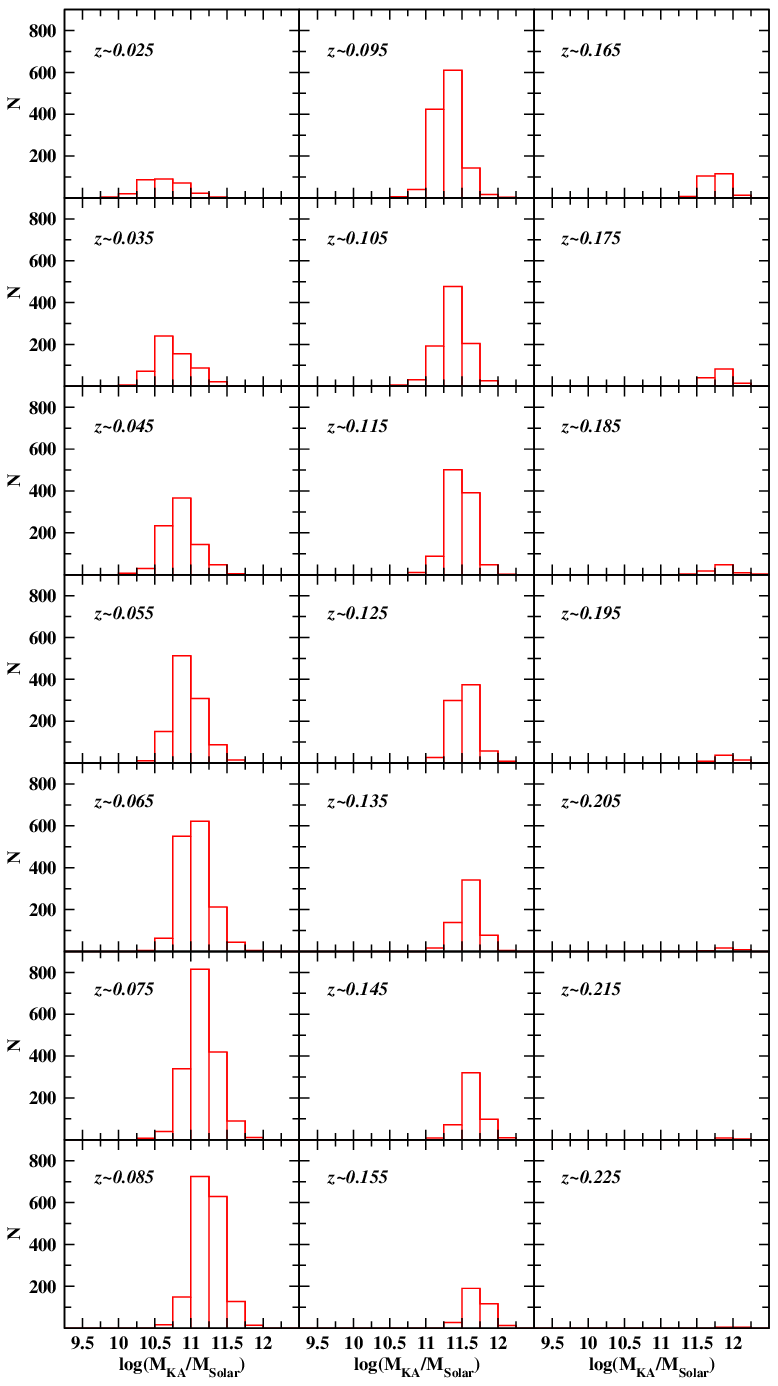}
      
         \caption{Frequency distributions of $\log(\mathbf{M_{KA}/M_{Solar}})$ at different quasi-constant redshifts, illustrating the Malmquist bias. Data are taken from column 1, row 2 of Figure~\ref{fig:oVSv:MED}.}
         \label{fig:histo}
  \end{center}
\end{figure*}

\clearpage

\begin{figure*}
   \begin{center}

    \includegraphics[angle=0,width=11cm]
    {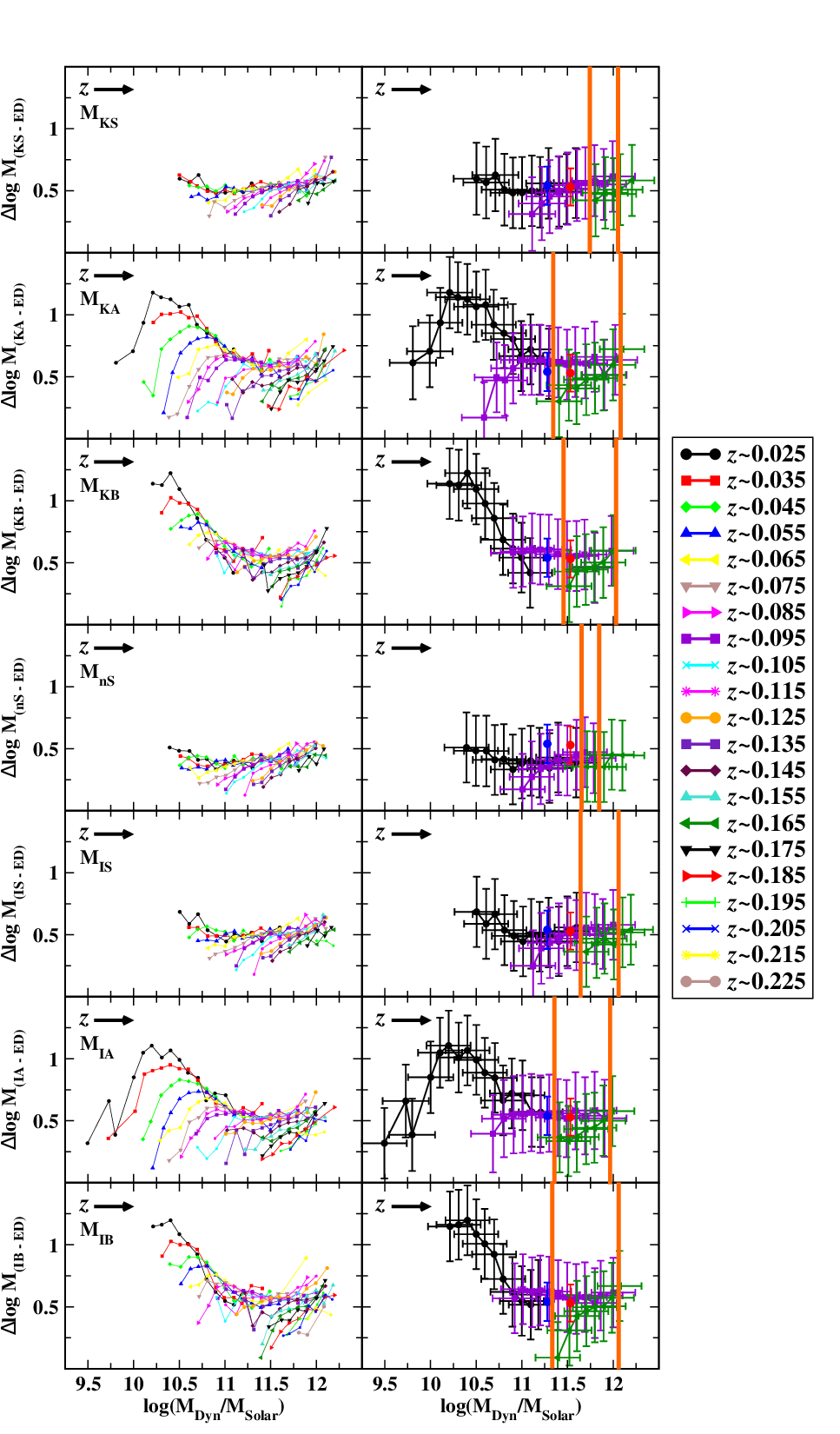}  
    \caption{Difference between dynamical and stellar mass 
    ($\Delta {\bf log M_{(Dyn - ED)}} $ = $\log({\bf M_{Dyn}/{\bf M_{Solar}}})$ - $\log({\bf M_{ED}/{\bf M_{Solar}}})$) as a function of the dynamical mass for the LTGs samples. Each row corresponds to a specific estimation of dynamical mass ($\mathbf{M_{KS}}$, $\mathbf{M_{KA}}$, $\mathbf{M_{KB}}$, $\mathbf{M_{nS}}$, $\mathbf{M_{IS}}$, $\mathbf{M_{IA}}$, $\mathbf{M_{IB}}$). Each color and symbol represents a quasi-constant redshift. The difference in redshift between consecutive symbols is approximately 0.01. The mean uncertainty of the difference between $\log({\bf M_{Dyn}/{\bf M_{Solar}}})$ and $\log({\bf M_{ED}/{\bf M_{Solar}}})$ is approximately 0.243 dex. The left column of the mosaic contains the full range of redshift, while the right column only contains three specific redshifts: low (black circles, $z \sim 0.025$), intermediate (purple squares, $z \sim 0.095$), and high (dark green, triangles $z \sim 0.165$), the latter with the aim of appreciating more clearly the differences in dynamical and stellar masses due to redshift and also to display the uncertainties for each point. The blue circle represents the Milky Way, and the red circle represents the Andromeda galaxy. The orange bars represent the boundaries of the quasi-constant mass bins shared by the $z = 0.095$ and $z = 0.165$ redshift samples.}

\label{fig:oVSv:MEDextra}
\end{center}
\end{figure*}

\clearpage

\begin{figure*}

   \begin{center}

    \includegraphics[angle=0,width=11cm]
    {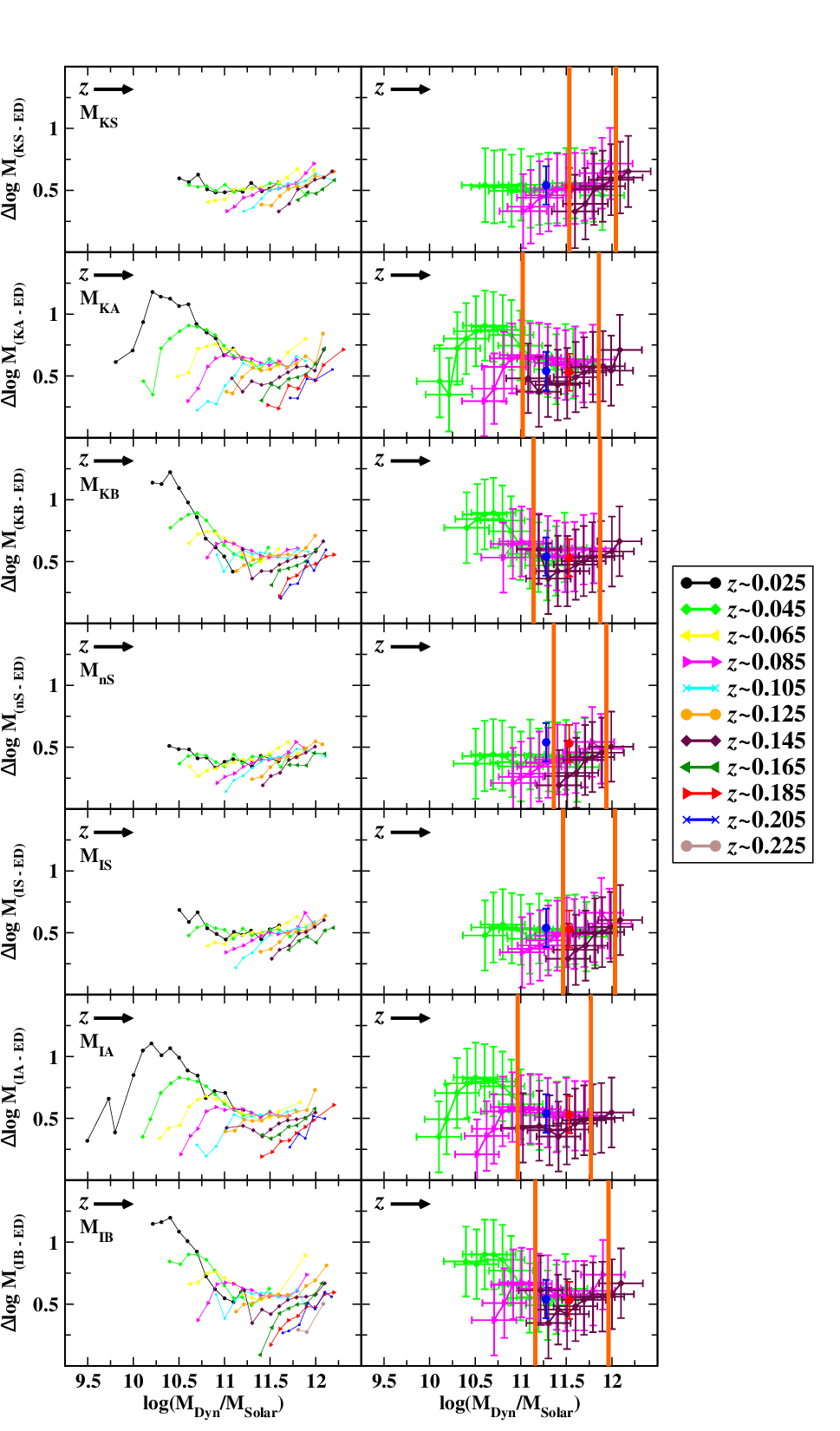}

         \caption{Mosaic of the difference between dynamical and stellar mass ($\Delta {\bf log M_{(Dyn - ED)}} $ = $\log({\bf M_{Dyn}/{\bf M_{Solar}}})$ - $\log({\bf M_{ED}/{\bf M_{Solar}}})$) as a function of the dynamical mass, analogous to Fig.~\ref{fig:oVSv:MED}. Each row corresponds to a specific estimation of dynamical mass ($\mathbf{M_{KS}}$, $\mathbf{M_{KA}}$, $\mathbf{M_{KB}}$, $\mathbf{M_{nS}}$, $\mathbf{M_{IS}}$, $\mathbf{M_{IA}}$, $\mathbf{M_{IB}}$). In the first column, the quasi-constant redshift trend lines are spaced every 0.02 instead of 0.01, which reduces crowding and makes the overall behavior easier to visualize. In the second column, three representative redshifts are shown: low (light green diamonds, $z \sim 0.045$), intermediate (magenta triangles, $z \sim 0.085$), and high (brown diamonds, $z \sim 0.145$). These examples complement those shown in Fig.~\ref{fig:oVSv:MED}, reinforcing the robustness of the behaviors discussed and facilitating the identification of completeness ranges shared by the different trends.
         The blue circle represents the Milky Way, and the red circle represents the Andromeda galaxy. The orange bars represent the boundaries of the quasi-constant mass bins shared by the $z = 0.085$ and $z = 0.145$ redshift samples.}
\label{fig:oVSv:MED2extra}
\end{center}
\end{figure*}

\begin{figure*}
  \begin{center}
     \includegraphics[angle=0,width=11cm]{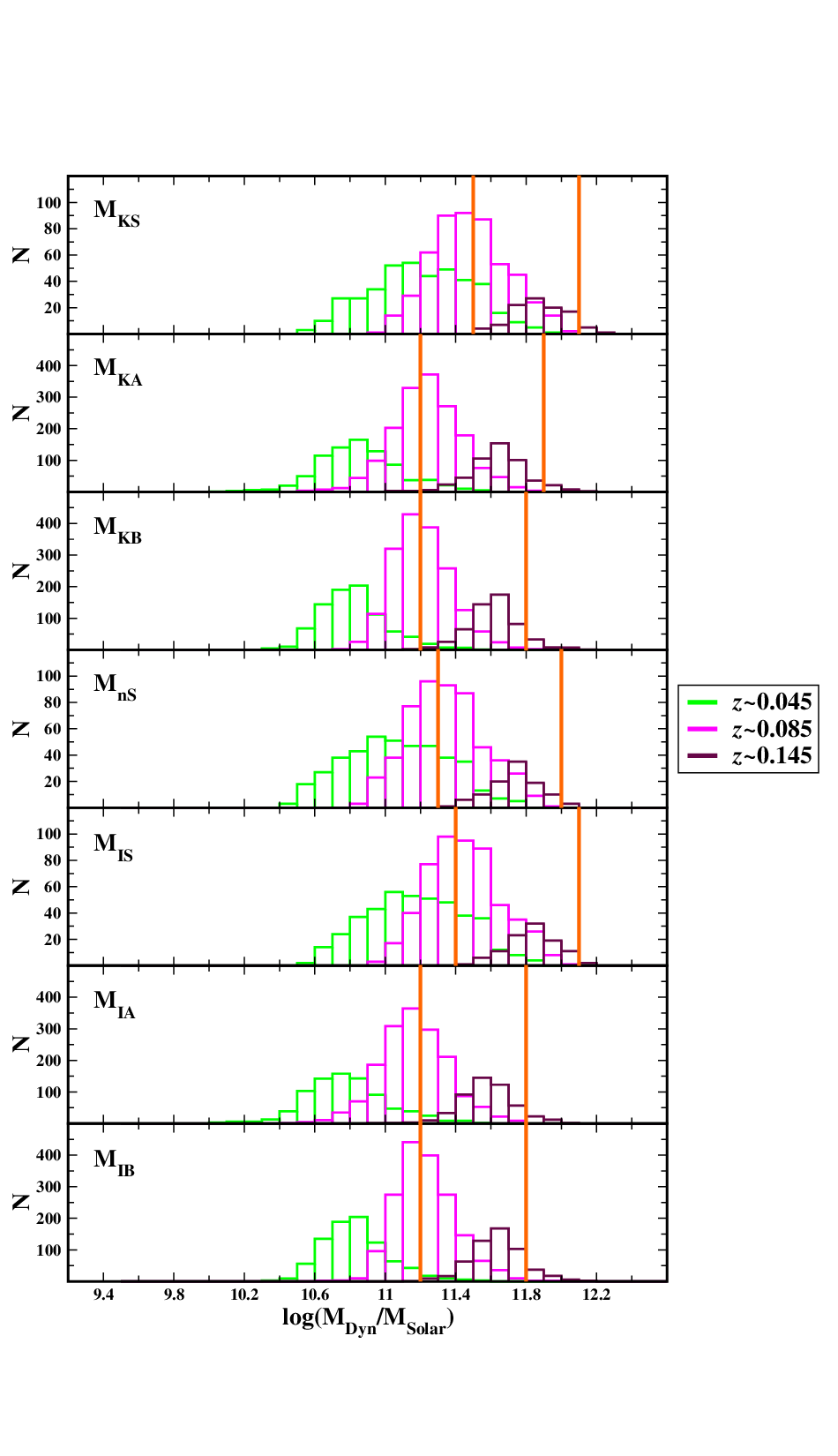}
      
         \caption{Frequency distributions of $\log(\mathbf{M_{Dyn}/M_{Solar}})$ at three quasi-constant redshifts. Each row corresponds to a specific estimation of dynamical mass ($\mathbf{M_{KS}}$, $\mathbf{M_{KA}}$, $\mathbf{M_{KB}}$, $\mathbf{M_{nS}}$, $\mathbf{M_{IS}}$, $\mathbf{M_{IA}}$, $\mathbf{M_{IB}}$). These histograms are derived from the same data as in the second column of Figure~\ref{fig:oVSv:MED2} and share the same color scheme for consistency. They show galaxy counts within quasi-constant dynamical mass bins at $z = 0.045$ (light green), $0.085$ (magenta), and $0.145$ ( brown). The orange bars mark the boundaries of the mass range common to the $z = 0.085$ and $z = 0.145$ samples, matching the representation in the second column of Figure~\ref{fig:oVSv:MED2extra}.}
         \label{fig:histo3}
  \end{center}
\end{figure*}


\begin{figure*}

   \begin{center}

    \includegraphics[angle=0,width=12cm]
    {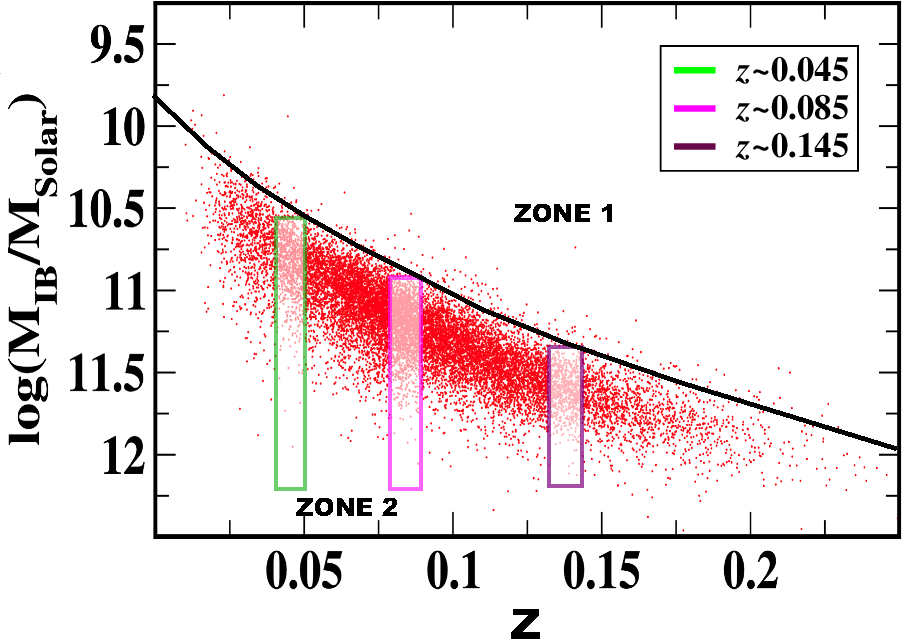}
         \caption{Distribution of dynamical mass ($\log(\mathbf{M_{IB}/M_{Solar}})$) as a function of redshift. Extracted from the last row of Figure~\ref{fig:vVSr:MED}, this panel visually outlines our binning strategy in dynamical mass and redshift. It illustrates the range of dynamical masses at three different quasi-constant redshifts: light green for $z = 0.045$, magenta for $z = 0.085$, and brown for $z = 0.145$ (the color scheme is consistent with Figures~\ref{fig:oVSv:MED2} and~\ref{fig:oVSv:MED2extra}). Two key regions are indicated: Zone 1: The region dominated by Malmquist bias, where lower-mass galaxies become progressively undetectable at higher redshifts. Zone 2: The region where the survey is sensitive enough to detect galaxies.}
\label{fig:smal-1}
\end{center}
\end{figure*}


\begin{figure*}

   \begin{center}

    \includegraphics[angle=0,width=12
    cm]
    {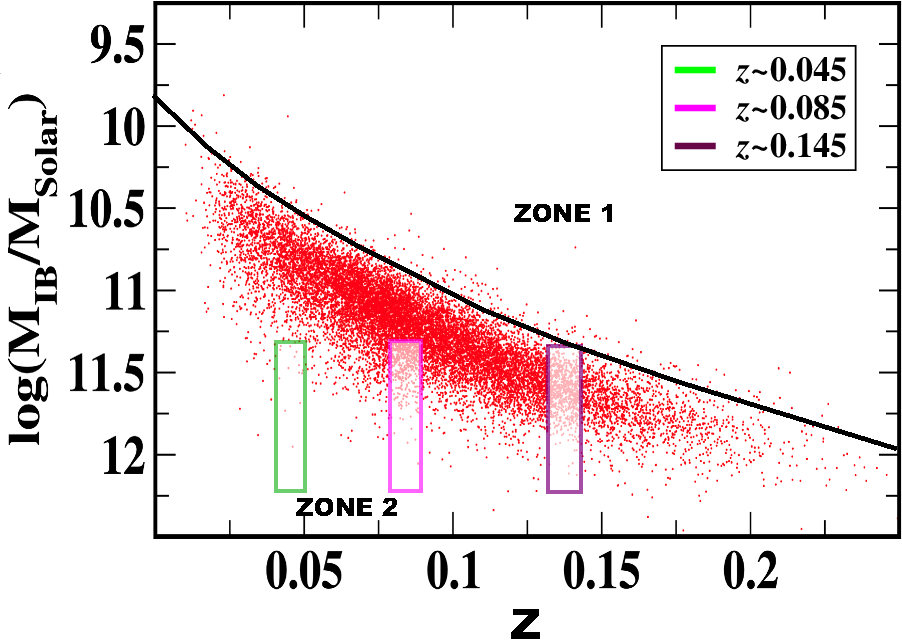}
         \caption{Distribution of dynamical mass ($\log(\mathbf{M_{IB}/M_{Solar}})$) as a function of the redshift. Extracted from the last row of Figure~\ref{fig:vVSr:MED}, this panel visually outlines our binning strategy in dynamical mass and redshift. It illustrates the overlapping mass range for three quasi-constant redshifts: light green for $z = 0.045$, magenta for $z = 0.085$, and brown for $z = 0.145$ (the color scheme is consistent with Figures~\ref{fig:oVSv:MED2} and~\ref{fig:oVSv:MED2extra}). Two key regions are indicated. Zone 1: the region dominated by Malmquist bias, where lower-mass galaxies become progressively undetectable at higher redshifts. Zone 2: the region where the survey is sensitive enough to detect galaxies.}
\label{fig:smal-2}
\end{center}
\end{figure*}

\begin{figure*}

   \begin{center}

    \includegraphics[angle=0,width=12
    cm]
    {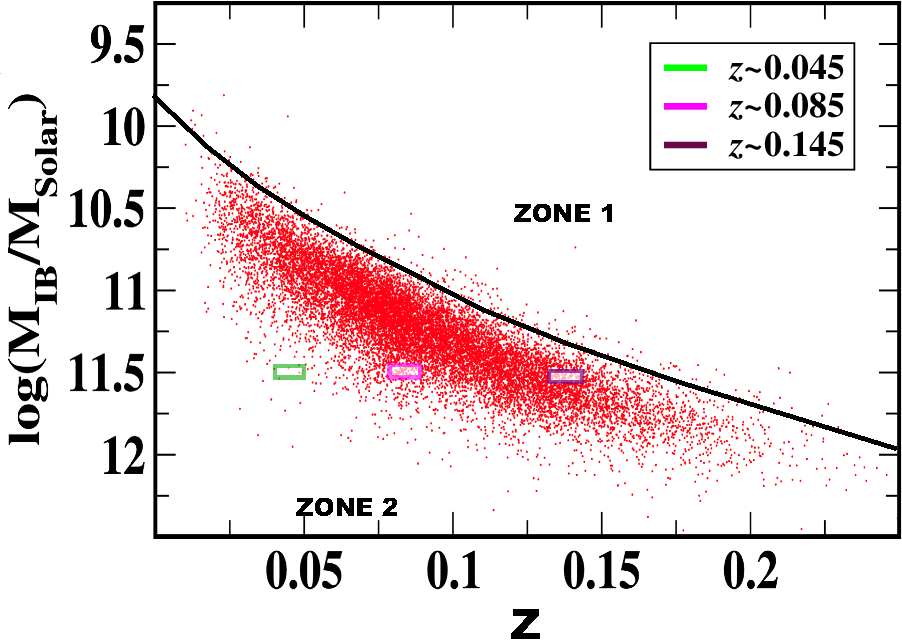}
         \caption{Distribution of dynamical mass ($\log(\mathbf{M_{IB}/M_{Solar}})$) as a function of the redshift. Extracted from the last row of Figure~\ref{fig:vVSr:MED}, this panel visually outlines our binning strategy. It highlights a specific quasi-constant dynamical mass bin (centered at $\log(\mathbf{M_{IB}/M_{Solar}})=11.5$, width of $0.1$ dex) across three quasi-constant redshifts: $z = 0.045$ (light green), $z = 0.085$ (magenta), and $z = 0.145$ (brown). The color scheme is consistent with Figures~\ref{fig:oVSv:MED2} and~\ref{fig:oVSv:MED2extra}. Two key regions are indicated: Zone 1: The area dominated by Malmquist bias, where lower-mass galaxies become progressively undetectable at higher redshifts. Zone 2: The region where the survey is sensitive enough to detect galaxies.}
\label{fig:smal-3}
\end{center}
\end{figure*}

\clearpage
\section{Stellar mass versus redshift}\label{sec:app:b}
\setcounter{figure}{0}
\renewcommand{\thefigure}{B\arabic{figure}}

This appendix presents a complete set of mosaics illustrating the behavior of stellar mass as a function of the redshift. These mosaics provide supplementary evidence regarding the physical properties of the galaxies analyzed in this study (see Figures \ref{fig:b1}--\ref{fig:b7}).

\clearpage


\begin{figure*}
   \begin{center}

      \includegraphics[angle=0,width=9cm]{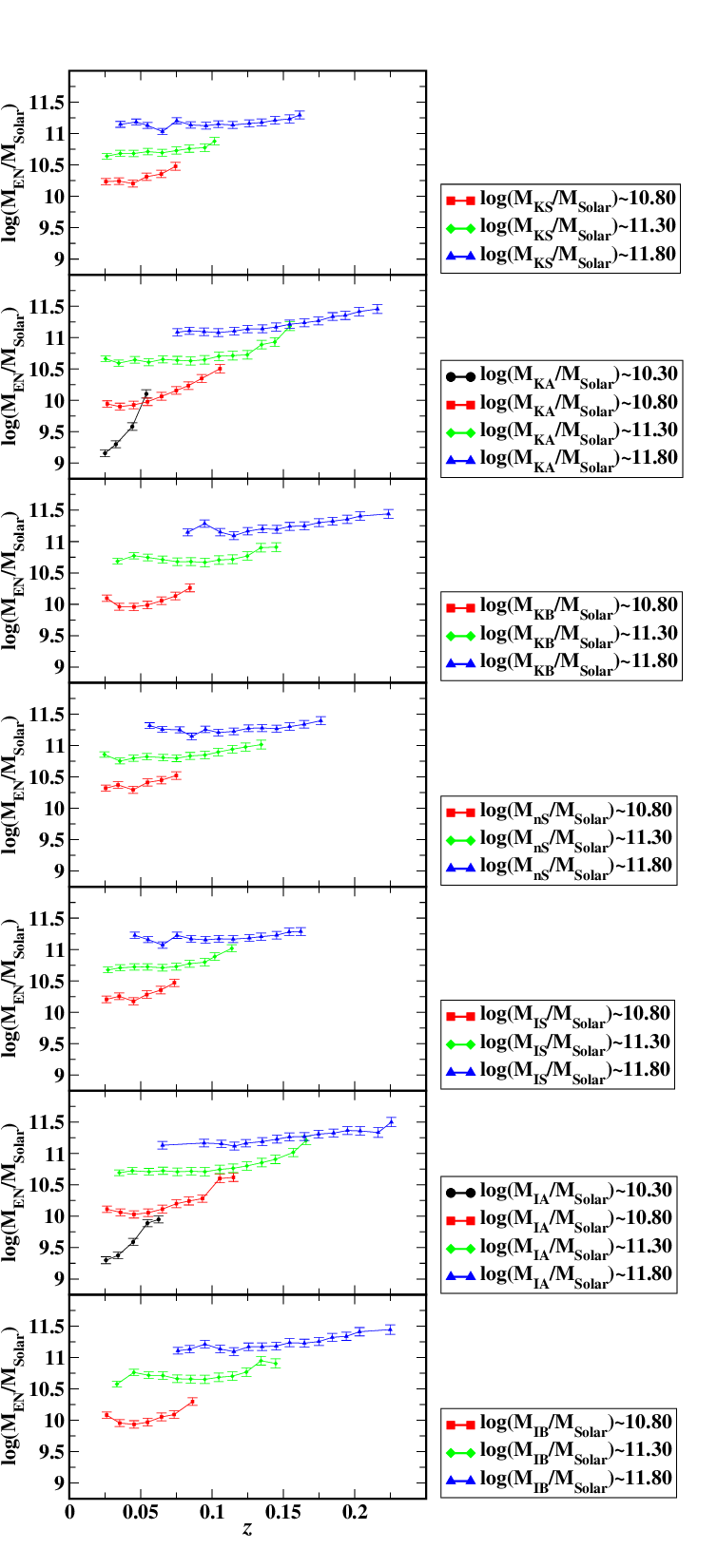}
      
         \caption{Behavior of stellar mass $({\bf M_{EN}})$ as function of the redshift for quasi-constant dynamical mass. Each color and symbol represents quasi-constant dynamical mass. The lower-left part of the graph (black dots) corresponds to $\log({\bf M_{Dyn}/{\bf M_{Solar}}})$ $\sim$ 10.30 while the upper-right part of the graph (blue triangles) corresponds to $\log({\bf M_{Dyn}/{\bf M_{Solar}}})$ $\sim$ 11.80. The difference in $\log({\bf M_{Dyn}/{\bf M_{Solar}}})$ between consecutive symbols is approximately 0.5. The mean uncertainty of the $\log({\bf M_{EN}/{\bf M_{Solar}}})$ is approximately 0.055 dex.}

         \label{fig:b1}
         \end{center}
   \end{figure*}

\clearpage


\begin{figure*}
   \begin{center}

      \includegraphics[angle=0,width=9cm]{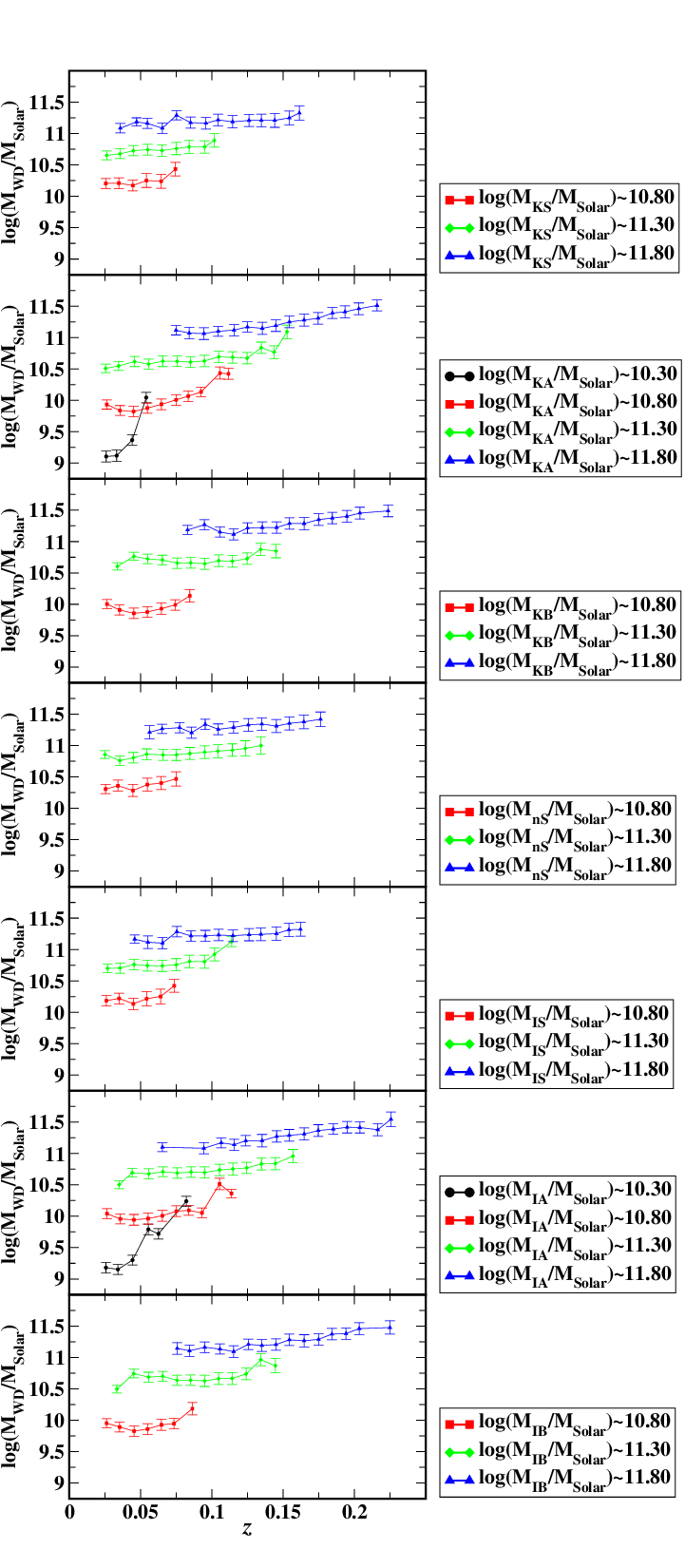}
      
         \caption{Behavior of stellar mass $({\bf M_{WD}})$ as function of the redshift for quasi-constant dynamical mass. Each color and symbol represents quasi-constant dynamical mass. The lower-left part of the graph (black dots) corresponds to $\log({\bf M_{Dyn}/{\bf M_{Solar}}})$ $\sim$ 10.30 while the upper-right part of the graph (blue triangles) corresponds to $\log({\bf M_{Dyn}/{\bf M_{Solar}}})$ $\sim$ 11.80. The difference in $\log({\bf M_{Dyn}/{\bf M_{Solar}}})$ between consecutive symbols is approximately 0.5. The mean error of the $\log({\bf M_{WD}/{\bf M_{Solar}}})$ is approximately 0.092 dex.}

         \label{fig:b2}
         \end{center}
   \end{figure*}

\clearpage


\begin{figure*}
   \begin{center}

      \includegraphics[angle=0,width=9cm]{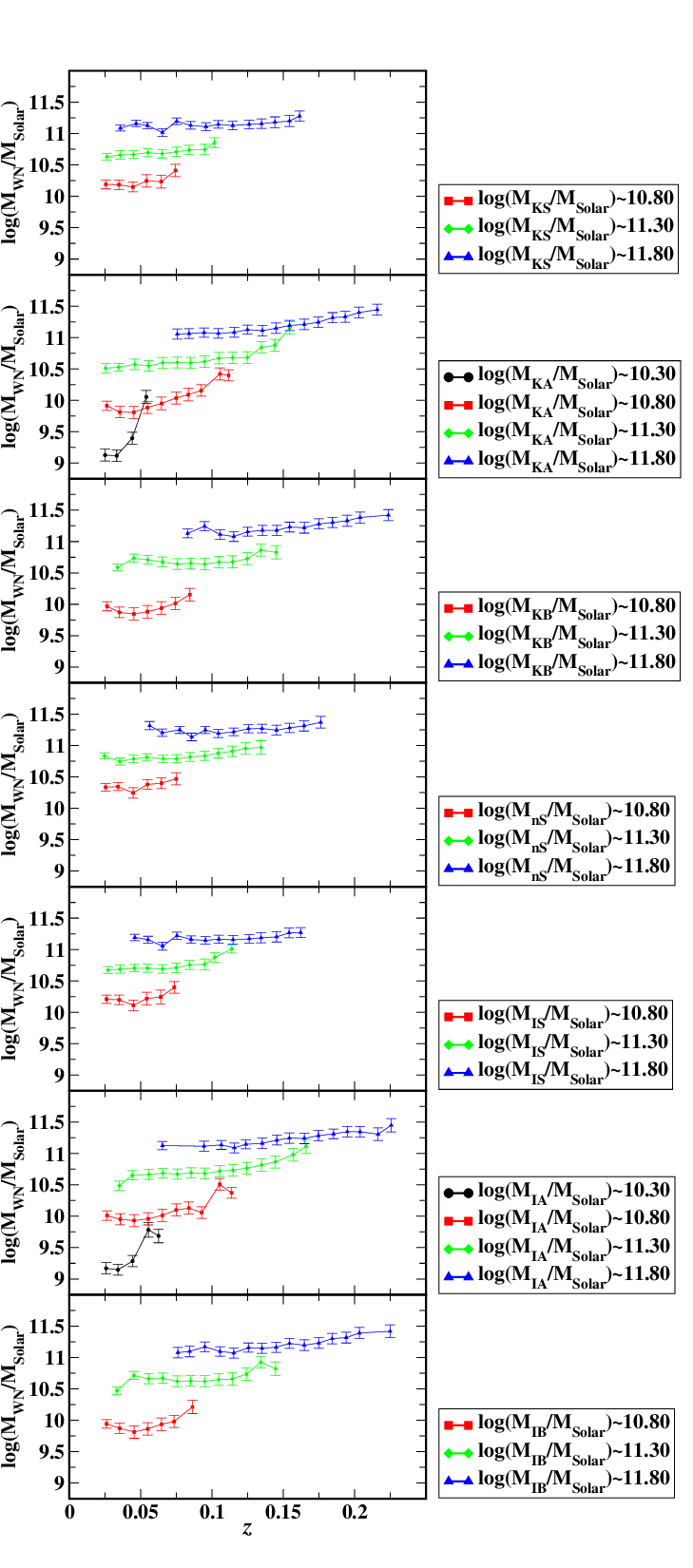}
      
         \caption{Behavior of stellar mass $({\bf M_{WN}})$ as function of the redshift for quasi-constant dynamical mass. Each color and symbol represents quasi-constant dynamical mass. The lower-left part of the graph (black dots) corresponds to $\log({\bf M_{Dyn}/{\bf M_{Solar}}})$ $\sim$ 10.30 while the upper-right part of the graph (blue triangles) corresponds to $\log({\bf M_{Dyn}/{\bf M_{Solar}}})$ $\sim$ 11.80. The difference in $\log({\bf M_{Dyn}/{\bf M_{Solar}}})$ between consecutive symbols is approximately 0.5. The mean uncertainty of the $\log({\bf M_{WN}/{\bf M_{Solar}}})$ is approximately 0.070 dex.}

         \label{fig:b3}
         \end{center}
   \end{figure*}

\clearpage


\begin{figure*}
   \begin{center}

      \includegraphics[angle=0,width=9cm]{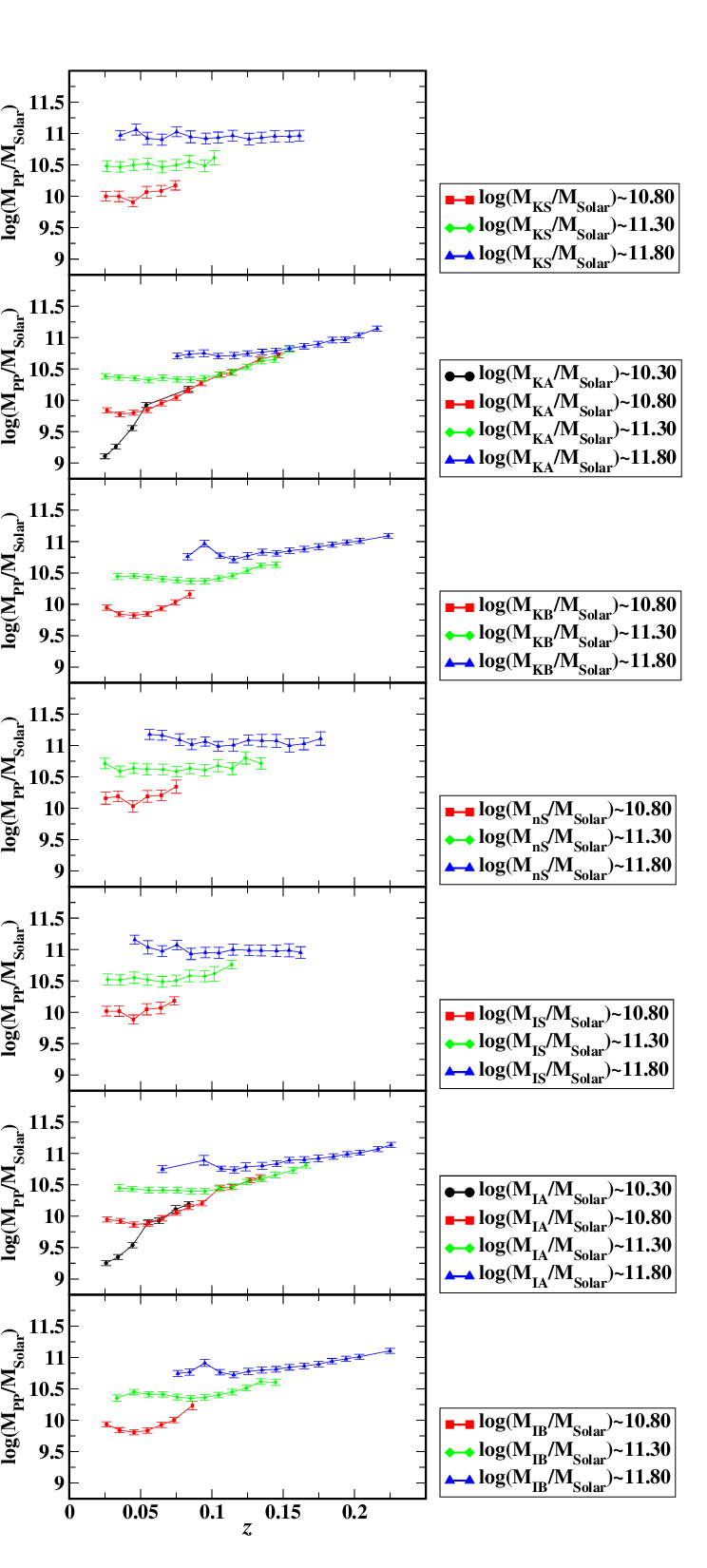}
      
         \caption{Behavior of stellar mass $({\bf M_{PP}})$ as function of the redshift for quasi-constant dynamical mass. Each color and symbol represents quasi-constant dynamical mass. The lower-left part of the graph (black dots) corresponds to $\log({\bf M_{Dyn}/{\bf M_{Solar}}})$ $\sim$ 10.30 while the upper-right part of the graph (blue triangles) corresponds to $\log({\bf M_{Dyn}/{\bf M_{Solar}}})$ $\sim$ 11.80. The difference in $\log({\bf M_{Dyn}/{\bf M_{Solar}}})$ between consecutive symbols is approximately 0.5. The mean error of the $\log({\bf M_{PP}/{\bf M_{Solar}}})$ is approximately 0.089 dex.}

         \label{fig:b4}
         \end{center}
   \end{figure*}

\clearpage


\begin{figure*}
   \begin{center}

      \includegraphics[angle=0,width=9cm]{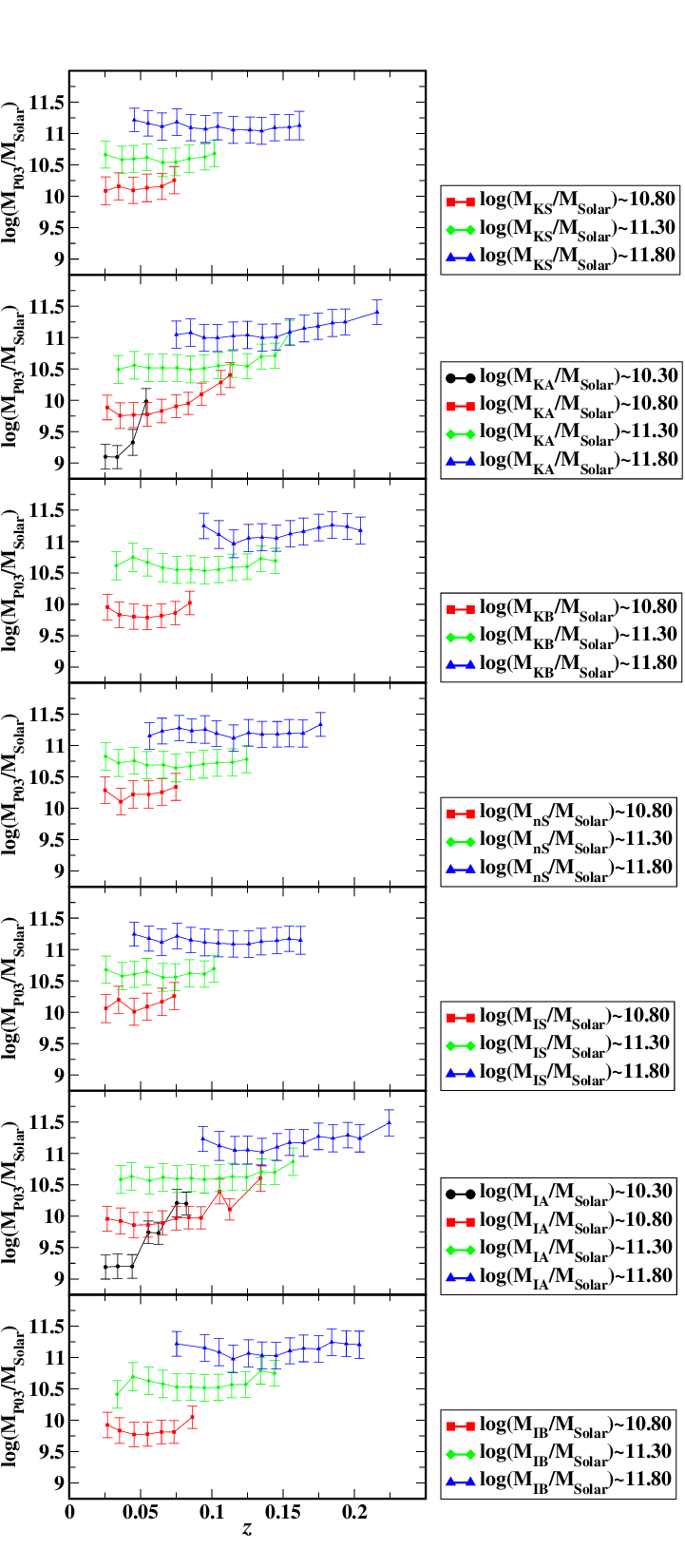}
      
         \caption{Behavior of stellar mass $({\bf M_{P03}})$ as function of the redshift for quasi-constant dynamical mass. Each color and symbol represents quasi-constant dynamical mass. The lower-left part of the graph (black dots) corresponds to $\log({\bf M_{Dyn}/{\bf M_{Solar}}})$ $\sim$ 10.30 while the upper-right part of the graph (blue triangles) corresponds to $\log({\bf M_{Dyn}/{\bf M_{Solar}}})$ $\sim$ 11.80. The difference in $\log({\bf M_{Dyn}/{\bf M_{Solar}}})$ between consecutive symbols is approximately 0.5. The mean uncertainty of the $\log({\bf M_{P03}/{\bf M_{Solar}}})$ is approximately 0.214 dex.}

         \label{fig:b5}
         \end{center}
   \end{figure*}

\clearpage


\begin{figure*}
   \begin{center}

      \includegraphics[angle=0,width=9cm]{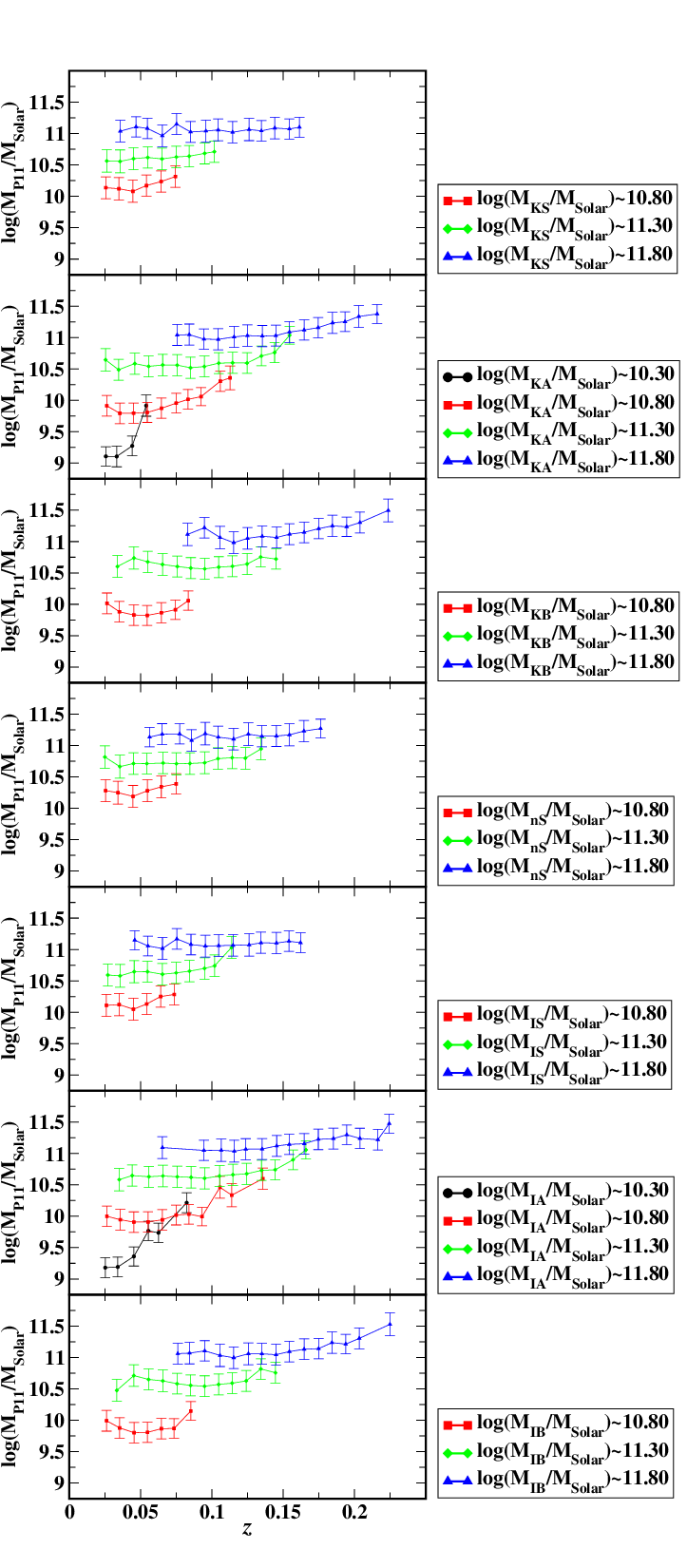}
      
         \caption{Behavior of stellar mass $({\bf M_{P11}})$ as function of the redshift for quasi-constant dynamical mass. Each color and symbol represents quasi-constant dynamical mass. The lower-left part of the graph (black dots) corresponds to $\log({\bf M_{Dyn}/{\bf M_{Solar}}})$ $\sim$ 10.30 while the upper-right part of the graph (blue triangles) corresponds to $\log({\bf M_{Dyn}/{\bf M_{Solar}}})$ $\sim$ 11.80. The difference in $\log({\bf M_{Dyn}/{\bf M_{Solar}}})$ between consecutive symbols is approximately 0.5. The mean error of the $\log({\bf M_{P11}/{\bf M_{Solar}}})$ is approximately 0.174 dex.}

         \label{fig:b6}
         \end{center}
   \end{figure*}

\clearpage


\begin{figure*}
   \begin{center}

      \includegraphics[angle=0,width=9cm]{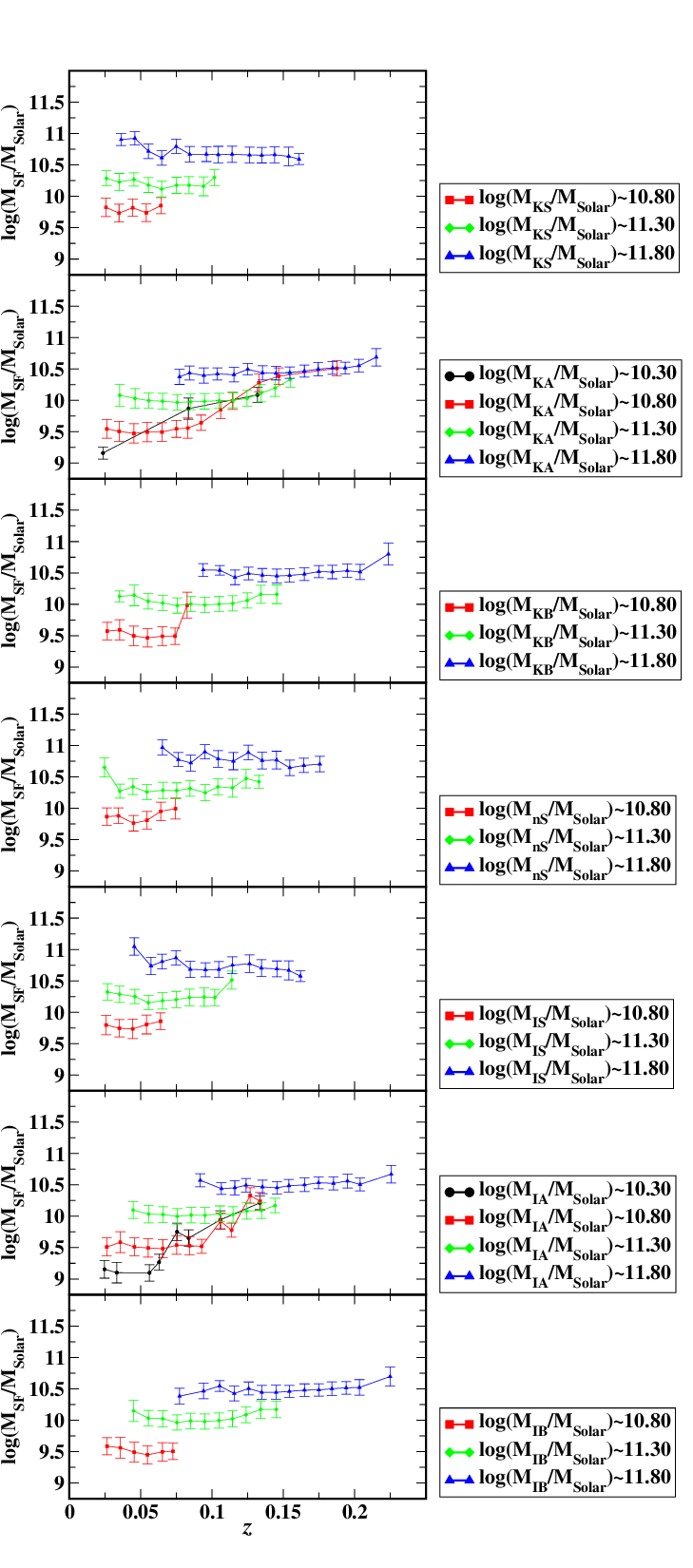}
      
         \caption{Behavior of stellar mass $({\bf M_{SF}})$ as function of the redshift for quasi-constant dynamical mass. Each color and symbol represents quasi-constant dynamical mass. The lower-left part of the graph (black dots) corresponds to $\log({\bf M_{Dyn}/{\bf M_{Solar}}})$ $\sim$ 10.30 while the upper-right part of the graph (blue triangles) corresponds to $\log({\bf M_{Dyn}/{\bf M_{Solar}}})$ $\sim$ 11.80. The difference in $\log({\bf M_{Dyn}/{\bf M_{Solar}}})$ between consecutive symbols is approximately 0.5. The mean uncertainty of the $\log({\bf M_{SF}/{\bf M_{Solar}}})$ is approximately 0.128 dex.}

         \label{fig:b7}
         \end{center}
   \end{figure*}

\clearpage

\section{Dynamical mass - stellar mass versus dynamical mass} 
\label{sec:app:c}
\setcounter{figure}{0}
\renewcommand{\thefigure}{C\arabic{figure}}

This appendix contains the complete set of mosaics for the logarithmic mass difference between dynamical and stellar mass ($\Delta \log \mathbf{M}$) as a function of the dynamical mass. These figures demonstrate the consistency of the observed evolutionary trends across different mass estimation models (see Figures \ref{fig:c1}--\ref{fig:c7}).

\begin{figure*}
   \begin{center}

      \includegraphics[angle=0,width=11cm]{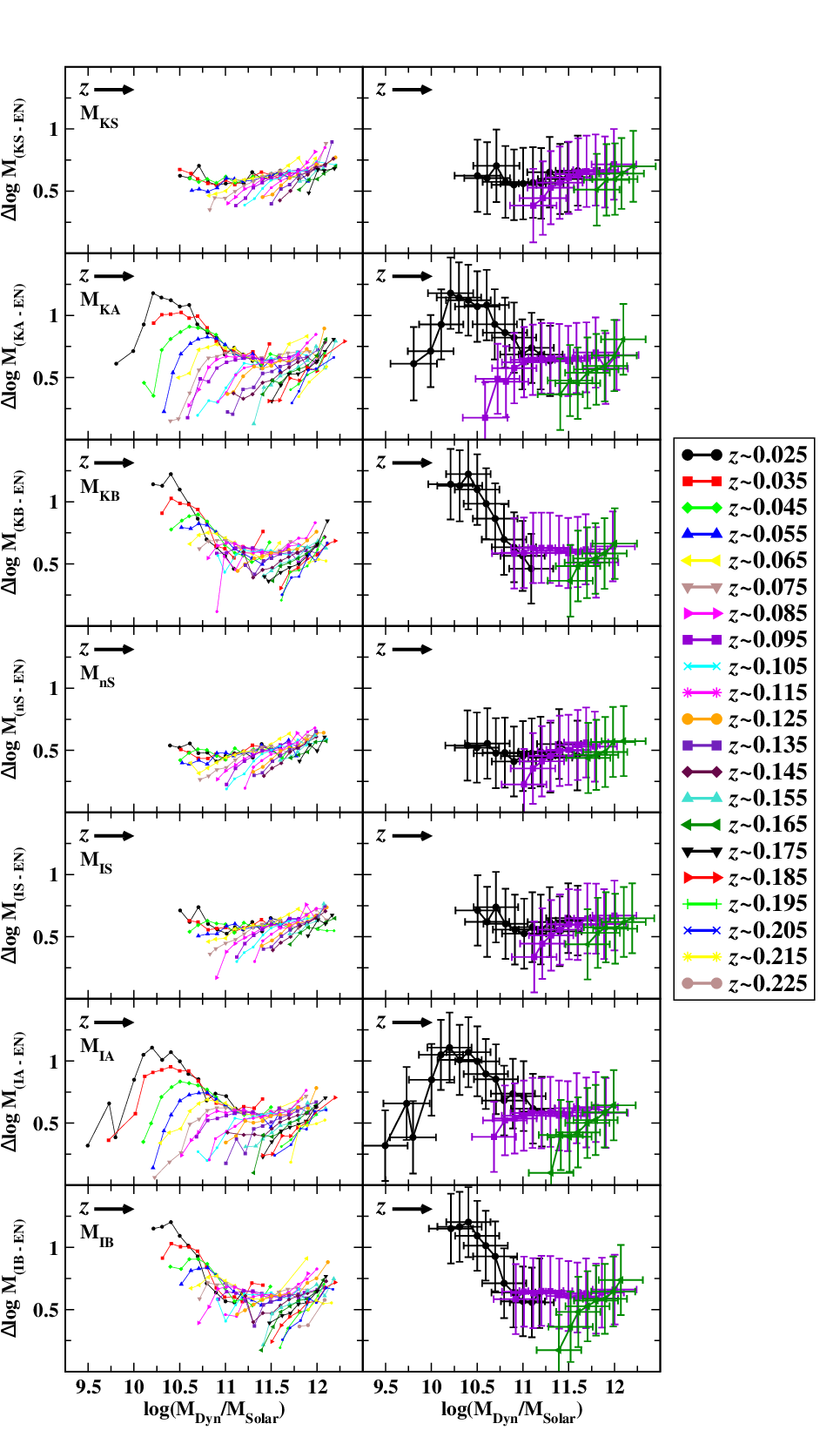}
      
         \caption{Difference between dynamical and stellar mass $({\bf M_{EN}})$ as function of the dynamical mass for the LTGs samples. Each color and symbol represents a quasi-constant redshift. The upper-left part of the graph (black dots) corresponds to $z\sim0.025$ while the lower-right part of the graph (dark green triangles) corresponds to $z\sim0.165$. The difference in redshift between consecutive symbols is approximately 0.01. The mean uncertainty of the difference between $\log({\bf M_{Dyn}/{\bf M_{Solar}}})$ and $\log({\bf M_{EN}/{\bf M_{Solar}}})$ is approximately 0.279 dex. The left column of the mosaic contains the full range of redshift, while the right column only contains three specific redshifts: low (black dots, $z \sim 0.025$), intermediate (purple squares, $z \sim 0.095$), and high (dark green triangles, $z \sim 0.165$), the latter with the aim of appreciating more clearly the differences in dynamical and stellar masses due to redshift and also to display the uncertainties for each point.}

         \label{fig:c1}
         \end{center}
   \end{figure*}

\clearpage


\begin{figure*}
   \begin{center}

      \includegraphics[angle=0,width=11cm]{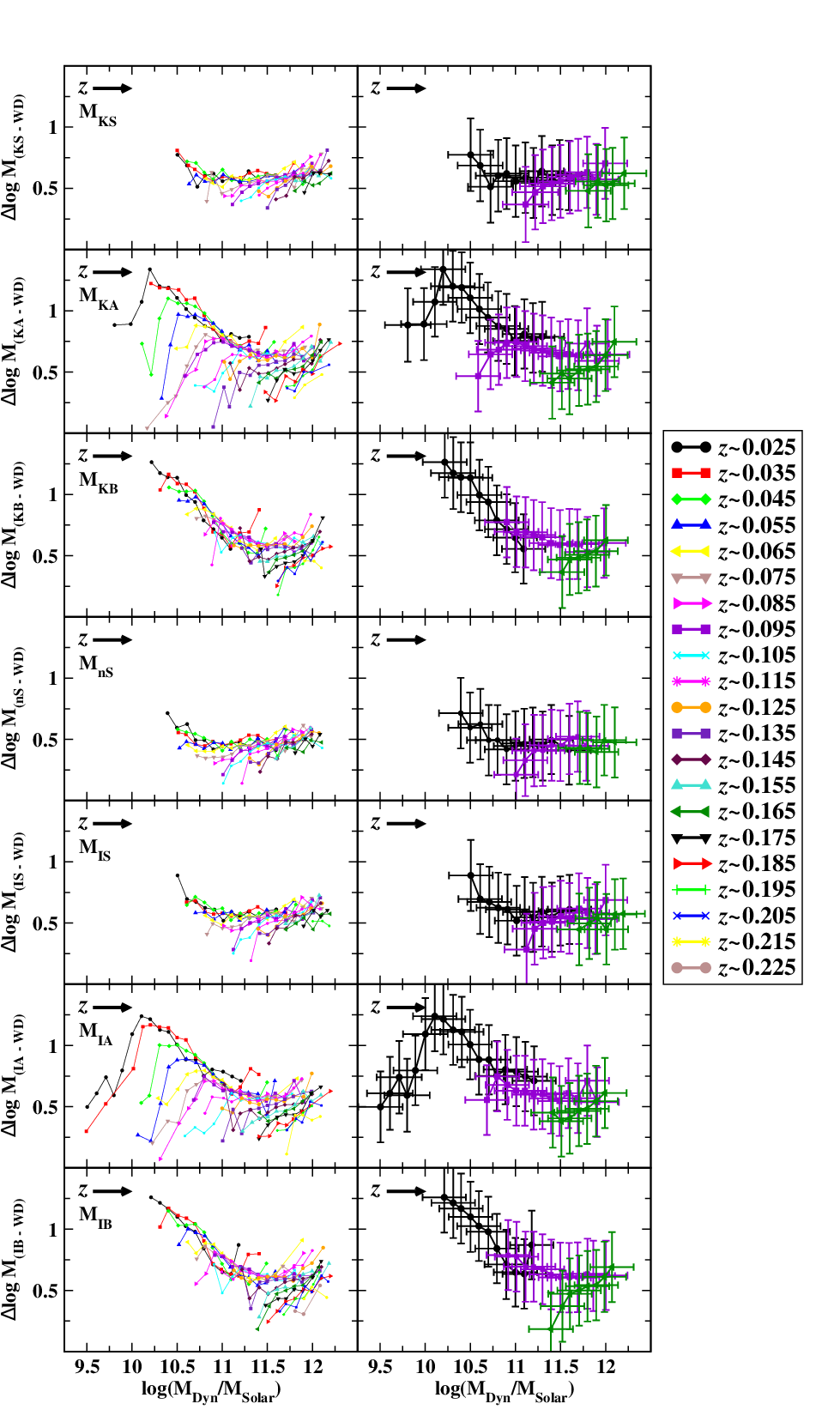}
      
         \caption{Difference between dynamical and stellar mass $({\bf M_{WD}})$ as function of dynamical mass for the LTGs samples. Each color and symbol represents a quasi-constant redshift. The upper-left part of the graph (black dots) corresponds to $z\sim0.025$ while the lower-right part of the graph (dark green triangles) corresponds to $z\sim0.165$. The difference in redshift between consecutive symbols is approximately 0.01. The mean uncertainty of the difference between $\log({\bf M_{Dyn}/{\bf M_{Solar}}})$ and $\log({\bf M_{WD}/{\bf M_{Solar}}})$ is approximately 0.285 dex. The left column of the mosaic contains the full range of redshift, while the right column only contains three specific redshifts: low (black dots $z \sim 0.025$), intermediate (purple squares $z \sim 0.095$), and high (dark green triangles $z \sim 0.165$), the latter with the aim of appreciating more clearly the differences in dynamical and stellar masses due to redshift and also to display the uncertainties for each point.}

         \label{fig:c2}
         \end{center}
   \end{figure*}

\clearpage


\begin{figure*}
   \begin{center}

      \includegraphics[angle=0,width=11cm]{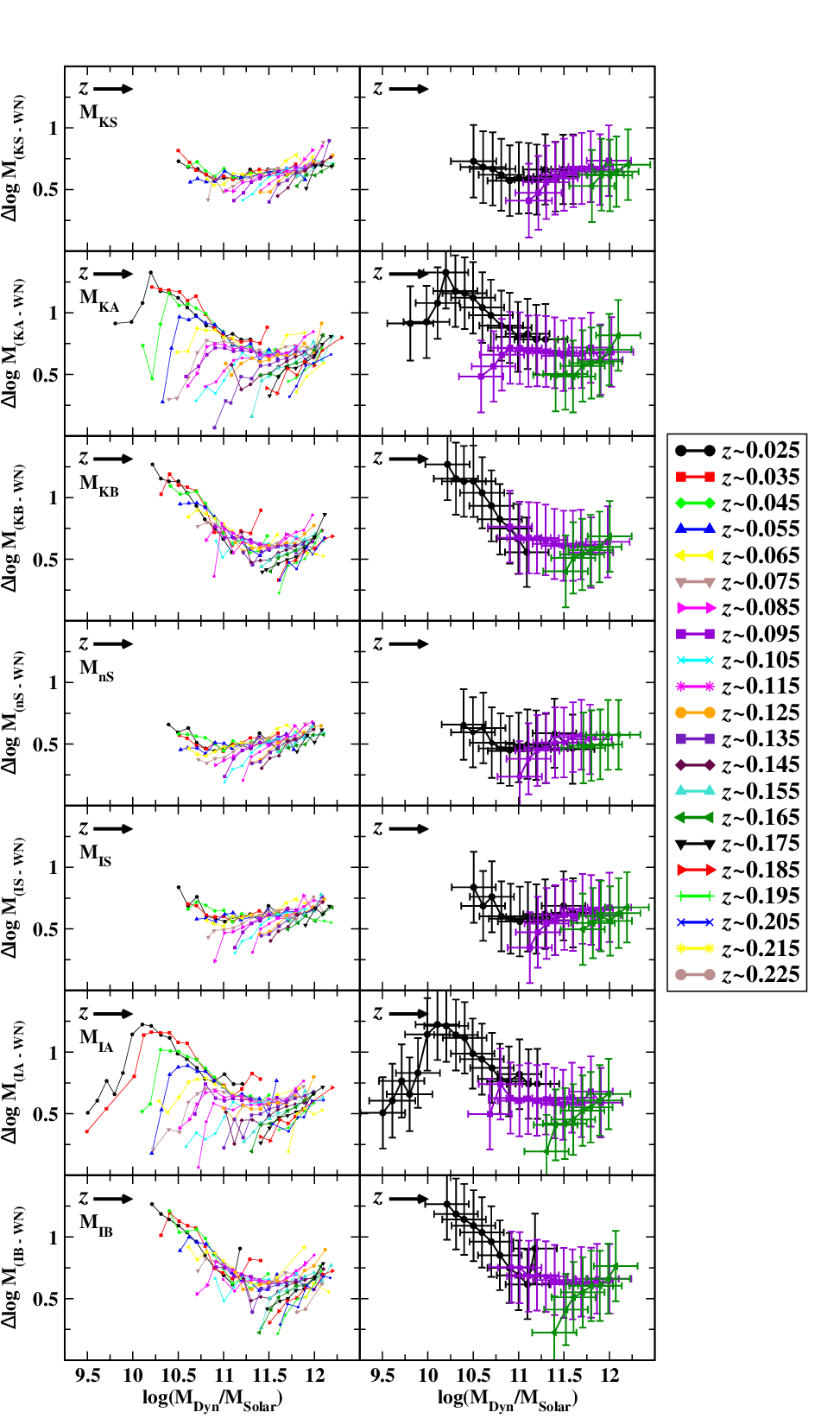}
      
         \caption{Difference between dynamical and stellar mass $({\bf M_{WN}})$ as function of dynamical mass for the LTGs samples. Each color and symbol represents a quasi-constant redshift. The upper-left part of the graph (black dots) corresponds to $z\sim0.025$ while the lower-right part of the graph (dark green triangles) corresponds to $z\sim0.165$. The difference in redshift between consecutive symbols is approximately 0.01. The mean uncertainty of the difference between $\log({\bf M_{Dyn}/{\bf M_{Solar}}})$ and $\log({\bf M_{WN}/{\bf M_{Solar}}})$ is approximately 0.281 dex. The left column of the mosaic contains the full range of redshift, while the right column only contains three specific redshifts: low (black dots $z \sim 0.025$), intermediate (purple squares $z \sim 0.095$), and high (dark green triangles $z \sim 0.165$), the latter with the aim of appreciating more clearly the differences in dynamical and stellar masses due to redshift and also to display the uncertainties for each point.}
         \label{fig:c3}
         \end{center}
   \end{figure*}

\clearpage


\begin{figure*}
   \begin{center}

      \includegraphics[angle=0,width=11cm]{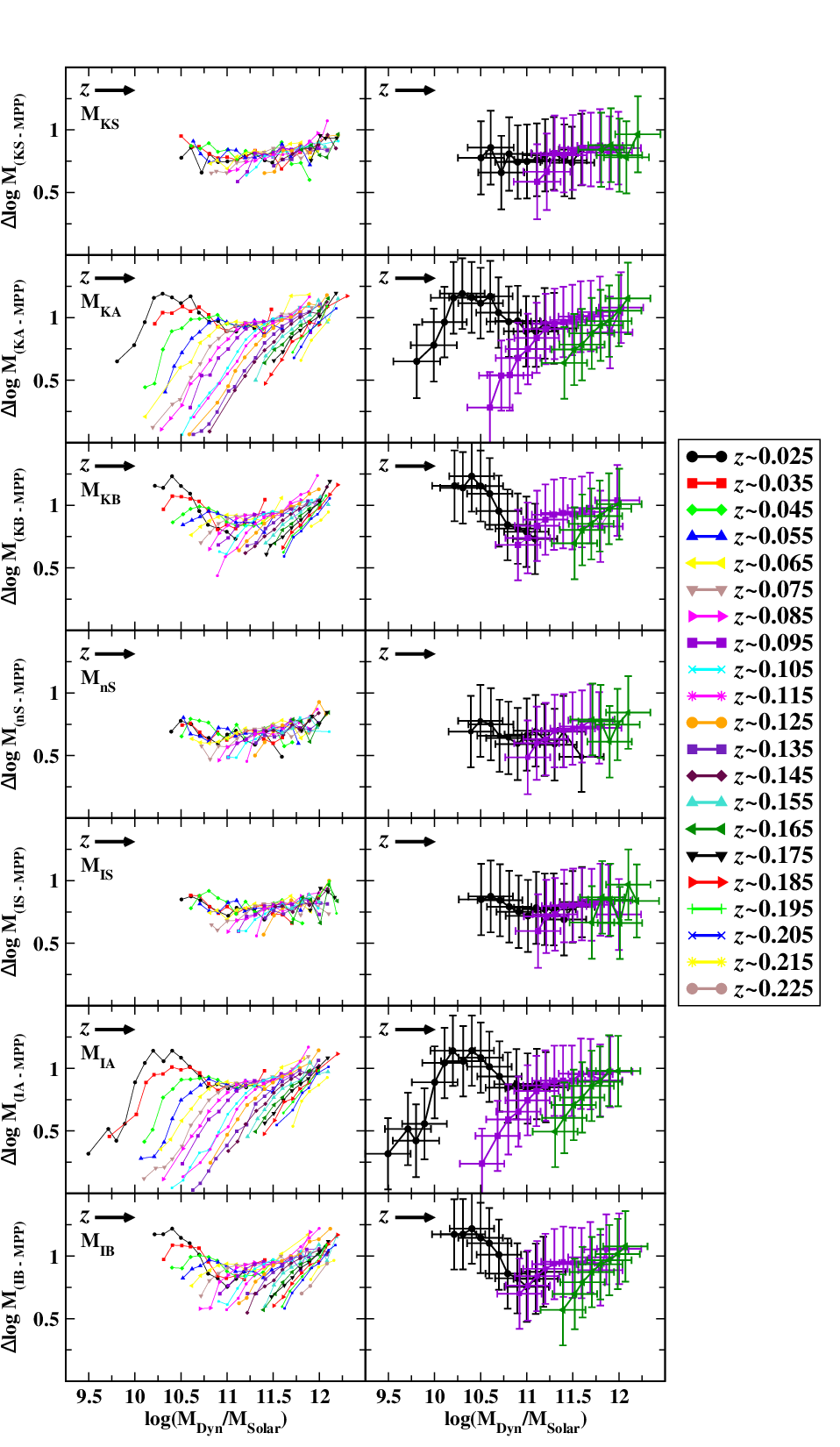}
      
         \caption{Difference between dynamical and stellar mass $({\bf M_{PP}})$as function of dynamical mass for the LTGs samples. Each color and symbol represents a quasi-constant redshift. The upper-left part of the graph (black dots) corresponds to $z\sim0.025$ while the lower-right part of the graph (dark green triangles) corresponds to $z\sim0.165$. The difference in redshift between consecutive symbols is approximately 0.01. The mean uncertainty of the difference between $\log({\bf M_{Dyn}/{\bf M_{Solar}}})$ and $\log({\bf M_{PP}/{\bf M_{Solar}}})$ is approximately 0.284 dex. The left column of the mosaic contains the full range of redshift, while the right column only contains three specific redshifts: low (black dots $z \sim 0.025$), intermediate (purple squares $z \sim 0.095$), and high (dark green triangles $z \sim 0.165$), the latter with the aim of appreciating more clearly the differences in dynamical and stellar masses due to redshift and also to display the uncertainties for each point.}

         \label{fig:c4}
         \end{center}
   \end{figure*}

\clearpage


\begin{figure*}
   \begin{center}

      \includegraphics[angle=0,width=11cm]{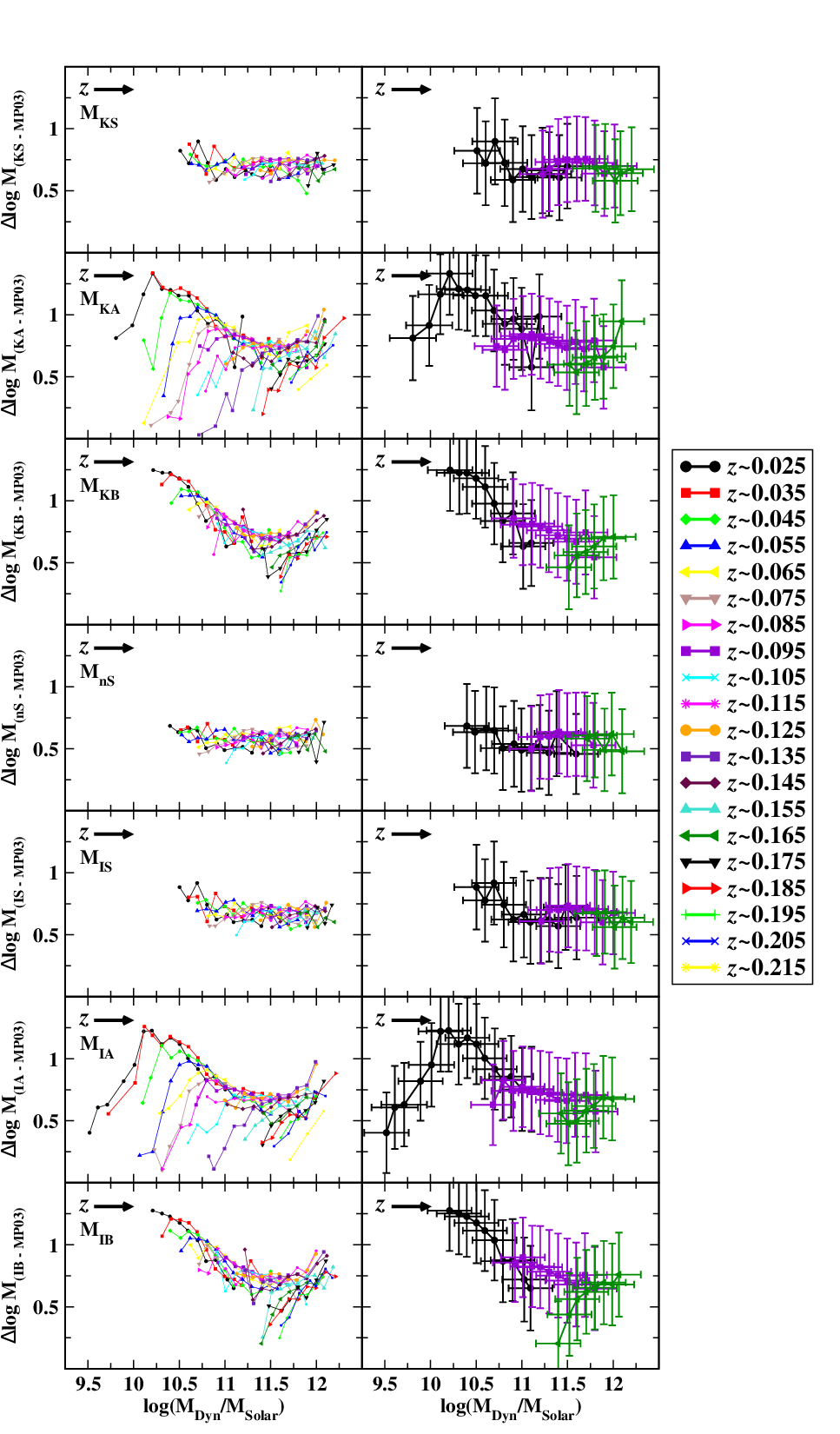}
      
         \caption{Difference between dynamical and stellar mass $({\bf M_{P03}})$as function of dynamical mass for the LTGs samples. Each color and symbol represents a quasi-constant redshift. The upper-left part of the graph (black dots) corresponds to $z\sim0.025$ while the lower-right part of the graph (dark green triangles) corresponds to $z\sim0.165$. The difference in redshift between consecutive symbols is approximately 0.01. The mean uncertainty of the difference between $\log({\bf M_{Dyn}/{\bf M_{Solar}}})$ and $\log({\bf M_{P03}/{\bf M_{Solar}}})$ is approximately 0.331 dex. The left column of the mosaic contains the full range of redshift, while the right column only contains three specific redshifts: low (black dots $z \sim 0.025$), intermediate (purple squares $z \sim 0.095$), and high (dark green triangles $z \sim 0.165$), the latter with the aim of appreciating more clearly the differences in dynamical and stellar masses due to redshift and also to display the uncertainties for each point.}
         \label{fig:c5}
         \end{center}
   \end{figure*}

\clearpage


\begin{figure*}
   \begin{center}

      \includegraphics[angle=0,width=11cm]{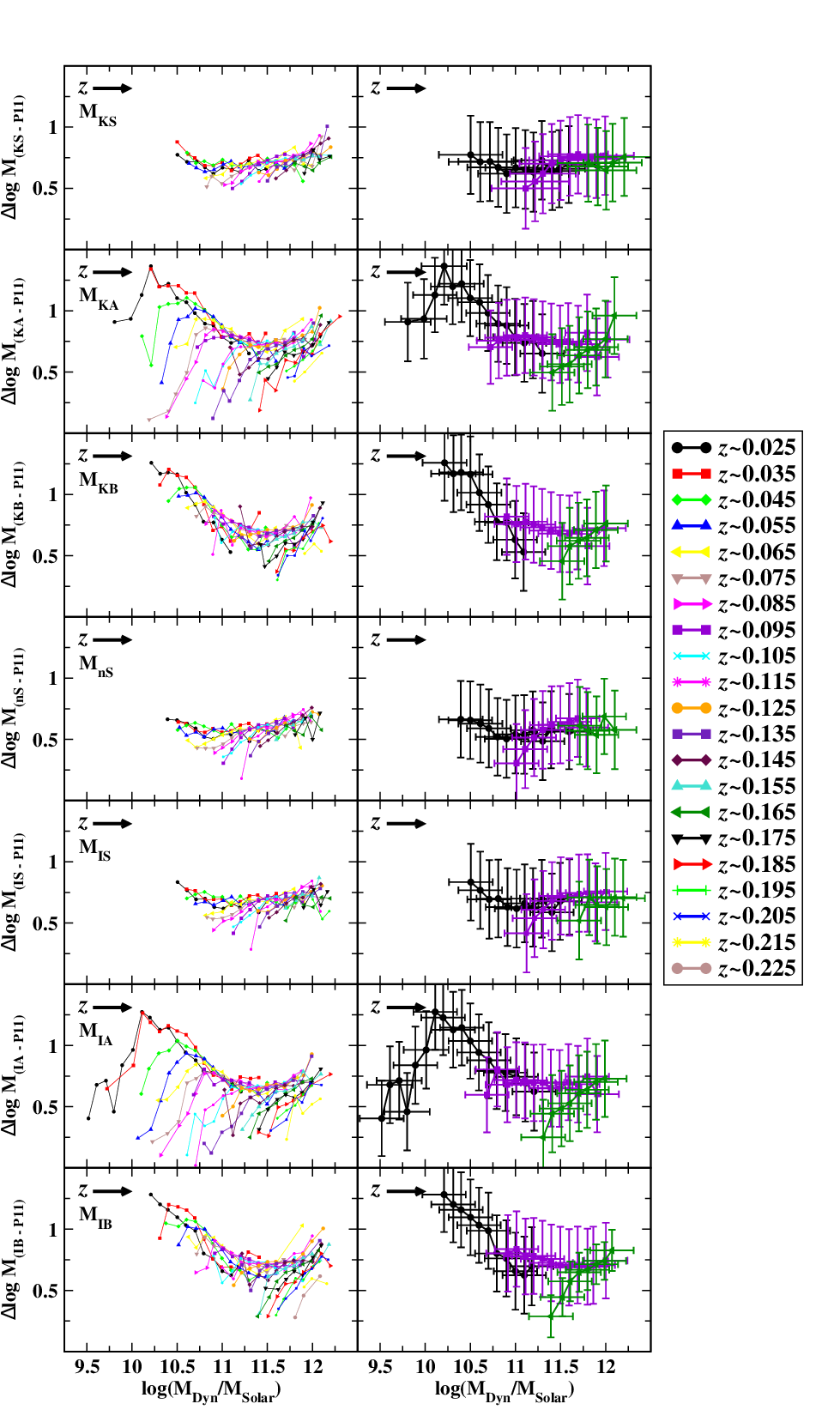}
      
         \caption{Difference between dynamical and stellar mass $({\bf M_{P11}})$as function of dynamical mass for the LTGs samples. Each color and symbol represents a quasi-constant redshift. The upper-left part of the graph (black dots) corresponds to $z\sim0.025$ while the lower-right part of the graph (green triangles) corresponds to $z\sim0.165$. The difference in redshift between consecutive symbols is approximately 0.01. The mean uncertainty of the difference between $\log({\bf M_{Dyn}/{\bf M_{Solar}}})$ and $\log({\bf M_{P11}/{\bf M_{Solar}}})$ is approximately 0.312 dex. The left column of the mosaic contains the full range of redshift, while the right column only contains three specific redshifts: low (black dots $z \sim 0.025$), intermediate (purple squares $z \sim 0.095$), and high (dark green triangles $z \sim 0.165$), the latter with the aim of appreciating more clearly the differences in dynamical and stellar masses due to redshift and also to display the uncertainties for each point.}

         \label{fig:c6}
         \end{center}
   \end{figure*}

\clearpage


\begin{figure*}
   \begin{center}

      \includegraphics[angle=0,width=11cm]{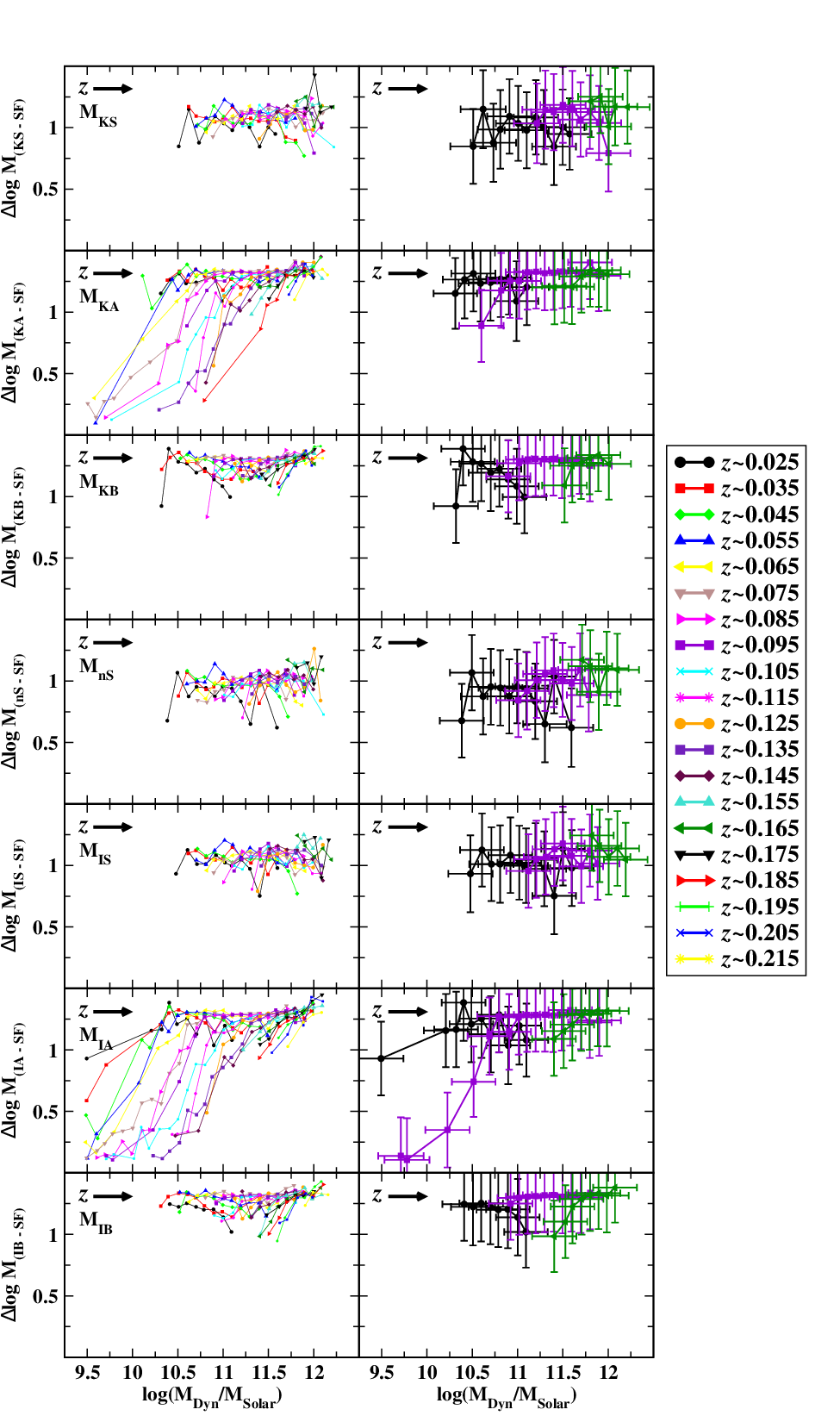}
      
         \caption{Difference between dynamical and stellar mass $({\bf M_{SF}})$as function of dynamical mass for the LTGs samples. Each color and symbol represents a quasi-constant redshift. The upper-left part of the graph (black dots) corresponds to $z\sim0.025$ while the lower-right part of the graph (green triangles) corresponds to $z\sim0.165$. The difference in redshift between consecutive symbols is approximately 0.01. The mean uncertainty of the difference between $\log({\bf M_{Dyn}/{\bf M_{Solar}}})$ and $\log({\bf M_{SF}/{\bf M_{Solar}}})$ is approximately 0.294 dex. The left column of the mosaic contains the full range of redshift, while the right column only contains three specific redshifts: low (black dots $z \sim 0.025$), intermediate (purple squares $z \sim 0.095$), and high (dark green triangles $z \sim 0.165$), the latter with the aim of appreciating more clearly the differences in dynamical and stellar masses due to redshift and also to display the uncertainties for each point.}
         \label{fig:c7}
         \end{center}
   \end{figure*}

\clearpage










\label{lastpage}
\end{document}